\documentclass[a4paper,10pt]{article}
\usepackage{amsfonts}
\usepackage{amsmath}
\usepackage{graphicx}
\usepackage{pict2e}
\usepackage{epsfig}
\usepackage[
    font      = footnotesize,
    labelfont = bf
]{caption}
\usepackage[
    font      = footnotesize,
    labelfont = bf
]{subcaption}

\usepackage[colorlinks=true]{hyperref}

\begin{document}
\newcommand{\ds}{\displaystyle}
\newcommand{\be}{\begin{equation}}
\newcommand{\ee}{\end{equation}}
\newcommand{\ba}{\begin{array}}
\newcommand{\ea}{\end{array}}
\newcommand{\bea}{\begin{eqnarray}}
\newcommand{\eea}{\end{eqnarray}}

\def\R{{\mathbb R}}
\def\C{{\mathbb C}} 
\newcommand{\bi}{\begin{itemize}}
\newcommand{\ei}{\end{itemize}}
\newcommand{\sn}{\mbox{sn}}
\newcommand{\cn}{\mbox{cn}}
\newcommand{\dn}{\mbox{dn}}
\newcommand{\x}{{\ensuremath{\times}}}
\newcommand{\bb}[1]{\makebox[16pt]{{\bf#1}}}
\newtheorem{theorem}{Theorem}
\newtheorem{definition}{Definition}
\newtheorem{lemma}{Lemma}
\newtheorem{comment}{Comment}
\newtheorem{corollary}{Corollary}
\newtheorem{example}{Example}
\newtheorem{examples}{Examples}
 \newtheorem{conjecture}{Conjecture}

\author{Ernest G. Kalnins\\
{\sl Department of Mathematics,
University
of Waikato,}\\
{\sl Hamilton, New Zealand}\\
W.\ Miller, Jr.
\\
{\sl School of Mathematics, University of Minnesota,}\\
{\sl Minneapolis, Minnesota,
55455, U.S.A.}\\
{\sl miller@ima.umn.edu}\\
and  
Eyal Subag\\
{\sl Department of Mathematics,
Pennsylvania State University,} \\
{\sl State College, Pennsylvania, 16802, U.S.A.}}
\date{\today}
\title{B\^ocher contractions of conformally superintegrable Laplace equations: Detailed computations}

\maketitle

\abstract{These supplementary notes in the ArXiv are a companion to our paper ``B\^ocher contractions of conformally superintegrable Laplace equations", http://arxiv.org/abs/1512.09315.
They contain background material and the details of some of the extensive computations that couldn't be put in the paper, due to space limitations.}
\tableofcontents
\newpage

\section{2D conformal superintegrability of the 2nd order}

Systems of  Laplace type are of the form 
 \be\label{Laplace} H\Psi\equiv \Delta_n\Psi+V\Psi=0.\ee
 Here $\Delta_n $ is the Laplace-Beltrami operator on a real or complex conformally flat $nD$ Riemannian or pseudo-Riemannian manifold. We assume that all functions occurring in this paper are 
 locally analytic, real or complex.)
A conformal symmetry of this equation is a partial differential operator  $ S$ in the variables ${\bf x}=(x_1,\cdots,x_n)$  such 
that $[ S, H]\equiv SH-HS=R_{ S} H$ for some differential operator  $R_{S}$.
A conformal symmetry maps any solution $\Psi$ of (\ref{Laplace}) to another solution. Two conformal symmetries ${ S}, { S}'$ are 
identified if $S=S'+RH$ for some differential operator $R$, since they agree on the solution space of (\ref{Laplace}). (For short we will say 
that $S=S', \mod (H)$ 
and that $S$ is a symmetry if $[S,H]=0,\mod(H)$.)
The system is {\it conformally superintegrable} for $n>2$  if there are $2n-1$ functionally independent conformal symmetries, 
${ S}_1,\cdots,{ S}_{2n-1}$ with ${ S}_1={ H}$. It is second order conformally superintegrable if each  
symmetry $S_i$ can be chosen to be a  differential operator of at most second order.

For $n=2$ the definition must be restricted, since for a  potential $V=0$ there will be an infinite dimensional space of conformal
symmetries when $n=2$; every analytic function induces such symmetries.
\begin{comment}
Indeed  necessary and sufficient  conditions that $S=u(x,y)\partial_x+v(x,y)\partial_y$ is a 1st order conformal symmetry  for $H=\Delta_2$ are  that $u$ and $v$ satisfy the Cauchy-Riemann equations  
\[ \partial_x u=\partial_y v,\   \partial_y u=-\partial_x v.\] \end{comment}

 However, in this paper we are interested in multiparameter Laplace equations, i.e.,
those with potentials of the form $V=\sum_{j=0}^sc_jV^{(j)}$ where the set  $\{V^{(j)}\}$ is linearly independent, $V^{(0)}=1$ and the 
$c_j$ are arbitrary parameters. Thus we require that each symmetry be conformal for arbitrary choice of parameters $c_j$ and, in 
particular for the special case $V_0=c_0$ where $c_0$ is arbitrary. With this restriction we say that a 2D multiparameter Laplace equation 
is superintegrable if it admits 3 algebraically independent symmetries.

Every $2D$ Riemannian manifold is conformally flat, so we can  always find a Cartesian-like coordinate system with coordinates ${\bf x}=(x,y)\equiv (x_1,x_2)$ 
such that the Laplace equation takes the form
\be\label{Laplace4} {\tilde H}=\frac{1}{\lambda(x,y)}(\partial_x^2+\partial_y^2)+{\tilde V}({\bf x})=0.\ee
However, this equation is equivalent to the flat space equation
\be\label{Laplace5}{ H}\equiv \partial_x^2+\partial_y^2+ V({\bf x})=0,\quad V({\bf x})=\lambda({\bf x}){\tilde V}({\bf x}).\ee
In particular, the conformal symmetries of (\ref{Laplace4}) are identical with the conformal symmetries of (\ref{Laplace5}).
Indeed, denoting by $\Lambda$ the operator of multiplication by the function $\lambda(x,y)$ and using the operator identity $[A,BC]=B[A,C]+[A,B]C$ we have
\[ [S,H]=[S,\Lambda{\tilde H}] =\Lambda[S,{\tilde H}]+[S,\Lambda]{\tilde H}=
\Lambda R{\tilde H}+[S,\Lambda]{\tilde H}=(\Lambda R \Lambda^{-1}+[S,\Lambda]\Lambda^{-1})H.\]
Thus without loss of generality  we can assume the manifold is flat space with $\lambda\equiv 1$. 

Since the Hamiltonians are formally self-adjoint, without loss of generality we can always assume that a 2nd order conformal symmetry $S$ is formally self-adjoint and that a 1st 
order conformal symmetry $L$ is skew-adjoint:
 \be
{ S}=\frac{1}{\lambda}\sum ^2_{k,j=1}\partial_k\cdot (\lambda a^{kj}({\bf x}))\partial_j +W({\bf x})\equiv
S_0+W,\quad a^{jk}=a^{kj} \label{2ndordersymm}
\ee
\be L=\sum_{k=1}^2\left(a^k({\bf x})\partial_k+\frac{\partial_k(\lambda a^k)}{2\lambda}\right).\label{1stordersymm}\ee

\be\label{confsym2}
[S,H]=(R^{(1)}({\bf x})\partial_x+(R^{(2)}({\bf x})\partial_y)H,
\ee
\be\label{confsym1}
[L,H]=R({\bf x}))H,
\ee
for some functions $R^{(j)}({\bf x}), R({\bf x})$.

Equating coefficients of the partial derivatives on both sides of (\ref{confsym2}), we obtain the conditions

\bea\label{killingtensors}
a_i^{ii}&=&2a_j^{ij}+a_i^{jj}, i\ne j
\eea
and
\be
W_j=\sum_{s=1}^2a^{sj}
V_s+a_j^{jj}V,\quad k,j=1,2
.\label{potc}
\ee
(Here a subscript $j$ on $a^{\ell m}$, $V$ or $W$ denotes differentiation with respect to $x_j$.)
The requirement that $\partial_{x} W_2=\partial_{y}W_1$ leads from
(\ref{potc}) to the
second order (conformal) Bertrand-Darboux  partial differential equations for the potential:
\[a^{12}(V_{11}-V_{22})+(a^{22}-a^{11})V_{12}+(a^{12}_1+a^{22}_2-a^{11}_2)V_1+(a^{22}_1-a^{11}_1-a^{12}_2)V_2\]
\be\label{BertrandDarboux}
+2A^{12}_{12}V=0.
\ee
Furthermore, we can always add the trivial conformal symmetry $\rho({\bf x})H$ to $S$.

Equating coefficients of the partial derivatives on both sides of (\ref{confsym1}), we obtain the conditions

\bea\label{killingvectors}
a^2_1+a^1_2=0; \frac{R({\bf x})}{2}=a^1_1=a^2_2,\ 2a^1_1V+a^1V_1+a^2V_2=0.
.\label{potc1}
\eea

In general the spaces of 1st and 2nd order symmetries could be infinite dimensional. However, the requirement that $H$ have a multiparameter potential 
reduces the possible symmetries to a finite dimensional space. Indeed each such symmetry must necessarily be a symmetry for the potential $V=c_0$ where $c_o$ is an 
arbitrary parameter. Thus the conformal Bertrand-Darboux condition for a 2nd order symmetry yields the requirement $\partial_{xy}(a^{11}-a^{22})=0$. 
Furthermore we can always assume, 
say, $a^{11}=0$. The result is that  the pure derivative terms $S_0$  belong to the space 
spanned by symmetrized products of the conformal Killing vectors
\be\label{conformalKV} P_1=\partial_x,\ P_2=\partial_y,\  J=x_1\partial_y-y\partial_x,\
D=x\partial_x+y\partial_y,\ee
\[K_1=(x^2-y^2)\partial_x +2xy\partial_y,\ K_2=(y^2-x^2)\partial_y+2xy\partial_x.\ 
\]
and terms $g({\bf x})(\partial_x^2+\partial_y^2))$ where $g$ is an
arbitrary function.   For a given multiparameter potential only a subspace   of these conformal tensors occurs.  
This is for two reasons. 
First the conformal Bertrand-Darboux equations restrict the allowed Killing tensors.
Second, on the hypersurface ${\cal H}=0$ in phase space all symmetries $g({\bf x}){\cal H}$ vanish, so any 
two symmetries  differing by $g({\bf x}){\cal H}$ can be identified.

Similarly the requirement that a 1st order conformal symmetry $L$ be a symmetry for the potential $V=c_0$ leads to the requirements
$a^1_x=a^2_y=R=0$ so, in particular, $L$ is a true (not just conformal) symmetry. Therefore its pure derivative part must be a linear combination of the 
Euclidean Killing vectors $\partial_x,\ \partial_y,\ x\partial_y-y\partial_x$.

The following results are easy modifications of results for 3D conformal superintegrable systems proved in \cite{KKMP2011}.
We give them for completeness. For a conformal superintegrable system with 3 2nd order symmetries there will be 2 independent conformal Bertrand-Darboux equations 
(the equation for the symmetry $H$ is trivial) and the assumption of algebraic independence means that we can solve these equations for 
$V_{22}$ and $V_{12}$:

\be\label{veqn1a}\ba{lllll}
V_{22}&=&V_{11}&+&A^{22}V_1+B^{22}V_2+C^{22}V,\\
V_{12}&=& && A^{12}V_1+B^{12}V_2+C^{12}V\\
\ea\ee
 Here the $A^{ij},B^{ij},C^{ij}$ are
functions of $\bf x$ that can be calculated explicitly. 
Indeed if ${\cal S}_1=\sum_{k,j=1}^2\partial_k\cdot (\ell^{kj}(x,y))\partial_j)+W^{(1)}(x,y)$, ${\cal S}_2=\sum_{k,j=1}^2\partial_k\cdot(b^{kj}(x,y)
\partial_j)+W^{(2)}(x,y)$, ${\cal H}$, is
 a basis for the symmetries then
\be\label{canoneqns1} A^{12}=\frac{D_{(2)}}{D},\quad A^{22}=\frac{D_{(3)}}{D},\quad B^{12}=-\frac{D_{(0)}}{D},\quad B^{22}=-\frac{D_{(1)}}{D},\ee
\be\label{canoneqns2} C^{12}=-\frac{D_{(5)}}{D},\quad C^{22}=-\frac{D_{(4)}}{D},\ee
\[ D=\det \left(\ba{cc} \ell^{11}-\ell^{22},& \ell^{12}\\ b^{11}-b^{22},& b^{12}\ea\right),\quad D_{(0)}=\det \left(\ba{cc} 3\ell^{12}_2,& -\ell^{12}\\ 3b^{12}_2,& -b^{12}
\ea\right), \]
\[ D_{(1)}=\det \left(\ba{cc} 3\ell^{12}_2,& \ell^{11}-\ell^{22}\\ 3b^{12}_2,& b^{11}-b^{22}\ea\right),\quad D_{(2)}=\det \left(\ba{cc} 3\ell^{12}_1,& \ell^{12}\\
3b^{12}_1,& b^{12}\ea\right), \]
\[ 
D_{(3)}=\det \left(\ba{cc} 3\ell^{12}_1,& \ell^{11}-\ell^{22}\\ 3b^{12}_1,& b^{11}-b^{22}\ea\right),\]
\[ D_{(4)}=\det \left(\ba{cc} 2\ell^{12}_{12},& \ell^{11}-\ell^{22}\\ 2b^{12}_{12},& b^{11}-b^{22}\ea\right),\ D_{(5)}=
\det \left(\ba{cc} 2\ell^{12}_{12},& -\ell^{12}\\ 2b^{12}_{12},& -b^{12}
\ea\right).\]
The functions $A^{22},B^{22},A^{12},B^{12},C^{22},C^{12}$ are defined independent of the choice of basis for the 2nd order symmetries.

\subsection{The integrability conditions for the potential}
To determine the integrability conditions for the system (\ref{veqn1a}) we first  introduce the dependent variables
$Z^{(0)}=V$, $Z^{(1)}=V_1$, $Z^{(2)}=V_2$, 
$Z^{(3)}=V_{11}$,  the vector
\be\label{wvector1a}{\bf z}^{\rm tr}=( Z^{(0)},Z^{(1)},Z^{(2)}, Z^{(3)}),
\ee
and the matrices
\be
{\bf A}^{(1)}=\left(\ba{rrrr} 0&1&0&0\\ 0&0&0&1\\ C^{12}&A^{12}&B^{12}&0\\  C^{13}&A^{13}&B^{13}&B^{12}-A^{22}\ea\right),
\ee
\be
{\bf A}^{(2)}=\left(\ba{rrrr} 0&0&1&0\\ C^{12}&A^{12}&B^{12}&0 \\ C^{22}&A^{22}&B^{22}&1 \\ C^{23}&A^{23}&B^{23}&A^{12}\ea\right),
\ee
where 
\bea
 A^{13}&=&A^{12}_2-A^{22}_1+B^{12}A^{22}+A^{12}A^{12}-B^{22}A^{12}-C^{22},\nonumber\\
%&=&A^{13}_3-A^{33}_1+B^{13}A^{23}+A^{13}A^{13}-B^{33}A^{12}-C^{33}A^{13}+C^{13}A^{33}-D^{33},\nonumber\\
B^{13}&=&B^{12}_2-B^{22}_1+A^{12}B^{12}+C^{12},\nonumber\\
%&=&B^{13}_3-B^{33}_1+B^{13}B^{23}+A^{13}B^{13}-B^{33}B^{12}-C^{33}B^{13}+C^{13}B^{33}+D^{13},\nonumber\\
 C^{13}&=&C^{12}_2-C^{22}_1+A^{12}C^{12}-B^{22}C^{12}+B^{12}C^{22},\nonumber\\
%&=&C^{13}_3-C^{33}_1+B^{13}C^{23}+A^{13}C^{13}-B^{23}C^{12}-C^{33}C^{13}+C^{13}C^{33},\nonumber\\
 A^{23}&=& A^{12}_1+B^{12}A^{12}+C^{12},\quad  B^{23}=B^{12}_1+B^{12}B^{12},\nonumber\\
 \quad C^{23}&=&B^{12}C^{12}+C^{12}_1.\nonumber\\
 \eea
Then  the integrability conditions for the system
\be\label{int21}
\partial_{x_j}{\bf z}={\bf A}^{(j )}{\bf z}\qquad  j=1,2,
\ee
must hold. They are
\be\label{int31}
{\bf A}^{(j)}_i-{\bf A}^{(i)}_j={\bf A}^{(i)}{\bf A}^{(j)}-{\bf A}^{(j)}{\bf A}^{(i)}\equiv [{\bf A}^{(i)},{\bf A}^{(j)}].
\ee

Suppose the integrability conditions for system  (\ref{veqn1a})  are satisfied identically. 
In this case we say that the potential is {\it nondegenerate}. Otherwise the potential is {\it degenerate}. 
If $V$ is nondegenerate then at any point ${\bf x}_0$, where the $A^{ij}, B^{ij}, C^{ij}$ are 
defined and analytic, there is a unique solution $V({\bf x})$ with
arbitrarily prescribed values of $V({\bf x}_0)$, $V_1({\bf x}_0)$, $V_2({\bf x}_0)$,
 $V_{11}({\bf x}_0)$. The points ${\bf x}_0$ are called {\it regular}. 
The points of singularity for the $A^{ij},B^{ij},C^{ij}, D^{ij}$ form a
manifold of dimension $<2$. Degenerate potentials depend on fewer
parameters. (For example, we could have 
 that the integrability conditions are not satisfied identically. 
 Or a first order conformal symmetry might exist and this would imply a linear condition on the first derivatives of $V$ alone.) 
 
 Note that for a nondegenerate potential the solution space of (\ref{veqn1a}) is exactly 4-dimensional, i.e.
 the potential depends on 4 parameters. Degenerate potentials depend on $<$ 4 parameters. Note also that the 
 integrability conditions depend only on the free parts 
 $\ell^{jk},b^{jk}$ of the conformal symmetry basis, not on the potential terms $V,W^{(1)},W^{(2)}$.
 If the integrability conditions are satisfied identically, 
 then  the equations for the potential terms can be solved.

\subsection{The conformal St\"ackel transform}
We quickly review the concept of the   St\"ackel transform \cite{KMP2010} and extend it to conformally superintegrable systems.
Suppose we have a second order {\it conformal}  superintegrable system 
\be\label{confl} { H}=\frac{1}{\lambda(x,y)}(\partial_{xx}+\partial_{yy})+V(x,y)=0,\quad  { H}={ H}_0+V.
\ee
and suppose $U(x,y) $ is a particular solution of equations (\ref{veqn1a}), nonzero in an open set.   The {\it conformal St\"ackel 
transform} of (\ref{confl}), induced by $U$, is the (Helmholtz) system  \be\label{helms} {\tilde {  H}}=E,\quad 
{\tilde { H}}=\frac{1}{{\tilde \lambda}}(\partial_{xx}+\partial_{yy})+{\tilde V},\quad {\tilde \lambda}=\lambda U,\  {\tilde V}=\frac{V}{U}
\ee
\begin{theorem}\label{stackelt}
The transformed (Helmholtz) system  (\ref{helms})
is {\it truly}  superintegrable. \end{theorem}
\medskip\noindent
{\bf Proof}
: Let ${S}={S}_0+W$  be a second order conformal symmetry 
of $H$ and ${S}_U={S}_0+W_U$  be the special case  that is in conformal involution with 
$\frac{1}{\lambda}(\partial_{xx}+\partial_{yy})+ U$. Then  $$[ {S}, H]=R_{{ S}_0} H,\quad  
[{S}_U, H_0+U]=R_{{ S}_0}({ H}_0+U),\quad [S_0,H_0]=R_{S_0}H_0$$
and ${\tilde{ S} }={ S}-\frac{W_U}{U}{ H}$
is a corresponding true symmetry of $\tilde { H}$. Indeed, 
$$[{\tilde{ S}},{\tilde { H}}]=[{ S},U^{-1} H]-[\frac{W_U}{U} H,\frac{1}{U} H]=U^{-1}R_{{ S}_0}H-U^{-1}[S_0,U]U^{-1}H$$
$$-U^{-1}[W_U,H_0]U^{-1}H=U^{-1}R_{{ S}_0}H-U^{-1}R_{S_0}H=0.$$
 This transformation of second order symmetries preserves linear and algebraic independence. 
 Thus the transformed system is Helmholtz  superintegrable. $\Box$
 
 Note that if $H\Psi=0$ then ${\tilde S}\Psi =S\Psi$ and $H(S\Psi)=0$ so $S$ and $\tilde S$ agree on the null space of $H$ and they preserve this null space.

There is a similar result for first order conformal symmetries $ L$.
\begin{corollary} Let $ L$ be a first order conformal symmetry  of the superintegrable system (\ref{confl}) and 
suppose $U({\bf x}) $ is a particular solution of equations (\ref{veqn1a}), nonzero in an open set. Then $ L$ is a true 
symmetry of the Helmholtz superintegrable system (\ref{helms}): $[{ L},{\tilde { H}}]=0$.
 \end{corollary}

\medskip\noindent   {\bf Proof}: By assumption, $[{ L},{ H}]=R_{ L}({\bf x}){ H}=R_{ L}({ H}_0+V)$ where $R_{ L}$ is a function.
Thus, $[{ L}, { H}_0]=R_{ L} { H}_0, [{ L},V]=R_{ L} V$, so also $[{ L},U]=R_{L} U$. Then 
\[[{ L},{\tilde{ H}}]=[{ L},U^{-1} H]=U^{-1}[{ L},{ H}]-U^{-1}[L,U]U^{-1}H\]
\[=U^{-1}RH-U^{-1}RUU^{-1}H=U^{-1}RH-U^{-1}RH=0.\]
$\Box$

These results show that any second order conformal Laplace superintegrable system admitting a nonconstant potential $U$ can be 
St\"ackel transformed to a Helmholtz
superintegrable system. This operation is  invertible, but the inverse is not a St\"ackel transform.  
By choosing all possible special potentials $U$ associated with the  fixed Laplace system (\ref {confl}) we generate the equivalence class of all Helmholtz 
superintegrable systems (\ref{helms}) obtainable through this process. As is easy to check,  any two Helmholtz superintegrable systems lie in the same equivalence class
if and only if they are St\"ackel equivalent in the standard sense. All Helmholtz superintegrable systems are related to conformal Laplace systems in this way, 
so the study of all Helmholtz superintegrability on conformally flat manifolds can be reduced to the study of all conformal Laplace superintegrable systems on flat space.
\begin{theorem} There is a one-to-one relationship between 
 flat space  conformally superintegrable Laplace systems with nondegenerate potential   and St\"ackel equivalence classes of superintegrable Helmholtz systems with nondegenerate potential on  conformally flat spaces.
 \end{theorem}
 Indeed, let 
 \be\label{nonconf} (H_1-E_1)\Psi =0,\ (H_2-E_2)\Psi =0,\ee be Schr\"odinger eigenvalue equations where
 \[ H_j-E_j=\frac{1}{\lambda_j(x,y)}(\partial_{xx}+\partial_{yy}+V^{(j)})-E_j,\quad j=1,2,\]
 and  
 \be\label{Vident}V=V^{(1)}+E_1\lambda_1=V^{(2)}+E_2\lambda _2\ee is a nondegenerate potential for the conformally superintegrable system 
 \be\label{confsup}\partial_{xx}+\partial_{yy}+V=0.\ee
 Suppose $\{ \lambda_1,\lambda_2\}$ is a linearly independent set (otherwise there is nothing to prove).
 Then we can find a potential basis for $V$ of the form 
 \[ V(x,y)=-E_1\lambda_1(x,y)-E_2\lambda_2(x,y)+k_3U^{(3)}(x,y)+k_4U^{(4)}(x,y)\]
 \[=-E_1\lambda_1-E_2\lambda_2+{\tilde V}\]
 where $\{ \lambda_1,\lambda_2,U^{(3)},U^{(4)}\}$ is a linearly independent set.
 Dividing (\ref{confsup}) by $\lambda_1,\lambda_2$, respectively, we see that systems (\ref{nonconf}) are 
 regular superintegrable with nondegenerate (3-parameter) potentials. Furthermore, multiplying the first system (\ref{nonconf}) by 
 $\lambda^{(1)}/\lambda^{(2)}$ 
 we see that it is St\"ackel equivalent to the second system. Conversely, if systems (\ref{nonconf}) are regular superintegrable 
 and equality (\ref{Vident}) holds, then it is easy to verify that system (\ref{confsup}) is conformally superintegrable with 
 nondegenerate (4-parameter) potential.
 
 Even for true Helmholtz superintegrable systems there are good  reasons to add a seemingly trivial constant to the potentials. 
 Thus, for a St\"ackel transform induced by the function 
 $U^{(1)}$, we can take the original system to have Hamiltonian
 \be\label{parameter} H=H_0+V=H_0+U^{(1)}\alpha_1+U^{(2)}\alpha_2+U^{(3)}\alpha_3+\alpha_4\ee
 where $\{U^{(1)},U^{(2)},U^{(3)},1\}$ is a basis for the 4-dimensional potential space. A 2nd order symmetry $S$ would have the form
 \[ S=S_0+W^{(1)}\alpha_1+W^{(2)}\alpha_2+W^{(3)}\alpha_3.\]
 The St\"ackel transformed Hamiltonian and symmetry take the form
 \[ {\tilde H}=\frac{1}{U^{(1)}}H_0+\frac{U^{(1)}\alpha_1+U^{(2)}\alpha_2+U^{(3)}\alpha_3+\alpha_4}{U^{(1)}},\ {\tilde S}=S-W^{(1)}{\tilde H}.\]
 Note that the parameter $\alpha_1$ cancels out of the expression for $\tilde S$; it is replaced by $-\alpha_4$. Now suppose that 
 $\Psi$ is a formal  eigenfunction of 
 $H$ (not required to be normalizable): $H\Psi=E\Psi$. If we choose the parameter $\alpha_4=-E$ in (\ref{parameter}) then, in terms of this redefined $H$, we have $H\Psi =0$.
 It follows immediately
 that ${\tilde S}\Psi =S\Psi$.  Thus, for the 3-parameter system $H'$ and the St\"ackel transform ${\tilde H}'$,
  \[H'=H_0+V'=H_0+U^{(1)}\alpha_1+U^{(2)}\alpha_2+U^{(3)}\alpha_3,\]
  \[{\tilde H}'=\frac{1}{U^{(1)}}H_0
  +\frac{-U^{(1)}E+U^{(2)}\alpha_2+U^{(3)}\alpha_3}{U^{(1)}},\]
we have $H'\Psi=E\Psi $ and ${\tilde H}'\Psi=-\alpha_1\Psi$. It follows that The effect of the St\"ackel transform is to replace $\alpha_1$ by $-E$ and $E$
by $-\alpha_1$. Further, since $S$ and $\tilde S$ don't depend on the choice of $\alpha_4$ we see that these 
operators must agree on eigenspaces of $H'$

We know that the symmetry operators of all 2nd order nondegenerate superintegrable systems in 2D generate a quadratic algebra
of the form 
\[{} [R,S_1]=f^{(1)}(S_1,S_2,\alpha_1,\alpha_2,\alpha_3,H'),\ [R,S_2]=f^{(2)}(S_1,S_2,\alpha_1,\alpha_2,\alpha_3,H'),\]
\be\label{quadratic1} R^2=f^{(3)}(S_1,S_2,\alpha_1,\alpha_2,\alpha_3,H'),\ee
where $\{S_1,S_2,H\}$ is a basis for the 2nd order symmetries and $\alpha_1,\alpha_2,\alpha_3$ are the parameters 
for the potential, \cite{4,5,MPW2013}. It follows from the above considerations that the effect of a St\"ackel transform generated by the potential function $U^{(1)}$ is to determine a new superintegrable 
system with structure 
\be\label{quadratic2}{} [{\tilde R},{\tilde S}_1]=f^{(1)}({\tilde S}_1,{\tilde S}_2,-{\tilde H}',\alpha_2,\alpha_3,-\alpha_1),\ee
\[
[R,{\tilde S}_2]=f^{(2)}({\tilde S}_1,{\tilde S}_2,-{\tilde H}',\alpha_2,\alpha_3,-\alpha_1),\]
\[ R^2=f^{(3)}({\tilde S}_1,{\tilde S}_2,-{\tilde H}',\alpha_2,\alpha_3,-\alpha_1).\]
Of course, the switch of $\alpha_1$ and $H'$ is only for illustration; there is a St\"ackel transform that replaces any 
$\alpha_j$ by $-H'$ and $H'$ by $-\alpha_j$.

Formulas (\ref{quadratic1}) and (\ref{quadratic2}) are just instances of the quadratic algebras of the superintegrable systems belonging to the 
equivalence class of a single nondegenerate conformally superintegrable Hamiltonian
\be\label{confham}\hat{H}=\partial_{xx}+\partial_{yy}+\sum_{j=1}^4\alpha_j V^{(j)}(x,y).\ee
Let $\hat{S}_1,\hat{S}_2, \hat{H}$ be a basis of 2nd order conformal  symmetries of $\hat H$. From the above discussion we can conclude the following.
\begin{theorem} The  symmetries of the 2D nondegenerate conformal superintegrable Hamiltonian $\hat H$ generate a quadratic algebra
 \be\label{confquadalg} [{\hat R},{\hat S}_1]=f^{(1)}({\hat S}_1,\hat{S}_2,\alpha_1,\alpha_2,\alpha_3,\alpha_4),\ [{\hat R},{\hat S}_2]=f^{(2)}
 ({\hat S}_1,{\hat S}_2,\alpha_1,\alpha_2,\alpha_3,\alpha_4),\ee
\[ {\hat R}^2=f^{(3)}({\hat S}_1,\hat{S}_2,\alpha_1,\alpha_2,\alpha_3,\alpha_4),\]
where $\hat{R}=[{\hat S}_1,\hat{S}_2]$ and all identities hold $\mod({\hat H})$. A conformal St\"ackel transform generated by the potential 
$V^{(j)}(x,y)$ yields a nondegenerate Helmholtz superintegrable Hamiltonian $\tilde H$ with quadratic algebra relations identical to (\ref{confquadalg}),
except that we make the replacements ${\hat S}_\ell\to {\tilde  S}_\ell$ for $\ell=1,2$   and $\alpha_j\to -{\tilde H}$. These modified relations  
(\ref{confham}) are now true identities, not $\mod ({\hat H})$.
\end{theorem}
Note that expressions (\ref{confquadalg}) define a true quadratic algebra, interpreted $\mod ({\hat H})$. They differ from the quadratic algebra for a Helmholtz system in that the Hamiltonian
doesn't appear, whereas there is an extra parameter. The quadratic algebras of all Helmholtz systems obtained from $\hat H$ via conformal St\"ackel transforms
follow by simple substitution.

\begin{comment}
Every 2nd order conformal symmetry is of the form $S=S_0+W$ where $S_0$ is a 2nd order element of the enveloping algebra of $so(4,\C)$.
The dimension of this space of 2nd order elements is 21 but there is an 11-dimensional subspace of symmetries congruent to 0 $\mod H_0$ 
where $H_0=P_1^2+P_2^2$. A basis for this subspace is 
\[ P_1^2+P_2^2\sim 0,\  J^2+D^2\sim 0,\ K_1^2+K_2^2\sim 0,\ \{P_1,K_2\}+2JD\sim 0,\]
\[ \{P_1,J\}-\{P_2,D\}\sim 0,\ \{P_1,K_1\}-\{P_2,K_2\}\sim0,\  \{J,K_1\}+\{D,K_2\}\sim 0,\]
\[ \{P_1,D\}+\{P_2,J\}\sim 0,\ \{P_1,K_2\}+\{P_2,K_1\}\sim0,\  \{J,K_2\}-\{D,K_1\}\sim 0,\]
\[ 4J^2+\{P_1,K_1\}+\{P_2,K_2\}\sim 0.\]
Thus $\mod H_0$ the space of 2nd order symmetries is 10-dimensional.
\end{comment}

\subsection{Contractions of conformal superintegrable systems with potential induced by generalized In\"on\"u-Wigner  contractions}

The basis symmetries ${\cal S}^{(j)} ={\cal S}^{(j)}_0+W^{(j)}$,\ ${\cal H}={\cal H}_0+V$ of a nondegenerate 2nd order conformally 
superintegrable system determine a conformal quadratic algebra
(\ref{confquadalg}), and if the parameters of the potential are set equal to $0$, the free system $ {\cal S}^{(j)}_0, {\cal H}_0,\ j=1,2$ also determines a conformal
quadratic algebra without parameters, which we call a {\it free conformal quadratic algebra}. The elements of this free algebra belong 
to the enveloping algebra of $so(4,\C)$
with basis (\ref{conformalKV}). Since the system is nondegenerate the integrability conditions for the potential are satisfied identically and the 
full quadratic algebra can be computed from the free algebra,
modulo a choice of basis for the 4-dimensional potential space. Once we choose a basis for  $so(4,\C)$,  its enveloping algebra is uniquely 
determined by the structure constants. Structure relations in the enveloping
algebra are continuous functions of the structure constants, so a contraction of one $so(4,\C)$ to itself induces a  contraction of the
 enveloping algebras. Then the  free conformal quadratic algebra constructed in the enveloping algebra will contract
to another free quadratic algebra. (In \cite{KM2014} essentially the same argument was given in more detail for Helmholtz  superintegrable
systems on constant curvature spaces.)

In this paper we consider a family of contractions of $so(4,\C)$ to itself that we call B\^ocher contractions. All these contractions are implemented via coordinate
transformations. Suppose we have a conformal nondegenerate superintegrable system with free generators  ${\cal H}_0, {\cal S}^{(1)}_0, {\cal S}^{(2)}_0$ 
that determines the   conformal and free conformal  quadratic algebras $Q$ and $Q^{(0)} $
and has structure functions $A^{ij}({\bf x}),\ B^{ij}({\bf x}),\ C^{ij}({\bf x})$ in Cartesian coordinates ${\bf x}=(x_1,x_2)$. 
Further, suppose this system contracts to another nondegenerate system
${\cal H'}_0, {\cal S'}^{(1)}_0 ,{\cal S'}^{(2)}_0 $ with conformal quadratic algebra ${Q'}^{(0)}$. We show here that this contraction 
induces a contraction of the associated nondegenerate superintegrable system
${\cal H}={\cal H}_0+V$, ${\cal S}^{(1)}={\cal L}^{(1)}_0+W^{(1)}$,
 ${\cal S}^{(2)}={\cal S}^{(2)}_0+W^{(2)}$, $Q$ to
${\cal H}'={{\cal H}'}_0+V'$, ${\cal S'}^{(1)}={\cal S'}_0^{(1)}+{W^{(1)}}'$,
 ${\cal S'}^{(2)}={\cal S'}_0^{(2)}+{W^{(2)}}'$, $Q'$.
The point is that in  the contraction process the symmetries ${{\cal H}'}_0(\epsilon)$,
${\cal S'}^{(1)}_0(\epsilon)$,
$ {\cal S'}^{(2)}_0(\epsilon)$
remain continuous functions of $\epsilon$, linearly independent as quadratic forms, and
 $\lim_{\epsilon\to 0} {\cal H'}_0(\epsilon)={{\cal H'}}_0$,
$\lim_{\epsilon\to 0} {\cal S'}^{(j)}_0(\epsilon)={\cal S'}_0^{(j)}$.
Thus the associated functions $A^{ij}(\epsilon), B^{ij}(\epsilon), C^{(ij)}$ will also be continuous functions of $\epsilon$ and
$\lim_{\epsilon\to 0}A^{ij}(\epsilon)={A'}^{ij}$, $\lim_{\epsilon\to 0}B^{ij}(\epsilon)={B'}^{ij}$, $\lim_{\epsilon\to 0}C^{ij}(\epsilon)={C'}^{ij}$. 
Similarly, the integrability conditions for the potential equations
\be\label{nondegpot2} \ba{lllll}
 V^{(\epsilon)}_{22}&=& V^{(\epsilon)}_{11}&+&A^{22}(\epsilon) V^{(\epsilon)}_1+B^{22}(\epsilon) V^{(\epsilon)}_2+C^{22}(\epsilon) V^{(\epsilon)},\\
 V^{(\epsilon)}_{12}&=& &&A^{12}(\epsilon) V^{(\epsilon)}_1+B^{12}(\epsilon) V^{(\epsilon)}_2+C^{12}(\epsilon) V^{(\epsilon)},\ea
\ee
will hold for each $\epsilon$ and in the limit. 
This means that the 4-dimensional solution space for the potentials $V$ will deform continuously into the 4-dimensional solution space for 
the potentials $V'$. Thus the target space of solutions $V'$ (and of the functions $W'$) is uniquely determined by the free quadratic algebra contraction.

There is an apparent lack of uniqueness in this procedure, since for a nondegenerate superintegrable system one typically
chooses a basis $V^{(j)},\ j=1,\cdots,4$ for the potential space and expresses a general potential as $V=\sum_{j=1}^4a_jV^{(j)}$. 
Of course the choice of basis for the source system is arbitrary, as is the choice for the target system. 
Thus the structure equations for the quadratic algebras and the dependence $a_j(\epsilon)$ of the contraction constants 
on $\epsilon$ will vary depending on these choices. However, all such possibilities are related by a basis change matrix.

\section{Tetraspherical coordinates and relations with the 2-sphere and 2D flat space}
The tetraspherical coordinates $(x_1,\cdots,x_4)$ satisfy 
$x_1^2+x_2^2+x_3^2+x_4^2=0$.
They are projective coordinates on the null cone and have 3 degrees of freedom. Their principal advantage over flat space Cartesian coordinates is
that the action of the conformal algebra (\ref{conformalKV}) and of the conformal group $\sim SO(4,\C)$ is linearized in tetraspherical coordinates.

\medskip
\noindent{\bf Relation to Cartesian coordinates $(x,y)$ and coordinates on the 2-sphere $(s_1,s_2,s_3)$ }:
\[ x_1=2XT,\ x_2=2YT,\ x_3=X^2+Y^2-T^2,\ x_4=i(X^2+Y^2+T^2).\]
\[ x=\frac{X}{T}=-\frac{x_1}{x_3+ix_4},\  y=\frac{Y}{T}=-\frac{x_2}{x_3+ix_4},  \]
\[ x=\frac{s_1}{1+s_3},\  y=\frac{s_2}{1+s_3},\]
\[ s_1=\frac{2x}{x^2+y^2+1},\  s_2=\frac{2y}{x^2+y^2+1},\ s_3=\frac{1-x^2-y^2}{x^2+y^2+1},\]
\[  H=\partial_{xx}+\partial_{yy}+{\tilde V}=(x_3+ix_4)^2\left(\sum_{k=1}^4\partial_{x_k}^2+V\right)
=(1+s_3)^2\left(\sum_{j=1}^3p_{s_j}^2+V\right),\]
where ${\tilde V}=(x_3+ix_4)^2V$ and 
\[ (1+s_3)=-i\frac{(x_3+ix_4)}{x_4},\ (1+s_3)^2=-\frac{(x_3+ix_4)^2}{x_4^2},\]
\[s_1=\frac{ix_1}{x_4},\ s_2=\frac{ix_2}{x_4},\ s_3=\frac{-ix_3}{x_4}.\]
Also, $ \sum_{k=1}^4x_k\partial_{x_k}=0$
and, classically, $\sum_{k=1}^4x_k{p_k}=0$.

\noindent {\bf Relation to flat space and 2-sphere 1st order conformal constants of the motion}:
We define
\[ L_{jk}=x_j\partial_{x_k}-x_k \partial_{x_j}, \ 1\le j,k\le 4,\ j\ne k,\]
where $L_{jk}=-L_{kj}$. The generators for flat space conformal symmetries are related to these via
\be\label{identifications}P_1= \partial_x=L_{13}+iL_{14},\ P_2=\partial_y=L_{23}+iL_{24},\  D=iL_{34},\ee
\[ J=L_{12},\  K_1=L_{13}-iL_{14},\  K_2=L_{23}-iL_{24}.\]
Here 
\[ D=x\partial_x+y\partial_y,\ J=x\partial_y-y\partial_x,\ K_1=2xD-(x^2+y^2)\partial_x,\]
etc.

The generators for $2$-sphere conformal constants of the motion are related to the $L_{jk}$ via 
\[ L_{12}=J_{12}=s_1\partial_{s_2}-s_2\partial_{s_1},\ L_{13}=J_{13},\ L_{23}=J_{23},\]
\[ L_{14}=-i\partial_{s_1},\  L_{24}=-i\partial_{s_2},\  L_{34}=-i\partial_{s_3}.\]
Note that in identifying tetraspherical coordinates we can always permute the parameters $1,2,3,4$. More generally, we can apply an
arbitrary $SO(4,\C)$ transformation to the tetraspherical coordinates, so the above relations between Euclidean and tetraspherical coordinates are far from unique.

\medskip
\noindent {\bf 2nd order conformal symmetries $\sim H $}:
The 11-dimensional space of conformal symmetries $\sim H$ has basis
\[L_{12}^2-L_{34}^2,\ L_{13}^2-L_{24}^2,\ L_{23}^2-L_{14}^2,\ L_{12}^2+L_{13}^2+L_{23}^2,\]
\be\label{congH}  L_{12}L_{34}+L_{23}L_{14}-L_{13}L_{24},\ee
\[ \{L_{13},L_{14}\}+\{L_{23},L_{24}\}, \  \{L_{13},L_{23}\}+\{L_{14},L_{24}\},\ \{L_{12},L_{13}\}+\{L_{34},L_{24}\}, \]
\[ \{L_{12},L_{14}\}-\{L_{34},L_{23}\}, \  \{L_{12},L_{23}\}-\{L_{34},L_{14}\},\ \{L_{12},L_{24}\}+\{L_{34},L_{13}\}, \]

All of this becomes much clearer if we make use of the decomposition $so(4,\C)\equiv so(3,\C)\oplus so(3,\C)$ and the functional realization of the Lie algebra. 
Setting
\[ J_1=\frac12(L_{23}-L_{14}),\ J_2=\frac12(L_{13}+L_{24}),\ J_3=\frac12(L_{12}-L_{34}),\]
\[ K_1=\frac12(L_{23}+L_{14}),\ K_2=\frac12(L_{13}-L_{24}),\ K_3=\frac12(L_{12}+L_{34}),\]
we have
\[ [J_i,J_j]=\epsilon_{ijk}J_k,\ [K_i,K_j]=\epsilon_{ijk}K_k,\ [J_i,K_j]=0.  \]
In terms of the variable $z=x+iy,{\bar z}=x-iy$ we have
\[ J_1=\frac12(i\partial_z-iz^2\partial_z),\ J_2=\frac12(\partial_z+z^2\partial_z),\ J_3=iz\partial_z,\]
\[ K_1=\frac12(-i\partial_{\bar z}+i{\bar z}^2\partial_{\bar z}),\ K_2=\frac12(\partial_{\bar z}+{\bar z}^2\partial_{\bar z}),
\ K_3=-i{\bar z}\partial_{\bar z},\]
so the $J_i$ operators depend only on the variable $z$ and the $K_j$ operators depend only on the variable $\bar z$. Also
\be\label{Cas} J_1^2+J_2^2+J_3^2\equiv 0,\ K_1^2+K_2^2+K_3^2\equiv 0.\ee
The space of 2nd order elements in the enveloping algebra is thus 21-dimensional and decomposes as $A_z\oplus A_{\bar z}\oplus A_{z{\bar z}}$ where 
$A_z$ is 5-dimensional with basis $J_1^2$, $J_3^2$, $\{J_1,J_2\}$, $\{J_1,J_3\}$, $\{J_2,J_3\}$,\
$A_{\bar z}$ is 5-dimensional with basis $K_1^2$, $K_3^2$, $\{K_1,K_2\}$, $\{K_1,K_3\}$, $\{K_2,K_3\}$,
and $A_{z{\bar z}}$ is 9-dimensional with basis $J_iK_j$, $1\le i,j\le 3$. Note that all of the elements of $A_{z{\bar z}}$ are $\sim H$, 
whereas none of the nonzero elements of $A_z,A_{\bar z}$ 
have this property. The 11 elements (\ref{congH}) include the  relations (\ref{Cas}). Here, the transposition $J_i\leftrightarrow K_i$ is a conformal equivalence.

\subsection{Classification of 2nd order conformally superintegrable systems with nondegenerate potential}
With this simplification it becomes feasable to classify all conformally 2nd order superintegrable systems with nondegenerate potential. Since  every such system
has generators ${ S}^{(1)}={ S}_0^{(1)}+W_1(z,{\bar z})$,  ${ S}^{(2)}={ S}_0^{(2)}+W_2(z,{\bar z})$, it is sufficient to classify, up to $
SO(4,\C)$ conjugacy,  all free conformal quadratic algebras 
with generators ${ S}_0^{(1)},\ { S}_0^{(2)}$, $\mod { H}_0$, and then to determine for which of these free conformal algebras the integrability conditions 
(\ref{int31}) hold identically, so that the system admits a nondegenerate potential ${\tilde V}(z,{\bar z})$ which can be computed. The classification breaks up into the 
following possible cases:
\begin{itemize} \item Case 1: ${ S}_0^{(1)},\ { S}_0^{(2)}\in A_z$. (This is conformally equivalent to ${ S}_0^{(1)},\ { S}_0^{(2)}\in A_{\bar z}$.)
The possible free conformal quadratic algebras of this type, classified up to $SO(3,\C)$ conjugacy $\mod J_1^2+J-2^2+J_3^2$ can easily be 
obtained from the computations in \cite{KM2014}. They are the pairs
\begin{enumerate} \item \[J_3^2,\ J_1^2\]
\item  \[J_3^2,\ \{J_1+iJ_2,J_3\}\]
\item \[ J_3^2,\ \{J_1,J_3\} \]
\item \[\{J_2,J_2+iJ_1\},\ \{J_2,J_3\}\]
\item \[J_3^2,\ (J_1+iJ_2)^2\]
\item \be \{J_1+iJ_2,J_3\},\ (J_1+iJ_2)^2.\label{conjugacyclasses}\ee
\end{enumerate}
 Checking pairs $1)-5)$ we find that they do not admit a nonzero potential, so they do not correspond to  nodegenerate conformal superintegrable systems. This is in
dramatic distinction to the results of  \cite{KM2014} where for Helmholtz systems on constant curvature spaces there was a 1-1 relationship between
free quadratic algebras and nondegenerate superintegrable systems. Pair $6)$, (\ref{conjugacyclasses}), does correspond to a superintegrable system, the singular case 
${\tilde V}=f(z)$ where $f(z)$ is arbitrary. (This system is conformally St\"ackel equivalent to the singular Euclidean system $E_{15}$.) 
Equivalently, the system in $A_{\bar z}$ with analogous 
$K$-operators yields the potential ${\tilde V}=f({\bar z})$, (\ref{Varb'}).
\item Case 2: ${ S}_0^{(1)}={ S}_J^{(1)}+{ S}_K^{(1)},\ { S}_0^{(2)}={ S}_J^{(2)}$ where ${ S}_J^{(1)},\ S_J^{(2)}$ are selected from one of the pairs $1)-6)$ above 
 and ${ S}_K^{(1)}$ is a nonzero element of $A_{\bar z}$. Again there is a conformally equivalent case where  the roles of $J_i$ and $K_i$ are switched. 
 To determine the possibilities for ${ S}_K^{(1)}$ we classify the 2nd order elements in the enveloping algebra of $so(3,\C)$ up to $SO(3,\C)$ conjugacy, 
 $\mod K_1^2+K_2^2+K_3^2$. From the computations in \cite{KM2014} we see easily that there are the following representatives for the equivalence classes:
 \begin{description}\item[a)]\[ K_3^2\]
 \item[b)]\[K_1^2+aK_2^2,\ a\ne 0,1\]
 \item[c)]\[(K_1+iK_2)^2\]
 \item[d)] \[ K_3^2+(K_1+iK_2)^2\]
 \item[e)] \[ \{K_3,K_1+iK_2\}.\]  
 \end{description}
 For pairs $1),3),4),5)$ above and all choices $a)-e)$ we find that the integrabilty conditions are never satisfied, so there are no corresponding nondegenerate 
 superintegrable systems. For pair $2)$, however, we find that any choice $a)-e)$ leads to the same nondegenerate superintegrable system 
 $[2,2]$, (\ref{V[22norm']}). While it appears that there are multiple generators for this one system, each set of generators maps to any other 
 set by a conformal St\"ackel transformation and a change of variable. For pair $6)$,  we find that any choice $a)-e)$ leads to the same nondegenerate superintegrable system 
 $[4]$, (\ref{V[4]norm'}). Again each set of generators maps to any other 
 set by a conformal St\"ackel transformation and a change of variable. 
\item Case 3: ${ S}_0^{(1)}={ S}_J^{(1)},\ { S}_0^{(2)}={ S}_J^{(2)}+{ S}_K^{(2)}$ where ${ S}_J^{(1)},\ S_J^{(2)}$ are selected from 
one the pairs $1)-6)$ above 
 and ${ S}_K^{(2)}$ is a nonzero element of $A_{\bar z}$. Again there is a conformally equivalent case where  the roles of $J_i$ and $K_i$ are switched. 
 To determine the possibilities for ${ S}_K^{(2)}$ we classify the 2nd order elements in the enveloping algebra os $so(3,\C)$ up to $SO(3,\C)$ conjugacy, 
 $\mod K_1^2+K_2^2+K_3^2$. They are $a)-e)$ above. 
 For pairs $1)-4),6)$ above and all choices $a)-e)$  the integrabilty conditions are never satisfied, so there are no corresponding nondegenerate 
 superintegrable systems. For pair $5)$, however, we find that any choice $a)-e)$ leads to the same nondegenerate superintegrable system 
 $[2,2]$, (\ref{V[22norm']}). Again each set of generators maps to any other 
 set (and to any $[2,2]$ generators in Case 2) by a conformal St\"ackel transformation and a change of variable. 
 \item Case 4:  ${ S}_0^{(1)}={ S}_J^{(1)},\ { S}_0^{(2)}={ S}_K^{(2)}$ where ${ S}_J^{(1)}$ is selected from 
one of the representatives $a)-e)$ above and  ${ S}_K^{(2)}$ is selected from one of the analogous representatives $a)-e)$ expressed as $K$-operators.
We find that each of the 25 sets of generators leads to the single conformally superintegrable system $[0]$, (\ref{V[0]norm'}), and each set of generators
maps to any other 
 set by a conformal St\"ackel transformation and a change of variable. 
 \item Case 5: ${ S}_0^{(1)}={ S}_J^{(1)}+{ S}_K^{(1)},\ { S}_0^{(2)}={ S}_J^{(2)}+{ S}_K^{(2)}$ where ${ S}_J^{(1)},\ S_J^{(2)}$ are selected from 
one of the pairs $1)-6)$ above and ${ S}_K^{(1)}$, $S_K^{(2)}$ are obtained from ${ S}_J^{(1)}$, $ S_J^{(2)}$, respectively, by replacing each 
$J_i$ by $K_i$. We find the following possibilities:
\begin{description}
 \item[i)] ${ S}_0^{(1)}=J_1^2+K_1^2,\ { S}_0^{(2)}=J_3^2+K_3^2$. This extends to the system $[1,1,1,1]$,  (\ref{V[1111norm']}).
 \item[ii)] ${ S}_0^{(1)}=J_3^2+K_3^2,\ { S}_0^{(2)}=\{J_3,J_1+iJ_2\}+\{K_3,K_1+iK_2\}$. This extends to the system $[2,1,1]$,  (\ref{V211norm'}).
 \item[iii)]  ${ S}_0^{(1)}=J_3^2+K_3^2,\ { S}_0^{(2)}=\{J_1,J_3\}+\{K_1,K_3\}$. This extends to the system $[1,1,1,1]$,  (\ref{V[1111norm']}) again,
  equivalent to the generators $i)$ by a conformal St\"ackel transformation and a change of variable. 
 \item[iv)]  ${ S}_0^{(1)}=\{J_1,J_2+iJ_1\}+\{K_1,K_2+iK_1\},\ { S}_0^{(2)}=\{J_2,J_3\}+\{K_2,K_3\}$. 
 This does not extend to a conformal superintegrable system.
\item[v)]  ${ S}_0^{(1)}=(J_1+iJ_2)^2+(K_1+iK_2)^2,\ { S}_0^{(2)}=J_3^2+K_3^2$. This extends to the system $[2,1,1]$,  (\ref{V211norm'}) again,
  equivalent to the generators $ii)$ by a conformal St\"ackel transformation and a change of variable. 
  \item[vi)]  ${ S}_0^{(1)}=\{J_3,J_1+iJ_2\}+\{K_3,K_1+iK_2\},\ { S}_0^{(2)}=(J_1+iJ_2)^2+(K_1+iK_2)^2$, 
  which extends to the system $[3,1]$,  (\ref{V[31]norm'}).
\end{description}
\end{itemize}
This completes the classification.
\begin{example}
 We describe how apparantly distinct superintegrable systems of a fixed type are actually the same.
 In Case 2 consider the system with generators $\{J_1+iJ_2,J_3\}+(K_1+iK_2)^2,\ (J_1+iJ_2)^2$. This extends to the conformally superintegrable system
 $[4]$ with 
 flat space Hamiltonian operator $H_1=\partial_{z{\bar z}}+ V^{(1)}$ where 
 \[ V^{(1)}=2k_3z{\bar z}+2k_4z+k_3{\bar z}^3+3k_4{\bar z}^2+k_1{\bar z}+k_2.\] The system  with generators 
 $\{J_1+iJ_2,J_3\}+K_3^2+(K_1+iK_2)^2,\ (J_1+iJ_2)^2$ again  extends to the conformally superintegrable system $[4]$. Indeed, replacing $z,{\bar z}$ by 
 $Z, {\bar Z}$ to distinguish the 
 two systems,  we find the 2nd 
 flat space Hamiltonian operator $H_2=\partial_{Z{\bar Z}}+ V^{(2)}$ where 
 \[ V^{(2)}=\frac{c_3\ {\rm arcsinh}^3({\bar Z})+3c_4\ {\rm arcsinh}^2({\bar Z})+(2c_3 Z+c_1)\ {\rm arcsinh}({\bar Z})+2c_4 Z+c_2}{\sqrt{1-{\bar Z}^2}}.\]
 Now we perform a conformal St\"ackel transform on $H_2$ to obtain the new flat space system
 \[ {\tilde H}_2=\sqrt{1-{\bar Z}^2}\ \partial_{Z{\bar Z}}+
  c_3\ {\rm arcsinh}^3({\bar Z})+3c_4\ {\rm arcsinh}^2({\bar Z})\]
  \[+(2c_3 Z+c_1)\ {\rm arcsinh}({\bar Z})+2c_4 Z+c_2.\]
  Making the change of variable ${\bar Z}=\sinh W $, we find
  \[ {\tilde H}_2= \partial_{ZW}+
  c_3 W^3+3c_4W^2+(2c_3 Z+c_1)W+2c_4 Z+c_2.\]
  Thus, with the identifications $Z=z$, $W={\bar z}$, $c_i=k_i$, we see that $H_1\equiv {\tilde H}_2$.
\end{example}

\subsection{Relation to separation of variables}

B\^ocher's analysis \cite{Bocher} involves symbols of the form $[n_1,n_2,..,n_p]$ where 
$n_1+...+n_p=4$. These symbols are used to define coordinate surfaces as 
follows. Consider the quadratic forms
\be\label{ellipsoidalcoords}\Omega =x^2_1+x^2_2+x^2_3+x^2_4=0,\   
\Phi =\frac{x^2_1}{ \lambda -e_1} + \frac{x^2_2}{ \lambda -e_2} + 
\frac{x^2_3}{ \lambda -e_3} + \frac{x^2_4}{ \lambda -e_4}=0.\ee
If $e_1,e_2,e_3,e_4$ are pairwise distinct, the elementary divisors of these two  forms are denoted by the symbol 
$[1,1,1,1]$. Given a point in 2D flat space with Cartesian coordinates $(x^0,y^0)$, there corresponds a set of tetraspherical coordinate 
$(x^0_1,x^0_2,x^0_3,x^0_4)$, unique up to multiplication by a nonzero constant. If we substitute these coordinates into expressions (\ref{ellipsoidalcoords}) 
we can verify that there are exactly 2 roots $\lambda=\rho,\mu$ such that $\Phi=0$. 
These are elliptic coordinates. It can be verified that they are orthogonal with respect to the metric $ds^2=dx^2+dy^2$ and that 
they are $R$-separable for the 
Laplace equations $(\partial^2_x+\partial^2_y)\Theta=0$ or $(\sum_{j-1}^4\partial_{x_j}^2)\Theta=0$. 
Now consider the  potential 
\[V_{[1,1,1,1]}=\frac{a_1}{ x^2_1} + \frac{a_2}{ x^2_2} + 
\frac{a_3}{ x^2_3} + \frac{a_4}{ x^2_4}.\]
It turns out to be the only possible potential $V$ such that the Laplace equation $(\sum_{j-1}^4\partial_{x_j}^2+V)\Theta=0$ is $R$-separable in elliptic 
coordinates for {\it all} choices of  the parameters $e_j$. The separation is characterized by 2nd order conformal symmetry operators that are 
linear in the parameters 
$e_j$. In particular the symmetries span a  3-dimensional subspace of symmetries, so the system $(\sum_{j-1}^4\partial_{x_j}^2+V_{[1,1,1,1]})\Theta=0$ must be conformally 
superintegrable. We can write this as 
\[H=(x_3+ix_4)^2(\partial ^2_{x_1}+ \partial ^2_{x_2}+ \partial ^2_{x_3}+ 
\partial ^2_{x_4} +\frac{a_1}{ x^2_1} +\frac {a_2}{ x^2_2} + \frac{a_3}{ x^2_3} + 
\frac{a_4}{ x^2_4}),\]
or in  terms of flat space coordinates $x,y$  as  
\[ H= \partial_x^2+\partial_y^2+\frac{a_1}{x^2}+\frac{a_2}{y^2}+\frac{4a_3}{(x^2+y^2-1)^2}-\frac{4a_4}{(x^2+y^2+1)^2}.\]
For the coordinates $s_i,i=1,2,3$  we obtain 
\[H=(1+s_3)^2(\partial ^2_{s_1}+\partial ^2_{s_2}+\partial ^2_{s_3} 
-\frac{a_1}{ s^2_1} - \frac{a_2}{ s^2_2} - \frac{a_3}{ s^2_3} -a_4).\]
The coordinate curves are  described by $[1,1,1,\stackrel{\infty}{1}
]$ (because we can always transform to equivalent coordinates for which $e_4=\infty$) and  the corresponding 
$H\Theta=0$ system is proportional to  $S_9$, the eigenvalue equation for the generic potential on the 2-sphere,
which separates variables in elliptic coordinates 
$s^2_i=\frac{(\rho -e_i)(\mu -e_i)}{ (e_i-e_j)(e_i-e_k)}$
where $(e_i-e_j)(e_i-e_k)\neq 0$ and $i,j,k=1,2,3$.
The quantum Hamiltonian  when written using these coordinates is equivalent to  
\[{\cal H}=\frac{1}{ \rho -\mu }[P_\rho^2-P_\mu^2] -\sum ^3_{i=1} 
a_i\frac{(e_i-e_j)(e_i-e_k)}{ (\rho -e_i)(\mu -e_i)}],\]
where $P_\lambda=\sqrt{\Pi ^3_{i=1}(\lambda -e_i)}\ \partial_\lambda$.

\section{B\^ocher contractions} These are contractions of $so(4,\C)$ to itself that are induced by coordinate transformations on the 
null cone that B\^ocher used to derive the separable coordinate systems for the flat space Laplace and wave equations,
\cite{Bocher,KMR1984}.
In the following notes we shall usually list 6 symmetries for each superintegrable system $[1,1,1,1]-[4]$, which is strictly the case for the analogous systems on the 2-sphere.
However, these systems are defined on the null cone, which implies extra constraints, Therefore  instead of 6 linearly independent symmetries we have only 3.

 We start with the potential
\be\label{V[1111]} V_{[1,1,1,1]}=\frac{a_1}{x_1^2}+\frac{a_2}{x_2^2}+\frac{a_3}{x_3^2}+\frac{a_4}{x_4^2},\ee
and the system $[1,1,1,1]$ and use successive B\^ocher contractions to derive the systems $[2,1,1],[2,2],[3,1], [4]$ and $[0]$.

\subsection{The $[1,1,1,1]$ to $[2,1,1]$ contraction}
If two of the $e_i$ in eqns (\ref{ellipsoidalcoords}) become equal, B\^ocher shows that the process of 
making $e_1\rightarrow e_2$ together with suitable transformations of the 
$a_i's$ produces a conformally equivalent $H$. This corresponds to the choice of 
coordinate curves  obtained by the B\^ocher limiting process $[1,1,1,1]\to [2,1,1]$, i.e.,
\[ e_1=e_2+\epsilon ^2,\  
x_1\rightarrow \frac{iy_1}{ \epsilon },\
x_2\rightarrow \frac{y_1}{ \epsilon } + \epsilon y_2, \ 
x_j\rightarrow y_j,j=3,4,\]
which  results in the pair of quadratic forms 
\[\Omega =2y_1y_2+y^2_3+y^2_4=0,\  \Phi  =\frac{y^2_1}{ (\lambda -e_2)^2}+\frac{2y_1y_2}{(\lambda -e_2)} + 
\frac{y^2_3}{ (\lambda -e_3)} +\frac {y^2_4}{ (\lambda -e_4)} =0.\]
The coordinate curves with $e_4= \infty $ correspond to cyclides with 
elementary divisors $[2,1,\stackrel{\infty}{1} ]$, \cite{Bromwich}, i.e.,
$\Phi  =\frac{y^2_1}{ (\lambda -e_2)^2}+\frac{2y_1y_2}{ (\lambda -e_2)} +
\frac{y^2_3}{ (\lambda -e_3)}=0$.
\begin{comment}
Indeed, making the substitution $\lambda=\frac{\alpha \lambda' +\beta}{\gamma\lambda'+\delta}$, $ e_i=\frac{\alpha e_i'+\beta}{\gamma 
e_i'+\delta}$
 we do not change the family of surfaces described (see \cite{Bocher},  page 59). In particular the second quadratic form becomes 
\[ \Phi=\frac{y_1^2(\gamma e_1'+\delta)^2}{(\lambda'-e_1')^2(\alpha\delta-\beta\gamma)}+\frac{2y_1y_2}{\lambda'-e_1'}+\frac{y_3^2}
{\lambda'-e_3'}+\frac{y_4^2}{\lambda'-e_4'}=0.\]
 Now if we let $e_1'=\infty$  we obtain essentially 
$\Phi=\frac{y_1^2\gamma^2}{(\alpha\delta-\beta\gamma)}+\frac{y_3^2}{\lambda'-e_3'}+\frac{y_4^2}{\lambda'-e_4'}=0$, which means 
that we have  degenerate elliptic coordinates of type 1 in the plane with coordinate curves  denoted by $[\stackrel{\infty}{2},1,1]$. 
If we took $e_4'=\infty$  we would obtain the coordinate curves of degenerate elliptic cordinates on the sphere with coordinate curves 
denoted by  $[2,1,\stackrel{\infty}{1}]$. If
we take $e_4'= \infty$  in generic tetracyclic coordinates we obtain elliptic coordinates on the 3-sphere with cordinate curves 
denoted by $[1,1,1,\stackrel{\infty}{1}]$. Our subsequent studies
elaborate on these observations. \end{comment}
Note that the composite linear coordinate mapping 
\[x_1+ix_2=\frac{i\sqrt{2}}{\epsilon}(x'_1+ix'_2)+\frac{i\epsilon}{\sqrt{2}}(x'_1-ix'_2),\  
x_1-ix_2=-\frac{i\epsilon}{\sqrt{2}}(x'_1-ix'_2),\]
\[x_3=x'_3,\ x_4=x'_4,\]
satisfies $\lim_{\epsilon\to 0} \sum_{j=1}^4 x_j^2= \sum_{j=1}^4 {x'}_j^2=0$, and induces a contraction of the Lie algebra $so(4,\C)$ to itself. An explicit computation yields
 \[ L'_{12}=L_{12},\ L'_{13}=-\frac{i}{\sqrt{2}\ \epsilon}(L_{13}-iL_{23})-\frac{i\ \epsilon}{\sqrt{2}}L_{13},\ 
 L'_{23}=-\frac{i}{\sqrt{2}\ \epsilon}(L_{13}-iL_{23})-\frac{\ \epsilon}{\sqrt{2}}L_{13}\]
  \[ L'_{34}=L_{34},\ L'_{14}=-\frac{i}{\sqrt{2}\ \epsilon}(L_{14}-iL_{24})-\frac{i\ \epsilon}{\sqrt{2}}L_{14},\ 
 L'_{24}=-\frac{i}{\sqrt{2}\ \epsilon}(L_{14}-iL_{24})-\frac{\ \epsilon}{\sqrt{2}}L_{14}.\]
This is the B\^ocher contraction $[1,1,1,1]\to [2,1,1]$.

\subsubsection{Conformal St\"ackel transforms of the [1,1,1,1] system} \label{3.1.1} We write the parameters $a_j$ defining the potential $V_{[1,1,1,1]}$ as a vector: $(a_1,a_2,a_3,a_4)$. 
\begin{enumerate} 
\item The potentials $(1,0,0,0)$, and any permutation of the indices $a_j$  generate conformal St\"ackel transforms to $S9$. 
\item  The potentials $(1,1,0,0)$ and $(0,0,1,1)$   generate conformal St\"ackel transforms to $S7$. 
\item The potentials $(1,1,1,1)$, $(0,1,0,1)$, $(1,0,1,0)$, $(0,1,1,0)$ and $(1,0,0,1)$    generate conformal St\"ackel transforms to $S8$. 
\item The potentials $(a_1,a_2,0,0),\  a_1a_2\ne 0,a_1\ne a_2$, and any permutation of the indices $a_j$.
generate conformal St\"ackel transforms to $D4B$. 
\item The potentials $(1,1,a,a), a\ne 0,1$, and any permutation of the indices $a_j$.
generate conformal St\"ackel transforms to $D4C$.
\item Each potential  not proportional to one of these must generate a conformal St\"ackel transform to a  superintegrable system on a Koenigs space in the 
family $K[1,1,1,1]$.
     \end{enumerate}

Now  under the contraction $[1,1,1,1]\to [2,1,1]$ we have 
\[  V_{[1,1,1,1]}
\stackrel{\epsilon\ \to\ 0}{\Longrightarrow}\ V_{[2,1,1]} \]
where
\be \label{V[211]} V_{[2,1,1]}=\frac{b_1}{(x'_1+ix'_2)^2}+\frac{b_2(x'_1-ix'_2)}{(x'_1+ix'_2)^3}+\frac{b_3}{{x'_3}^2}+\frac{b_4}{{x'_4}^2},\ee
and
\[ a_1=-\frac12(\frac{b_1}{\epsilon^2}+\frac{b_2}{2\epsilon^4}),\ a_2=- \frac{b_2}{4\epsilon^4},\ a_3=b_3,\ a_4=b_4.\]

\begin{comment}
 We established the potential limit by direct computation. However this contraction can be understood in terms of the generic elliptic 
 coordinates.
We put $e_1=0 
, e_2=\epsilon ^2$ and $e_3=e_3$  in the coordinates      

\[s^2_1 = \frac{(x_1-e_1)(x_2-e_1)}{ (e_1-e_2)(e_1-e_3)},\       
s^2_2 = \frac{(x_1-e_2)(x_2-e_2)}{ (e_2-e_1)(e_2-e_3)},\      
s^2_3 = \frac{(x_1-e_3)(x_2-e_3)}{ (e_3-e_2)(e_3-e_1)}.\]
We take as the generic potential $V=\frac{ a_1}{ s^2_1} + \frac{a_2}{ s^2_2} + 
\frac{a_3}{ s^2_3}$ and subject the coefficents $a_i$ to the contraction 
transformations       
\[ a_1\rightarrow  -\frac{b_1}{ \epsilon ^2} + \frac{b_2}{\epsilon ^4},\    
a_2\rightarrow  \frac{b_2}{ \epsilon ^4},\]
Under $\epsilon \rightarrow 0$ we obtain        
\[V\rightarrow  \frac{b_1e_3}{ x_1x_2} +b_2 
[\frac{-e_3(x_1+x_2)+x_1x_2}{ x^2_1x^2_2}] + \frac{e^2_3a_3}{ (x_1-e_3)(x_2-e_3)}.\]
This form of the potential is separable in type $[2,1,1]$ coordinates, as can be seen from the relations         
\[\frac{(x_1-x_2)}{ x_1x_2} = \frac{1}{ x_2}-\frac {1}{x_1},\         
\frac{(x^2_1-x^2_2)}{ x^2_1 x^2_2} = \frac{1}{x^2_2} - \frac{1}{x^2_1},\]
\[
\frac{(x_1 -x_2)}{ (x_1-e_3)(x_2-e_3)} =\frac{1}{ x_2-e_3} -\frac{1}{ x_1-e_3}.\]

Consider another case, viz $e_1=0,\ e_2=\epsilon$ and $e_3=A\epsilon$. From
 the transformations        
\[a_1\rightarrow  \frac{c_1}{ \epsilon ^4} + \frac{c_2}{ \epsilon ^6} + 
\frac{b_3}{ \epsilon ^8},\       
a_2\rightarrow  \frac{c_2}{ \epsilon ^6(A-1)} + 
\frac{c_3A^2}{ \epsilon ^8(A-1)^2},\       
a_3\rightarrow  \frac{c_3}{ \epsilon ^8(A-1)^2},\]
we obtain the limit of $V$ as       
\[V \rightarrow  \frac{Ac_1}{ x_1x_2} - \frac{Ac_2(x_1+x_2)}{ x^2_1x^2_2} + 
\frac{A^2c_3(x^2_1+x_1x_2+x^2_2)}{ x^3_1x^3_2}.\]

What we deduce from these examples is that if $e_1$ is a root of the 
polynomial of the corresponding degenerate elliptic system on the complex sphere
and it has multplicity $p$, there are terms in the potential of the form 
\[\frac{1}{ (x_1-x_2)}[\frac{1}{ (x_1-e_1)^s} -\frac {1}{ (x_2-e_1)^s}],\ s=1\cdots p\]
and the transformation of $a_i$ can be determined.
\end{comment}

\begin{examples}
 
Using Cartesian coordinates
$x,y$, we consider the Hamiltonian
\[ H=\partial^2_x+\partial^2_y+ \frac{a_1}{ x^2} + \frac{a_2}{ y^2} + \frac{4a_3}{ (x^2+y^2-1)^2} + 
\frac{4a_4}{ (x^2+y^2+1)^2}.\]
Multiplying on the left by 
$x^2$ we obtain 
\[\hat H=x^2(\partial^2_x+\partial^2_y)+a_1 + a_2 \frac{x^2}{ y^2}+ 4a_3\frac{x^2}{ (x^2+y^2-1)^2}
- 4a_4\frac{x^2}{ (x^2+y^2+1)^2},\]
 the case ${\bf a}=(1,0,0,0)$. This becomes more transparent if we introduce variables 
$x=e^{-a},y=r$. The Hamiltonian $\hat H$  can be written 
\[\hat H=\partial^2_a+\partial_a+e^{-2a}\partial^2_r + a_1+ a_2\frac {e^{-2a}}{ r^2}
+
a_3 \frac{4}{ (e^{-a}+e^a(r^2-1))^2} - a_4\frac{4}{ (e^{-a}+e^a(r^2+1))^2}.\]
Recalling horospherical coordinates on the complex two sphere,
viz. 
\[s_1=\frac{i}{ 2}(e^{-a}+(r^2+1)e^a),\ 
s_2=re^a,\ 
s_3=\frac{1}{ 2}(e^{-a}+(r^2-1)e^a)\]
we see that the Hamiltonian $\hat H$ can be written as  
\[ \hat H=\partial^2_{s_1}+ \partial^2_{s_2}+ \partial^2_{s_3}+ a_1+ \frac{a_2}{ s^2_2} + 
\frac{a_3}{ s^2_3} + \frac{a_4}{ s^2_1},\]
and this is explicitly the superintegrable system $S_9$.

Now consider the case ${\bf a}=(0,1,0,1)$ which for
$x=e^a\sin\varphi,\ y=e^a\cos\varphi $ and conformal  St\"ackel multiplier 
\[ 
(\frac{1}{ y^2} - \frac{4}{ (x^2+y^2+1)^2})=e^{-2a}(\frac{1}{ \cos ^2\varphi } -\frac {1}{ \cosh^2a})\]
yields the  Hamiltonian
\[\frac{1}{ (\frac{1}{ \cos ^2\varphi } - \frac{1}{ \cosh^2a})}\left [\partial^2_a+\partial^2_\varphi  + 
\frac{a_1}{ \sin ^2\varphi } + \frac{a_2+a_4}{2}(\frac{1}{ \cos ^2\varphi } +
\frac{1}{ \cosh^2a}) + \frac{a_3}{ \sinh^2a}\right]+ \]
\[
\frac{a_2-a_4}{ 2},\]
which is just $S_8$ in elliptic coordinates of type 1,  the coordinates on the 
2-sphere  being taken as  
\[s_1+is_2 = \frac{1}{ \cos\varphi \cosh a},\ 
s_1-is_2 = \frac{\cos\varphi }{ \cosh a} + \frac{\cosh a}{ \cos\varphi } - 
\frac{1}{ \cos\varphi \cosh a},\ 
s_3=i\tan\varphi \tanh a,\]
where $s^2_1+s^2_2+s^2_3=1$. 

Now consider the case ${\bf a}=(1,1,0,0)$,  with
\[ x=e^{ia/2}\cos b,\ y=e^{ia/2} \sin b.\]
If instead we use the variable $B$ where 
$\sin 2b= \frac{1}{ \cosh B}$
then the  Hamiltonian can be written 
\[\partial^2_B+\tanh B\, \partial_B- \frac{1}{ \cosh^2 B} \partial^2_a + b_1\tanh B + b_2 
\frac{1}{\sinh^2a\cosh^2B} + b_3 \frac{1}{ \cosh^2a\cosh^2B} +b_0\]
which is directly St\"ackel equivalent  to $S_7$. A suitable choice of coordinates on the 
complex 2-sphere is 
\[
s_1=\cosh a\cosh B,\ s_2=i\cosh a\sinh B,\ s_3=i\sinh a.\]

For the case ${\bf a}=(b_1,b_2,0,0)$ the St\"ackel multiplier (potential that induces the St\"ackel transform) is $b_1/x^2+b_2/y^2$. In terms of the coordinates
$x=e^v\cos\theta,y=e^v\sin\theta$ the Hamiltonian takes the form
\[ H=\frac{\sin^22\theta}{2[(b_2-b_1)\cos 2\theta+(b_1-b_2)]}\left[\partial_\theta^2+\partial_v^2+k+\frac{a_3}{\sinh^2v}+\frac{a_4}{\cosh^2v}\right]\]
for $k$ a parameter.  This is equivalent to $D4B$.

For the case ${\bf a}=(0,0,b_3,b_4)$ the St\"ackel multiplier is $b_3/(x^2+y^2-1)^2+b_4/(x^2+y^2+1)^2$. In terms of the coordinates
$x=-ie^{iu}\cosh v,y=e^{iu}\sinh v$ the Hamiltonian again takes a form equivalent to $D4B$.

For the case ${\bf a}=(1,1,a,a)$,  using polar coordinates as directly above,  we see that  the Hamiltonian takes the form
\[ H=\frac{1}{[\frac{1}{\sin^22\theta}+\frac{a}{\sinh^22v}]}\left[\partial_\theta^2+\partial_v^2+\frac{a_1}{\cos^2\theta}+\frac{a_2}{\sin^2\theta}+
\frac{a_3}{\sinh^2v}+\frac{a_4}{\cosh^2v}\right],\]
 equivalent to $D4C$.

From these examples we note that it is always possible to choose coordinates 
for which the entire Hamiltonian is a rational function.

\end{examples}

\subsubsection{[1,1,1,1] to [2,1,1] contraction and St\"ackel transforms}\label{3.1.2}
For fixed $A_j$, $B_j$, $D_j$ we have the expansions
\[ V^A_{[1,1,1,1]}=\frac{A_1}{x_1^2}+\frac{A_2}{x_2^2}+\frac{A_3}{x_3^2}+\frac{A_4}{x_4^2}\]
\[ =\frac{A_3}{{x'}_3^2}+\frac{A_4}{{x'}_4^2}+\frac{2(A_2-A_1)\epsilon^2}{(x'_1+ix'_2)^2}
+\frac{4A_2(-x'_1+ix'_2)\epsilon^4}{(x'_1+ix'_2)^3}+O(\epsilon^6),\]
\[ V^A_{[2,1,1]}=\frac{A_1}{(x_1+ix_2)^2}+\frac{A_2(x_1-ix_2)}{(x_1+ix_2)^2}+\frac{A_3}{x_3^2}+\frac{A_4}{x_4^2}\]
\[ =\frac{A_3}{{x'_3}^2}+\frac{A_4}{{x'}_4^2}-\frac{A_1}{2(x'_1+ix'_2)^2}\epsilon^2+\frac{(A_2+2A_1)(x'_1-ix'_2)}{4(x'_1+ix'_2)^3}
\epsilon^4+O(\epsilon^6),\]
\[ V^A_{[2,2]}=\frac{A_1}{(x_1+ix_2)^2}+\frac{A_2(x_1-ix_2)}{(x_1+ix_2)^3}+\frac{A_3}{(x_3+ix_4)^2}+\frac{A_4(x_3-ix_4)}{(x_3+ix_4)^3}\]
\[ =\frac{A_3}{(x'_3+ix'_4)^2}+\frac{A_4(x'_3-ix'_4)}{(x'_3+ix'_4)^3}-\frac{A_1}{2(x'_1+ix'_2)^2}\epsilon^2 \]
\[+\frac{(A_2+2A_1)(x'_1-ix'_2)}{4(x'_1+ix'_2)^3}\epsilon^4+O(\epsilon^6),\]
\[ V^B_{[3,1]}=\frac{B_1}{(x_1+ix_2)^2}+\frac{B_2x_1}{(x_1+ix_2)^3}+\frac{B_3(4x_3^2+x_4^2)}{(x_1+ix_2)^4}+\frac{B_4}{x_4^2}\]
\[=\frac{B_3(4{x'_3}^2+{x'_4}^2)}{(x'_1+ix'_2)^4}+\frac{B_4}{{x'_4}^2}-\frac{(B_2+2B_1)}{4(x'_1+ix'_2)^2}\epsilon^2+O(\epsilon^4),\]
\[ V^D_{[4]}=-\frac{D_1}{2(x'_1+ix'_2)^2}\epsilon^2+\frac{i\sqrt{2}}{4}\ \frac{D_2(x'_3+ix'_4)-2D_3(x'_3-ix'_4)}{(x'_1+ix'_2)^3}\epsilon^3\]
\[ +\left[ 
 \frac{3D_3(x'_3+ix'_4)^2}{(x'_1+ix'_2)^4}-\frac{(D_1+2D_4)({x'_3}^2+{x'_4}^2)}{2(x'_1+ix'_2)^4}\right]\epsilon^4+O(\epsilon^5),\ ({\rm see}\ 
 (\ref{V[4]})),\]

\subsubsection{Conformal St\"ackel transforms of the [2,1,1] system} \label{3.1.3} We write the potential in the normalized form
\be\label{V211norm} V'_{[2,1,1]}=\frac{a_1}{x_1^2}+\frac{a_2}{x_2^2}+\frac{a_3(x_3-ix_4)}{(x_3+ix_4)^3}+\frac{a_4}{(x_3+ix_4)^2},\ee
 and designate it via the vector $(a_1,a_2,a_3,a_4)$. 
\begin{enumerate} 
\item The potential $(1,1,0,0)$ generates a conformal St\"ackel transform to $S4$. 
\item The potentials $(1,0,0,0)$, $(0,1,0,0)$ generate  conformal St\"ackel transforms to $S2$.
\item The potential $(0,0,0,1))$  generates a  conformal St\"ackel transforms to $E1$.
\item  The potential $(0,0,1,0)$ generates a conformal St\"ackel transform to $E16$. 
\item Potentials $(a_1,a_2,0,0)$, with $ a_1a_2\ne 0$, $a_1\ne a_2$   generate  conformal St\"ackel transforms to $D4A$. 
\item Potentials $(0,0,a_3,a_4)$, with $a_3a_4\ne 0$  generate conformal St\"ackel transforms to $D3B$. 
\item Potentials $(a,0,0,1)$ and  $(0,a,0,1)$ with $a\ne 0$  generate conformal St\"ackel transforms to $D2B$. 
\item Potentials $(1,1,a,0)$ with $a\ne 0$  generate conformal St\"ackel transforms to $D2C$. 
\item Each potential not proportional to one of these must generate a conformal St\"ackel transform to a superintegrable system on a Koenigs space in the family $K[2,1,1]$.
\end{enumerate}

\noindent {\bf Basis of conformal symmetries for original system}: 
Let $H_0=\sum_{j=1}^4\partial_{x_j}^2$. A basis is 
\[ H_0+V_{[1,1,1,1]},\ Q_{12},\ Q_{13},\]
where
\[Q_{jk}=L_{jk}^2+a_j\frac{x_k^2}{x_j^2}+a_k\frac{x_j^2}{x_k^2},\  1\le j<k\le 4.\]

\noindent {\bf Contraction of basis}: Using the notation of (\ref{V[211]}) we have 
\[ H_0+V_{[1,1,1,1]}\to H'_0+V_{[2,1,1]},\]
\[ Q'_{12}=Q_{12}-\frac{b_1}{2\epsilon^2}-\frac{b_2}{2\epsilon^4}=(L'_{12})^2+b_1(\frac{x_1'-ix_2'}{x_1'+ix_2'})+b_2(\frac{x_1'-ix_2'}{x_1'+ix_2'})^2,\]
\[ Q'_{13}=2\epsilon^2 Q_{13}=(L'_{23}-iL'_{13})^2+\frac{b_2{x'_3}^2}{(x'_1+ix'_2)^2}-\frac{b_3(x'_1+ix'_2)^2}{{x'_3}^2},\]

If we apply the same $[1,1,1,1]\to [2,1,1]$  contraction to the $[2,1,1]$ system, the system contracts to itself, but with parameters 
$c_1,\cdots,c_4$ where
\[ b_1=-\frac{2c_1}{\epsilon^2},\ b_2=\frac{c_1}{\epsilon^2}+\frac{4c_2}{\epsilon^4},\ b_3=c_3,\ b_4=c_4.\]

If we apply the same contraction to the $[2,2]$ system, the system contracts to itself, but with altered parameters, and to $[0]$.

If we apply the same contraction to the $[3,1]$ system,  the system contracts to $(1)$ or to itself. 

If we apply the same contraction to the $[4]$ system  the system contracts to $(2)$ or to  a system with potential 
\be\label{V[0]} V[0]= \frac{c_1}{(x'_1+ix'_2)^2}+\frac{c_2x'_3+c_3x'_4}{(x'_1+ix'_2)^3}+c_4\frac{{x'}_3^2+{x'}_4^2}{(x'_1+ix'_2)^4}.\ee

If we apply this same contraction to the $[0]$ system, (\ref{V[0]norm}) it contracts to itself with altered parameters.

If we apply this same contraction to the $(1)$ system, (\ref{V(1)norm}) it contracts to $(2)$ or to itself with altered parameters.

If we apply this same contraction to the $(2)$, (\ref{V(2)norm}) it contracts to itself with altered parameters.

\subsubsection{Conformal St\"ackel transforms of the [0] system}  We write the potential $V[0]$ in the normalized form
\be\label{V[0]norm} V'_{[0]}=\frac{c_1}{(x_3+ix_4)^2}+\frac{c_2x_1+c_3x_2}{(x_3+ix_4)^3}+c_4\frac{x_1^2+x_2^2}{(x_3+ix_4)^4},\ee
 and designate it by the vector $(c_1,c_2,c_3,c_4)$. 
\begin{enumerate} 
\item The potentials $(\frac{c_2^2+c_3^2}{4},c_2,c_3,1)$ generate  conformal St\"ackel transforms to $E20$. 
\item The potentials $(c_1,1,\pm i,0)$ generate  conformal St\"ackel transforms to $E11$.
\item The potential $(1,0,0,0))$  generates a  conformal St\"ackel transform to $E3'$.
\item Potentials $(c_1,c_2,c_3,0)$, with $c_2^2+c_3^2\ne 0$   generate  conformal St\"ackel transforms to $D1C$. 
\item Potentials $(c_1,c_2,c_3,1)$, with $c_1\ne \frac{c_2^2+c_3^2}{4} $  generate conformal St\"ackel transforms to $D3A$. 
\item Each potential not proportional to one of these must generate a conformal St\"ackel transform to a superintegrable system on a Koenigs space in the family
$K[0]$.
\end{enumerate}

\subsection{[1,1,1,1] to [2,2]:} 
\[ L'_{12}=L_{12},\ L'_{34}=L_{34}, \ L'_{24}+L'_{13}=L_{24}+L_{13},\]
\[ L'_{24}-L'_{13}=(\epsilon^2+\frac{1}{\epsilon^2})L_{13}-\frac{1}{\epsilon^2}(iL_{14}-L_{24}-iL_{23}),\]
\[ L'_{23}-L'_{14}=2L_{23}+iL_{13}-iL_{24},\]
\[ L'_{23}+L'_{14}=i\left((\epsilon^2-\frac{1}{\epsilon^2})L_{13}+\frac{1}{\epsilon^2}(iL_{14}+L_{24}+iL_{23})\right).\]

\noindent Coordinate implementation 
\[ x_1=\frac{i}{\sqrt{2}\ \epsilon}(x'_1+ix'_2),\ x_2=\frac{1}{\sqrt{2}}\left(\frac{x'_1+ix'_2}{\epsilon}+\epsilon \ 
(x'_1-ix'_2)\right),\]
\[ x_3=\frac{i}{\sqrt{2}\ \epsilon}(x'_3+ix'_4),\ x_4=\frac{1}{\sqrt{2}}\left(\frac{x'_3+ix'_4}{\epsilon}+\epsilon \ 
(x'_3-ix'_4)\right),\]

\noindent
{\bf Limit of 2D potential}: \[ V_{[1,1,1,1]} \stackrel{\epsilon\ \to\ 0}{\Longrightarrow}\ V_{[2,2]}\]
where 
\be\label{V[22]} V_{[2,2]}=\frac{b_1}{(x'_1+ix'_2)^2}+\frac{b_2(x'_1-ix'_2)}{(x'_1+ix'_2)^3}
+\frac{b_3}{(x'_3+ix'_4)^2}+\frac{b_4(x'_3-ix'_4)}{(x'_3+ix'_4)^3},\ee
and 
\[ a_1=-\frac12\frac{b_1}{\epsilon^2}-\frac{b_2}{4\epsilon^4},\ a_2=- \frac{b_2}{4\epsilon^4},\ 
a_3=-\frac12\frac{b_3}{\epsilon^2}-\frac{b_4}{4\epsilon^4},
\ a_4=- \frac{b_4}{4\epsilon^4}.\]

\subsubsection{Conformal St\"ackel transforms of the [2,2] system}
We designate the potential (\ref{V[22]}) by the vector $(b_1,b_2,b_3,b_4)$.
\begin{enumerate}
\item The potential $(0,0,1,0)$ generates a conformal St\"ackel transform to $E8$. 
\item The potential $(0,0,0,1)$ generates a conformal St\"ackel transform to $E17$.
\item Potentials $(1,0,a,0))$  generate  conformal St\"ackel transforms to $E7$.
\item  Potentials $(0,1,0,a)$ generate  conformal St\"ackel transforms to $E19$. 
\item Potentials $(0,0,b_3,b_4)$, with $b_3b_4\ne 0$  generate conformal St\"ackel transforms to $D3C$. 
\item Potentials $(b_1,b_2,0,0)$ with $b_1b_2\ne 0$  generate conformal St\"ackel transforms to $D3D$. 
\item Each potential not proportional to one of these must generate a conformal St\"ackel transform to a superintegrable system on a Koenigs space
in the family $K[2,2]$.
\end{enumerate}

\noindent
{\bf Contracted basis}:
\[H_0+V_{[1,1,1,1]}\to H'_0+V_{[2,2]},\]
\[Q_{12}-\frac{b_2}{2\epsilon^4}-\frac{b_1}{2\epsilon^2}\to Q_1'={ L'}_{12}^2+b_1\frac{x'_1-ix'_2}{x'_1+ix'_2}+b_2\frac{(x'_1-ix'_2)^2}{(x'_1+ix'_2)^2},\]
\[ 4\epsilon^4 Q_{13}\to Q_2'=(L'_{13}+iL'_{14}+iL'_{23}-L'_{24})^2-b_2\frac{(x'_3+ix'_4)^2}{(x'_1+ix'_2)^2}-b_4\frac{(x'_1+ix'_2)^2}{(x'_3+ix'_4)^2},\]
Note also that
\[ \epsilon^2(Q_{23}-Q_{14})\to Q_3'=-\frac{i}{2}\{L'_{14}-L'_{23},iL'_{23}+L'_{13}-L'_{24}+iL'_{14}\}-\frac{b_1}{2}\frac{(x'_3+ix'_4)^2}{(x'_1+ix'_2)^2}\]
\[-b_2\frac{(x'_2x'_4 +x'_1x'_3)(x'_3+ix'_4)}{(x'_1+ix'_2)^3)}+\frac{b_3}{2}\frac{(x'_1+ix'_2)^2}{(x'_3+ix'_4)^2}+
b_4\frac{(x'_2x'_4 +x'_1x'_3)(x'_1+ix'_2)}{(x'_3+ix'_4)^3)}\]

If we apply the same $[1,1,1,1]\to [2,2]$ contraction to the $[2,1,1]$ system with potential parameters $k_1,\cdots,k_4$ , the system contracts to the $[2,2]$ potential with parameters 
$b_1,\cdots,b_4$, where,
\[ k_1=-\frac{2b_1}{\epsilon^2},\ k_2=\frac{4b_3}{\epsilon^4},\ k_3=-\frac{b_2}{2\epsilon^2}-\frac{b_4}{4\epsilon^4},\ k_4=-\frac{b_4}{4\epsilon^4},\]
 or to a special case of $E15$.

If we apply the same contraction to the $[2,2]$ system we recover the same system but with altered parameters, or $[0]$.
If we apply the same contraction to the  superintegrable  $[3,1]$  system in the form
\[V[3,1]'=\frac{k_1}{(x_1+ix_2)^2}+\frac{k_4x_3}{(x_1+ix_2)^3}+k_3\frac{(4x_3^2+x_4^2)}{(x_1+ix_2)^4}+\frac{k_4}{x_4^2},\]
the system contracts  to a special case of $E15$, or to one with potential of the form
\be\label{V(1)} V(1)=\frac{c_1}{(x'_1+ix'_2)^2}+\frac{c_2}{(x'_3+ix'_4)^2} +c_3\frac{x'_3+ix'_4}{(x'_1+ix'_2)^3}+c_4\frac{(x'_3+ix'_4)^2}{(x'_1+ix'_2)^4}.\ee

It admits 2 1st order symmetries and is St\"ackel equivalent to special cases of the Euclidean superintegrable system $E15$ via transforms $(x'_1+ix'_2)^2$ or $(x'_3+ix'_4)^2$.

If we apply the same contraction to the  superintegrable  $[4]$ system we get  system conformally  equivalent to  (\ref{V(2)norm}). 
This admits a 1st order symmetry and goes to a special case of $E15$ by a conformal St\"ackel transform.

If we apply this same contraction to the $[0]$ system, (\ref{V[0]norm}) it contracts to itself with altered parameters.

If we apply this same contraction to the $(1)$ system, (\ref{V(1)norm}) it contracts to itself with altered parameters, or to a special case of $E15$.

If we apply this same contraction to the $(2)$ system, (\ref{V(2)norm}) it contracts to itself with altered parameters.

\subsubsection{[1,1,1,1] to  [2,2] contraction and St\"ackel transforms}
For fixed $A_j$ we have the expansions

\[ V^A_{[1,1,1,1]}=\frac{A_1}{x_1^2}+\frac{A_2}{x_2^2}+\frac{A_3}{x_3^2}+\frac{A_4}{x_4^2}\]
\[ =\frac{2(A_2+A_4-A_1-A_3)\epsilon^2}{(x'_1+ix'_2)^2}+
 \left(\frac{4A_4(-x'_3+ix'_4)}{(x'_3+ix'_4)^3}+\frac{4A_2(-x'_1+ix'_2)}{(x'_1+ix'_2)^3}\right)\epsilon^4\]
\[ +\left(\frac{6A_4(-x'_3+ix'_4)^2}{(x'_3+ix'_4)^4}+\frac{6A_2(-x'_1+ix'_2)^2}{(x'_1+ix'_2)^4}\right)\epsilon^6+O(\epsilon^8),\]
\[ V^A_{[2,1,1]}=\frac{A_1}{x_1^2}+\frac{A_2}{x_2^2}+\frac{A_3(x_3-ix_4)}{(x_3+ix_4)^3}+\frac{A_4}{(x_3+ix_4)^2}\]
\[ = \left(\frac{2(A_2-A_1)}{(x'_1+ix'_2)^2}-\frac{A_4}{2(x'_3+ix'_4)^2}\right)\epsilon^2\]
\[+\left(-\frac{4A_2(x'_1-ix'_2)}{(x'_1+ix'_2)^3}+\frac{(A_3+2A_4)(x'_3-ix'_4)}{4(x'_3+ix'_4)^3}\right)\epsilon^4+O(\epsilon^6),\]
\[ V^A_{[2,2]}=\frac{A_1}{(x_1+ix_2)^2}+\frac{A_2(x_1-ix_2)}{(x_1+ix_2)^3}
+\frac{A_3}{(x_3+ix_4)^2}+\frac{A_4(x_3-ix_4)}{(x_3+ix_4)^3}\]
\[ =-\frac12\left(\frac{A_1}{(x'_1+ix'_2)^2}+\frac{A_3}{(x'_3+ix'_4)^2}\right)\epsilon^2\]
\[ +\frac14\left(\frac{(A_2+2A_1)(x'_1-ix'_2)}{(x'_1+ix'_2)^3}+\frac{(A_4+2A_3)(x'_3-ix'_4)}{(x'_3+ix'_4)^3}\right)\epsilon^4+O(\epsilon^6),\]

\subsection{[2,1,1] to [3,1]}

\noindent {\bf Coordinate implementation}:
\[ x_1+ix_2=-\frac{i\sqrt{2}\ \epsilon}{2}x_2' +\frac{(i x'_1- x'_3)}{\epsilon},\]
\[x_1-ix_2=-\epsilon \ (x_3'+ix'_1)
+\frac{3i\sqrt{2}x'_2}{4\epsilon}+\frac12 \frac{(ix'_1- x'_3)}{\epsilon^3},\]
\[ x_3=-\frac12 x'_2-\frac{\sqrt{2}}{2} \frac{(x'_1+i x'_3)}{\epsilon^2},\ x_4=x_4'.\]

\[ L'_{24}=\frac{\sqrt{2}i}{2\epsilon}(L_{14}+iL_{24})-L_{34},\ L'_{14}+iL'_{34}=-i\epsilon\ (L_{14}+iL_{24}),\]
\[ L'_{14}-iL'_{34}=\frac{1}{\epsilon}\left(iL_{14}(1+\frac{1}{2\epsilon^2})+L_{24}(1-\frac{1}{2\epsilon^2})-\frac{\sqrt{2}}{\epsilon}L_{34}\right),\]
\[ L'_{13}=-L_{12}-2\sqrt{2}\ L_{13}\,(\epsilon+2\epsilon^3),\]
\[ \ L'_{23}+iL'_{12}=4\epsilon^3 L_{13},\ L'_{23}-iL'_{12}=(2\sqrt{2}-\frac{\sqrt{2}}{\epsilon^2})L_{12}\]
\[+(8\epsilon^3+4\epsilon-\frac{2}{\epsilon}+\frac{1}{2\epsilon^3})L_{13}+\frac{i}{2\epsilon^3}L_{23}.\]

\noindent
{\bf Limit of 2D potential}: \[ V_{[2,1,1]}
\stackrel{\epsilon\ \to\ 0}{\Longrightarrow}\ V_{[3,1]},\]
where 
\be\label{V[31]} V_{[3,1]}=\frac{c_1}{(x'_1+ix'_3)^2}+\frac{c_2x'_2}{(x'_1+ix'_3)^3}
+\frac{c_3(4{x'_2}^2+{x'_4}^2)}{(x'_1+ix'_3)^4}+\frac{c_4}{{x'_4}^2},\ee
and 
\[ b_1=\frac{c_3}{\epsilon^6}+\frac{\sqrt{2}\ c_2}{4\epsilon^4}-\frac{c_1}{\epsilon^2},\ b_2=- 
\frac{c_3}{\epsilon^4}-\frac{\sqrt{2}\ c_2}{2\epsilon^2},\ b_3=\frac{c_3}{4\epsilon^8},\ b_4=c_4.\]

\subsubsection{Conformal St\"ackel transforms  of the [3,1] system}  We write potential $V_{[3,1]}$ in the normalized form
\be\label{V[31]norm} V'_{[3,1]}=\frac{a_1}{(x_3+ix_4)^2}+\frac{a_2x_1}{(x_3+ix_4)^3}
+\frac{a_3(4{x_1}^2+{x_2}^2)}{(x_3+ix_4)^4}+\frac{a_4}{{x_2}^2},\ee
and designate it  $(a_1,a_2,a_3,a_4)$.
\begin{enumerate}
\item The potential $(0,0,0,1)$ generates a conformal St\"ackel transform to $S1$. 
\item The potential $(1,0,0,0)$ generates a conformal St\"ackel transform to $E2$.
\item The potential $(a,1,0,0)$  generates a conformal St\"ackel transform to $D1B$.
\item The potential $(0,0,1,0)$ generates  a conformal St\"ackel transform to $D2A$. 
\item Each potential not proportional to one of these must generate a conformal St\"ackel transform to a superintegrable system on a Koenigs space
in the family $K[3,1]$.
\end{enumerate}

\noindent {\bf Basis of conformal symmetries for original system}: 
\[ H_0+V_{[2,1,1]},\]
\[ Q_{12}=(L_{12})^2+b_1(\frac{x_1-ix_2}{x_1+ix_2})+b_2(\frac{x_1-ix_2}{x_1+ix_2})^2,\]
\[ Q_{13}=(L_{23}-iL_{13})^2+\frac{b_2{x_3}^2}{(x_1+ix_2)^2}-\frac{b_3(x_1+ix_2)^2}{{x_3}^2},\]

\noindent {\bf Contraction of basis}:
\[ H_0+V_{[2,1,1]}\to H_0'+V_{[3,1]},\]
\[ Q'_{12}=-2\epsilon^4\ Q_{12}+\frac{c_3}{2\epsilon^4}-c_1=(L'_{12}-iL'_{23})^2+\frac{c_2x_2'}{x_1'+ix_3'}
+\frac{4c_3{x_2'}^2}{(x_1'+ix_3')^2},\]
\[ Q'_{13}=-\frac{\sqrt{2}}{4}(Q_{13}+2\epsilon^2 Q_{12}-\frac{3c_3}{2\epsilon^6}-\frac{\sqrt{2}\ c_2}{4\epsilon^4}+c_1)=\]
\[\frac12\{ L'_{13},L'_{23}+iL'_{12}\}
+\frac{ c_1x'_2}{x'_1+ix'_3}+\frac{\ c_2({x'_4}^2+4{x'_2}^2)}{4(x'_1+ix'_3)^2}+\frac{2 c_3x'_2({x'_4}^2+2{x'_2}^2)}{(x'_1+ix'_3)^3},\]

If we apply the same $[2,1,1]\to [3,1]$ contraction to the $[1,1,1,1]$ system, the system contracts to the $[3,1]$ potential, but with parameters 
$c_1,\cdots,c_4$ where
\[ a_1=\frac{c_1}{\epsilon^8}+c_2\ (\frac{16}{\epsilon^{10}}+\frac{1}{\epsilon^{12}}),\ a_2=\frac{c_2}{\epsilon^{12}},\ a_3=\frac{c_3}{\epsilon^4}+
\frac{8c_1-512c_2}{\epsilon^6}+\frac{64c_2}{\epsilon^8},\ a_4=c_4.\]

If we apply the same contraction to the $[2,2]$ system, the system contracts to one with   potential
\be\label{E3'} V=\frac{c_1}{(x'_1+ix'_3)^2}+\frac{c_2x'_2+c_3x'_4}{(x'_1+ix'_3)^3}+c_4\frac{{x'}_2^2+{x'}_4^2}{(x'_1+ix'_3)^4},\ee
where
\[ b_1=-\frac{\sqrt{2}\ c_2}{4\epsilon^4}+\frac{c_4}{4\epsilon^6},\ b_2=-\frac{c_4}{4\epsilon^4},\ 
b_3=\sqrt{2}\ \frac{(-2c_2+ic_3)}{8\epsilon^6}+\frac{c_4}{8\epsilon^8},\]
\[ b_4=\frac{c_1}{2\epsilon^4}+\sqrt{2}\ \frac{(-ic_3+c_2)}{8\epsilon^6}-\frac{c_4}{16\epsilon^8}.\]
This is conformally equivalent to (\ref{V[0]norm}).

If we apply this same contraction to the  system with $V[3,1]$ potential, the system contracts to one with  $V[3,1]$  potential again, but with different 
parameters, or to $[0]$.

If we apply this same contraction to the  system with $V[4]$ potential, the system contracts to one with   potential (\ref{E3'})  again, but with different parameters.

If we apply this same contraction to the $[0]$ system, (\ref{E3'}) it contracts to itself with altered parameters.

If we apply this same contraction to the $(1)$ system, (\ref{V(1)norm}) it becomes a  potential conformally equivalent to (\ref{V(2)norm}).

If we apply this same contraction to the $(2)$ system, (\ref{V(2)norm}) it contracts to itself with altered parameters.

\subsubsection{[2,1,1] to  [3,1] contraction and St\"ackel transforms}
For fixed $A_j$, $B_j$ we have the expansion
\[ V^A_{[1,1,1,1]}=\frac{A_1}{x_1^2}+\frac{A_2}{x_2^2}+\frac{A_3}{x_3^2}+\frac{A_4}{x_4^2}=\]
\[ \frac{A_4}{{x'_4}^2}+\frac{2A_3}{(x'_1+ix'_3)^2}\epsilon^4+\left(\frac{16(A_2-A_1)}
{(x'_1+ix'_3)^2}-\frac{2\sqrt{2}A_3x'_2}{(x'_1+ix'_3)^3}\right)\epsilon^6+O(\epsilon^8).\]
\[ V^A_{[2,1,1]}=\frac{A_1}{x_1^2}+\frac{A_2}{x_2^2}+\frac{A_3(x_3-ix_4)}{(x_3+ix_4)^3}+\frac{A_4}{(x_3+ix_4)^2} =\frac{2(A_3+A_4)}{(x'_1+ix'_3)^2}\epsilon^4\]
\[+\left(\frac{16(A_2-A_1)}{(x'_1+ix'_3)^2}+\frac{(3A_3+2A_4)\sqrt{2}(-x'_2+2ix'_4)}{(x'_1+ix'_3)^3}
+\frac{A_3\sqrt{2}(x'_2+2ix'_4)}{(x'_1+ix'_3)^3}\right)\epsilon^6+O(\epsilon^8).\]
\[ V^A_{[2,2]}=\frac{A_1}{(x_1+ix_2)^2}+\frac{A_2(x_1-ix_2)}{(x_1+ix_2)^3}+\frac{A_3}{(x_3+ix_4)^2}+\frac{A_4(x_3-ix_4)}{(x_3+ix_4)^3}\]
\[ =-\frac{A_2}{2(x'_1+ix'_3)^2}+\left(-\frac{A_1}{(x'_1+ix'_3)^2}-\frac{3\sqrt{2}A_2x'_2}{(x'_1+ix'_2)^3}\right)\epsilon^2\]
\[ \left(-\frac{\sqrt{2}A_1x'_2}{(x'_1+ix'_3)^3}-\frac{(4{x'_4}^2+19{x'_2}^2)A_2}{(x'_1+ix'_3)^4}\right)\epsilon^4+O(\epsilon^6),\]
\[V^B_{[3,1]}=\frac{B_1}{(x_1+ix_2)^2}+\frac{B_2x_2}{(x_1+ix_3)^3}+\frac{B_3(4x_2^2+x_4^2)}{(x_1+ix_3)^4}+\frac{B_4}{x_4^2}\]
\[=\frac{B_4}{{x'_4}^2}-16\frac{(B_1+iB_2-4B_3)}{(x'_1+ix'_3)^2}\epsilon^6+O(\epsilon^7),\]

\subsection{[1,1,1,1] to [4]:} In this case there is a 2-parameter family of contractions, but all lead to the same result. 
Let $A,B$ be constants such that $AB(1-A)(1-B)(A-B)\ne 0$.

\noindent Coordinate implementation
\[ x_1=\frac{i}{\sqrt{2AB}\ \epsilon^3}(x'_1+ix'_2),\] \[x_2=\frac{(x'_1+ix'_2)+\epsilon^2(x'_3+ix'_4)+\epsilon^4(x'_3-ix'_4)+\epsilon^6(x'_1-ix'_2)}{\sqrt{2(A-1)(B-1)}\ \epsilon^3},\]
\[ x_3=\frac{(x'_1+ix'_2)+A\epsilon^2(x'_3+ix'_4)+A^2\epsilon^4(x'_3-ix'_4)+A^3\epsilon^6(x'_1-ix'_2)}{\sqrt{2A(A-1)(A-B)}\ \epsilon^3},\]
\[ x_4=\frac{(x'_1+ix'_2)+B\epsilon^2(x'_3+ix'_4)+B^2\epsilon^4(x'_3-ix'_4)+B^3\epsilon^6(x'_1-ix'_2)}{\sqrt{2B(B-1)(B-A)}\ \epsilon^3},\]

 In this case:{\small
\bea
iL'_{14}+iL'_{23}+L'_{13}-L'_{24}&=&-2i\epsilon^4\sqrt{AB(A-1)(B-1)}\ L_{12},\\
iL'_{14}-iL'_{23}-L'_{13}-L'_{24}&=&2i\ \epsilon^2\left(\sqrt{B(A-1)(A-B)}\ L_{13}-\sqrt{AB(A-1)(B-1)}\ L_{12}\right),\nonumber\\
L'_{12}&=& \frac{\sqrt{AB}}{\sqrt{(A-1)(B-1)}}L_{12}+\frac{\sqrt{B}}{\sqrt{(A-1)(A-B)}}L_{13}\nonumber\\
&-&\frac{i\sqrt{A}}{\sqrt{(B-1)(A-B)}}L_{14},\nonumber \\
 L'_{34}&=& \frac{\sqrt{B(B-1)}}{\sqrt{A(A-1)}}L_{12}-\frac{\sqrt{B(A-B)}}{\sqrt{(A-1)}}L_{13}+i\frac{\sqrt{(B-1)(A-B)}}{\sqrt{A}}L_{23},\nonumber\\
 -iL'_{14}+iL'_{23}-L'_{13}-L'_{24}&=&\frac{2}{\epsilon^2}\left( \frac{i(A+B-1)}{\sqrt{AB(A-1)(B-1)}}L_{12}+\frac{i\sqrt{B}}{\sqrt{(A-1)(A-B)}}L_{13}\right.\nonumber\\
 &-&\frac{\sqrt{A}}{\sqrt{B(B-1)(A-B)}}L_{14}+\frac{\sqrt{(B-1)}}{\sqrt{A(A-B)}}L_{23}\nonumber\\
 &-&\left.\frac{i\sqrt{(A-1)}}{\sqrt{B(A-B)}}L_{24}\right),\nonumber\\
   iL'_{14}+iL'_{23}-L'_{13}+L'_{24}&=&\frac{2i}{\epsilon^4}
   \left(-\frac{1}{\sqrt{AB(A-1)(B-1)}}(L_{12}+L_{34})\right.\nonumber\\
 &+&\frac{i}{\sqrt{A(B-1)(A-B)}}(L_{14}+L_{23})\nonumber\\
 &-&\left.\frac{1}{\sqrt{B(A-1)(A-B)}}(L_{13}-L_{24}) \right). \nonumber \eea}

\noindent {\bf Limit of 2D potential}:
\[ V_{[1,1,1,1]}
\stackrel{\epsilon\ \to\ 0}{\Longrightarrow}\ V_{[4]},\]
where 
\be\label{V[4]} V_{[4]}=\frac{d_1}{(x'_1+ix'_2)^2}+\frac{d_2(x'_3+ix'_4)}{(x'_1+ix'_2)^3}\ee
\[+d_3\left(\frac{3({x'_3}+ix'_4)^2}{(x'_1+ix'_2)^4}-2\frac{(x'_1+ix'_2)(x'_3-ix'_4)}{(x'_1+ix'_2)^4}\right)+\]
\[d_4\frac{4(x'_1+ix'_2)({x'_1}^2+{x'_2}^2)+2(x'_3+ix'_4)^3
}{(x'_1+ix'_2)^5}    .\]  
and 
\[ a_1=-\frac{d_4}{4A^2B^2\epsilon^{12}}-\frac{d_3}{2AB^2\epsilon^{10}}-\frac{ d_2}{4AB\epsilon^8}-\frac{d_1}{2AB\epsilon^6},\]
\[ a_2=- 
\frac{d_4}{4(1-A)^2(1-B)^2\epsilon^{12}}+\frac{d_3}{2(1-A)(1-B)^2\epsilon^{10}}-\frac{d_2}{4(1-A)(1-B)\epsilon^8},\]
\[ a_3=-\frac{d_4}{4A^2(1-A)^2(A-B)^2\epsilon^{12}},\]
\[ a_4=-\frac{d_4}{4B^2(1-B)^2(A-B)^2\epsilon^{12}}-\frac{d_3}{2B^2(1-A)^2(A-B)\epsilon^{10}}.\]

\subsubsection{Conformal St\"ackel transforms of the [4] system}  We write potential $V_{[4]}$ in the normalized form
\be\label{V[4]norm} V'_{[4]}=\frac{a_1}{(x_3+ix_4)^2}+a_2\frac{x_1+ix_2}{(x_3+ix_4)^3}
+a_3\frac{3(x_1+ix_2)^2-2(x_3+ix_4)(x_1-ix_2)}{(x_3+ix_4)^4}\ee
\[+a_4\ \frac{4(x_3+ix_4)(x_3^2+x_4^2)+2(x_1+ix_2)^3}{(x_3+ix_4)^5},\]
and designate it  $(a_1,a_2,a_3,a_4)$.
\begin{enumerate}
\item The potentials $(1,a,0,0)$ generate  conformal St\"ackel transforms to $E10$. 
\item The potential $(0,1,0,0)$ generates a  conformal St\"ackel transform to $E9$.
\item The potential $(0,0,0,1))$  generates a conformal St\"ackel transform to $D1A$.
\item Each potential not proportional to one of these must generate a conformal St\"ackel transform to a superintegrable system on a Koenigs space
in the family $K[4]$.
\end{enumerate}
In these coordinates a basis for the conformal symmetry algebra is $H,Q_1,Q_2$ where
\[ Q_1=\frac14(L_{14}+L_{23}-iL_{13}+iL_{24})^2+4a_3(\frac{x_1+ix_2}{x_3+ix_4})+4a_4(\frac{x_1+ix_2}{x_3+ix_4})^2,\]
\[ Q_2=\frac12\{L_{23}+L_{14}-iL_{13}+iL_{24},L_{12}+L_{34}\}+\frac14(L_{14}-L_{23}+iL_{13}+iL_{24})^2\]
\[+2a_1(\frac{x_1+ix_2}{x_3+ix_4}) +a_2\left(2\frac{x_1-ix_2}{x_3+ix_4}-(\frac{x_1+ix_2}{x_3+ix_4})^2\right)\]
\[+2a_3\left(6(\frac{x_1^2+x_2^2}{(x_3+ix_4)^2})-(\frac{x_1+ix_2}{x_3+ix_4})^3\right)\]
 \[ -4a_4\left((\frac{x_1-ix_2}{x_3+ix_4})^2-3(\frac{(x_1+ix_2)^2(x_1-ix_2)}{(x_3+ix_4)^3}+\frac14(\frac{x_1+ix_2}{x_3+ix_4})^4\right).\]

\noindent {\bf Basis of conformal symmetries for original system}: 
\[ H_0+V_{[1,1,1,1]},\ Q_{12},\ Q_{13},\]
where
\[Q_{jk}=(x_j\partial_{x_k}-x_k\partial_{x_j})^2+a_j\frac{x_k^2}{x_j^2}+a_k\frac{x_j^2}{x_k^2},\  1\le j<k\le 4.\]

\noindent {\bf Contraction of basis}: 
\[H_0+V_{[1,1,1,1]}\to H_0'+V_{[4]},\]

\[\epsilon^8 Q_{12}\sim \frac{-1}{4(A-1)(B-1)AB}(L'_{13}-L'_{24}+iL'_{23}+iL'_{14})^2\]
\[+\frac{4d_3(x'_3+ix'_4)}{AB(A-1)(B-1)(x'_1+ix'_2)}\]
\[+\frac{d_4}{4AB(A-1)(B-1)}\left[\frac{(x'_3+ix'_4)^2}{(x'_1+ix'_2)^2}+2\frac{x_3'-ix_4'}{x_1'+ix_2'}\right],\]
In this case we do not obtain a basis of symmetries for the $[4]$ system. The basis can be computed from the contracted potential.

If we apply the same $[1,1,1,1]\to [4]$ contraction to the $[2,1,1]$ system, the system contracts to a modified $[4]$ potential, 
of the form
\[{\tilde V}_{[4]}=\frac{d'_1}{(x'_1+ix'_2)^2}+\frac{d'_2(x'_3+ix'_4)}{(x'_1+ix'_2)^3}\]
\[+d'_3\left(\frac{3 ({x'_3}+ix'_4)^2}{(x'_1+ix'_2)^4}-2\lambda \frac{(x'_1+ix'_2)(x'_3-ix'_4)}{(x'_1+ix'_2)^4}\right)+\]
\[d'_4\ \left(\frac{4\lambda (x'_1+ix'_2)({x'_1}^2+{x'_2}^2)}{(x'_1+ix'_2)^5}+\frac{2(x'_3+ix'_4)^3
}{(x'_1+ix'_2)^5}\right)
,\]  
where $\lambda$ is a nonzero function of $A$ and $B$. However, under an appropriate conformal transformation 
\[ x_1'+ix_2'\to \mu(x_1'+ix_2'), \ x'_1-ix'_2 \to \mu^{-1}(x_1'-ix_2'),\]
we obtain the potential $V_{[4]}$ exactly.

If we apply the same contraction to the $[2,2]$ system, the system contracts to 
\be\label{E3p} V=\frac{e_1}{(x'_1+ix'_2)^2}+e_2\frac{(x'_3+ix'_4)}{(x'_1+ix'_2)^3}+e_3\frac{ (x'_3-ix'_4)}{(x'_1+ix'_2)^3}
+e_4\frac{({x'}_3^2+{x'}_4^2)}{(x'_1+ix'_2)^4},\ee
conformally equivalent to (\ref{V[0]norm}).

If we apply the same contraction to the $[3,1]$ system, the system contracts to 
\[ V=\frac{f_1}{(x'_1+ix'_2)^2}+f_2\frac{(x'_3+ix'_4)}{(x'_1+ix'_2)^3}\]
\[+\frac{f_3}{(x'_1+ix'_2)^4}\left[3
\lambda(x'_3+ix'_4)^2+(x'_1+ix'_2)(x'_3-ix'_4)\right] \]
\[+\frac{f_4(x'_3+ix'_4)}{(x'_1+ix'_2)^5}\left[\lambda(x'_3+ix'_4)^2+(x'_1+ix'_2)(x'_3-ix'_4)\right],\]
where the nonzero scalar $\lambda$ depends on the choice of $A$ and $B$. It can be rescaled to any desired nonzero value by a conformal  transform 
\[ x_1'+ix_2'\to \mu(x_1'+ix_2'), \ x'_1-ix'_2 \to \mu^{-1}(x_1'-ix_2').\] This system  is conformally equivalent to (\ref{V[4]norm}) again.

If we apply the same contraction to the $[4]$ system, the system contracts to one with potential (\ref{E3p}) again, but with different parameters.

If we apply the same contraction to the $[0]$ system (\ref{E3p}) the system contracts to one with potential (\ref{E3p}) again, but with different parameters.

If we apply this same contraction to the $(1)$ system, (\ref{V(1)norm}) it becomes a  potential conformally equivalent to (\ref{V(2)norm}).

If we apply this same contraction to the $(2)$ system, (\ref{V(2)norm}) it contracts to itself with altered parameters.

\subsubsection{[1,1,1,1] to [4] contraction and St\"ackel transforms}
For fixed $A_j$ we have (in the special case $A=10,B=5$) the expansions
\[ V^A_{[1,1,1,1]}=\frac{A_1}{x_1^2}+\frac{A_2}{x_2^2}+\frac{A_3}{x_3^2}+\frac{A_4}{x_4^2}\]
\[=\frac{4(-5A_1+2A_2+30A_4+3A_3)}{(x'_1+ix'_2)^2}\epsilon^6+\frac{16(-A_2+3A_3-75A_4)(x'_3+ix'_4)}{(x'_1+ix'_2)^3}\epsilon^8+O(\epsilon^{10}).\]
\[ V^A_{[2,1,1]}=\frac{A_1}{x_1^2}+\frac{A_2}{x_2^2}+\frac{A_3(x_3-ix_4)}{(x_3+ix_4)^3}+\frac{A_4}{(x_3+ix_4)^2}\]
\[=\frac{-\frac{4}{127}(135A_1-54A_2+[110+2\sqrt{10}]A_4)-\frac{40}{243}(161+44\sqrt{10})A_3}{(x'_1+ix'_2)^2}\epsilon^6+O(\epsilon^8),\]

\subsection{[2,2] to [4]:}
\bea L'_{12}&=&i(1+\frac{2}{\epsilon}-\frac{1}{2\epsilon^2})L_{12}+\frac{1}{\epsilon}(1-\frac{3}{4\epsilon}+\frac{1}{4\epsilon^2})L_{13}
 +\frac{i}{4\epsilon^2}(3-\frac{1}{\epsilon})L_{14}\nonumber\\
 &+&\frac{i}{4\epsilon^2}(3-\frac{1}{\epsilon})L_{23}+(3-\epsilon+\frac{3}{4\epsilon^2}-
 \frac{1}{4\epsilon^3})L_{24}+i(\frac{3\epsilon}{2}-2+\frac{1}{\epsilon}-\frac{1}{2\epsilon^2})L_{34},\nonumber\\
 L'_{12}+iL'_{24}&=&\epsilon(L_{13}-iL_{14}),\\
 L'_{13}+iL'_{34}&=&\epsilon(L_{23}-iL_{24}),\nonumber\\
 L'_{14}&=&(-1+\epsilon)L_{12}+i(1-\epsilon)L_{13}+(1+\epsilon)L_{14},\nonumber\\
 L'_{23}-L'_{14}&=&-L_{14}+L_{23},\nonumber\\
L'_{13}+L'_{24}&=& (\frac12-\frac{1}{\epsilon})L_{12}+\frac{i}{\epsilon}L_{13}+\frac12 L_{14}+\frac12L_{23}+(2+\frac{i}{\epsilon})L_{24}+(\epsilon-\frac12+\frac{1}{\epsilon})L_{34}.\nonumber \eea

\noindent {\bf Coordinate implementation}:
 \[ x_1=\frac{1}{2}(\frac{1}{\epsilon}+\frac{1}{\epsilon^2})(x'_1-ix'_4)+\frac{\epsilon}{2}(x'_1+ix'_4)
 -(1+\frac{1}{2\epsilon})(x'_2-ix'_3)+\frac{1}{2}(\epsilon-1)(x'_2+ix'_3),\]   
 \[ x_2=\frac{i}{2}(\frac{1}{\epsilon}-\frac{1}{\epsilon^2})(x'_1-ix'_4)-\frac{i\epsilon}{2}(x'_1+ix'_4)
 -i(1-\frac{1}{2\epsilon})(x'_2-ix'_3)+\frac{i}{2}(\epsilon+1)(x'_2+ix'_3),\]
 \[ x_3=\frac{1}{2}(\frac{1}{\epsilon}-\frac{1}{\epsilon^2})(x'_1-ix'_4)+(-\frac12+\frac{1}{\epsilon})(x'_2-ix'_3),\]
 \[x_4=\frac{i}{2}(\frac{1}{\epsilon}+\frac{1}{\epsilon^2})(x'_1-ix'_4)-i(\frac12+\frac{1}{\epsilon})(x'_2-ix'_3).\]
 
\noindent
{\bf Limit of 2D potential}: \[ V_{[2,2]}
\stackrel{\epsilon\ \to\ 0}{\Longrightarrow}\ V'_{[4]},\]

\be\label{V[4]'} V'_{[4]}=\frac{e_1}{(x'_1-ix'_4)^2}+\frac{e_2(x'_2-ix'_3)}{(x'_1-ix'_4)^3}\ee
\[+e_3\left(\frac{3({x'_2}-ix'_3)^2}{(x'_1-ix'_4)^4}+2\frac{(x'_1-ix'_4)(x'_2+ix'_3)}{(x'_1-ix'_4)^4}\right)+\]
\[e_4\left(\frac{4(x'_1-ix'_4)({x'_2}^2+{x'_3}^2)+2(x'_2-ix'_3)^3
}{(x'_1-ix'_4)^5} \right)   ,\]  
where
\[ b_1=\frac{e_1}{\epsilon^4}+2\frac{e_4}{\epsilon^7},\ b_2=-\frac{e_2}{4\epsilon^6}-\frac{e_3}{2\epsilon^7}-\frac{e_4}{\epsilon^8},
\ b_3=2\frac{e_3}{\epsilon^6}-2\frac{e_4}{\epsilon^7},\ b_4=-\frac{e_2}{4\epsilon^6}+\frac{3e_3}{2\epsilon^7}-\frac{e_4}{\epsilon^8}.\]
This is conformally equivalent to $V[4]$.

\medskip\noindent

\noindent {\bf Basis of conformal symmetries for original system}: 
\[ H_0+V_{[2,2]},\ Q_1,\ Q_3\]

\medskip
\noindent{ \bf Contraction of basis}:

\[ H_0+V_{[2,2]}\to H'_0+V'_{[4]},\]
\[-4\epsilon^4( Q_1+\frac{k_4}{\epsilon^6}-\frac{k_3}{2\epsilon^5})\to (iL'_{13}-L'_{12}-iL'_{24}-L'_{34})^2\]
\[+k_2+4k_3\frac{x'_2-ix'_3}{x'_1-ix'_4}-4k_4\frac{(x'_2-ix'_3)^2}{(x'_1-ix'_4)^2},\]
\[ \epsilon^3 (Q_3-\frac{2k_4}{\epsilon^7}+\frac{k_3}{\epsilon^6}+
\frac{k_1}{2\epsilon^4})\to \]
\[\frac{i}{2}
\{L'_{23}-L'_{14},(L'_{12}-iL'_{13}+L'_{24}+L'_{34}\}+k_1\frac{(x'_2-ix'_3)}{(x'_1-ix'_4)}+
k_2\frac{(x'_2-ix'_3)^2}{(x'_1-ix'_4)^2}\]
\[ +k_3\frac{3(x'_2-ix'_3)^3+2({x'_2}^2+{x'_3}^2)(x'_1-ix'_4)}{(x'_1-ix'_4)^3}\]
\[-2k_4(x'_2-ix'_3)\frac{(x'_2-ix'_3)^3+2({x'_2}^2+{x'_3}^2)(x'_1-ix'_4)}{(x'_1-ix'_4)^4}.\]
However, the second limit here is equivalent to the contracted Hamiltonian, not an independent basis element.

\medskip
If we apply the $[2,2]\to [4]$  contraction to the $[1,1,1,1]$ system, the system contracts to 
\[ {V[4]}''=\frac{f_1}{(x'_1-ix'_4)^2}+f_2\frac{(x'_2-ix'_3)}{(x'_1-ix'_4)^3}\]
\[+\frac{f_3}{(x'_1-ix'_4)^4}
\left[3(x'_2-ix'_3)^2+2(x'_1-ix'_4)(x'_2+ix'_3)\right] \]
\be\label{V4pp}+\frac{f_4(x'_2-ix'_3)}{(x'_1-ix'_4)^5}\left[(x'_2-ix'_3)^2+2(x'_1-ix'_4)(x'_2+ix'_3)\right],\ee
where
\bea  b_1&=&\frac{f_1+2f_3}{4\epsilon^4}+\frac{f_2+10f_4}{64\epsilon^6}-\frac{f_3-4f_4}{32\epsilon^7}+\frac{f_4}{32\epsilon^8},\nonumber\\
 b_2&=&\frac{f_2+10f_4}{64\epsilon^6}-\frac{f_3+4f_4}{32\epsilon^7}+\frac{f_4}{32\epsilon^8},\nonumber\\
b_3&=& \frac{f_2-16f_3+10f_4}{64\epsilon^6}+\frac{f_3-4f_4}{32\epsilon^7}+\frac{f_4}{32\epsilon^8},\nonumber\\
b_4&=&\frac{f_2+16f_3+10f_4}{64\epsilon^6}+\frac{3f_3+4f_4}{32\epsilon^7}+\frac{f_4}{32\epsilon^8},\nonumber\eea
also conformally equivalent to $V_{[4]}$.

If we apply the same contraction to the $[2,1,1]$ system, the system contracts to potential $(2)$, or to (\ref{V4pp}) again, except that now
\bea  b_1&=&\frac{f_1-f_3}{\epsilon^4}-\frac{2f_2+5f_4}{2\epsilon^5}-\frac{2f_3}{\epsilon^6}+\frac{f_4}{\epsilon^7},\nonumber\\
 b_2&=&\frac{3f_3}{2\epsilon^7}-\frac{f_4}{2\epsilon^8},\nonumber\\
b_3&=& \frac{f_2+7f_4}{4\epsilon^5}-\frac{f_2+7f_4}{16\epsilon^6}+\frac{f_3-4f_4}{32\epsilon^7}+\frac{f_4}{32\epsilon^8},\nonumber\\
b_4&=-&\frac{f_2+7f_4}{4\epsilon^5}-\frac{f_2+7f_4}{16\epsilon^6}+\frac{f_3+4f_4}{32\epsilon^7}+\frac{f_4}{32\epsilon^8}.\nonumber\eea

If we apply the same contraction to the $[3,1]$ system, the system contracts to a system with potential
\be\label{V2'} V(2)=\frac{c_1}{(x'_1-ix'_4)^2}+\frac{c_2(x'_2-ix'_3)}{(x'_1-ix'_4)^3}+\frac{c_3(x'_2-ix'_3)^2}{(x'_1-ix'_4)^4}
 +\frac{c_4(x'_2-ix'_3)^3}{(x'_1-ix'_4)^5}.\ee
 This system admits a first order symmetry. It corresponds to a special case of the  flat space superintegrable system $E15$ via the
 transform $(x'_1-ix'_4)^2$.
 
 If we apply the same contraction to the $[4]$ system, the system contracts to a system with potential (\ref{V[4]norm}) again, but with different parameters.
 
 If we apply the same contraction to the $[0]$ system (\ref{E3p}) the system contracts to one with potential (\ref{E3p}) again, but with different parameters. or to $(2)$.

 If we apply this same contraction to the $(1)$ system, (\ref{V(1)norm}) it becomes a  potential conformally equivalent to (\ref{V(2)norm}).

If we apply this same contraction to the $(2)$ system, (\ref{V(2)norm}) it contracts to itself with altered parameters.
 
\subsubsection{[2,2] to  [4] contraction and St\"ackel transforms}
For fixed $A_j$ we have the expansions
\[ V^A_{[1,1,1,1]}=\frac{A_1}{x_1^2}+\frac{A_2}{x_2^2}+\frac{A_3}{x_3^2}+\frac{A_4}{x_4^2}
=4\frac{A_1-A_2+A_3-A_4}{(x'_1-ix'_4)^2}\epsilon^4\] \[-8\frac{(A_1+A_2-A_3-A_4)(x'_1-ix'_4)-(A_1-A_2+2A_3-2A_4)
(x'_2+ix'_3)}{(x'_1-ix'_4)^3}\epsilon^5\]
\[+O(\epsilon^6),\]
\[ V^A_{[2,1,1]}=\frac{A_1}{x_1^2}+\frac{A_2}{x_2^2}+\frac{A_3(x_3-ix_4)}{(x_3+ix_4)^3}+\frac{A_4}{(x_3+ix_4)^2}=
\frac{(4A_1-4A_2+A_4)}{(x'_1-ix'_4)^2}\epsilon^4\]
\[-\frac{[(8A_1+8A_2+A_3)(x'_1-ix'_4)+4(-2A_1+2A_2-A_4)(x'_2-ix'_3)}{(x'_1-ix'_4)^3}\epsilon^5+O(\epsilon^6),\]
\[  V^A_{[2,2]}=\frac{A_1}{(x_1+ix_2)^2}+\frac{A_2(x_1-ix_2)}{(x_1+ix_2)^3}
+\frac{A_3}{(x_3+ix_4)^2}+\frac{A_4(x_3-ix_4)}{(x_3+ix_4)^3}\]
\[ = \frac{A_1+A_3}{(x'_1-ix'_4)^2}\epsilon^4+\left(\frac{(2A_1+4A_3)(x'_2-ix'_3)}{(x'_1-ix'_4)^3}
+\frac{(A_2-A_4)}{(x'_1-ix'_4)^2}\right)\epsilon^5+O(\epsilon^6). \]

 \subsection{[3,1] to [4]} This specific contraction is not needed because already the $[1,1,1,1]\to[4]$ contraction takes the system $V[3,1]$ to $V[4]$.
 
\subsection{[2,1,1] to [4]} This specific contraction is not needed because already the $[1,1,1,1]\to[4]$ contraction takes the system $V[2,1,1]$ to $V[4]$.

\subsection{[1,1,1,1] to  [3,1]}
\bea -L'_{12}+iL'_{24}&=& -a\sqrt{2a^2-2}\ \epsilon L_{12}\\
L'_{13}&=& -\frac{i}{\sqrt{a^2-1}}(L_{13}+aL_{12}),\nonumber\\
L'_{14}+iL'_{34}&=&\sqrt{2}\ a\epsilon L_{14},\nonumber\\
-L'_{12}+iL'_{23}&=&i\sqrt{2}a\epsilon L_{23},\nonumber\\
L'_{24}&=&i(\sqrt{a^2-1}\ L_{24}-iaL_{14}),\nonumber\\
-L'_{14}+iL'_{34}&=& \frac{\sqrt{2}}{\epsilon\ a\sqrt{a^2-1}}\left( L_{34}-\sqrt{a^2-1}L_{14}-iaL_{24}\right).\nonumber\eea

\noindent {\bf Coordinate implementation}:
\[ x_1=\frac{1}{\sqrt{2}\ a\epsilon}(x'_1+ix'_3)+\frac{x'_2}{a}+\frac{a\epsilon}{\sqrt{2}}(x'_1-ix'_3),\] 
\[x_2=\frac{i(x'_1+ix'_2)}{\sqrt{2a^2-2}\ \epsilon},\]
\[ x_3=-\frac{(x'_1+ix'_3)}{\sqrt{2a^2-2}\ a\epsilon}+\frac{\sqrt{a^2-1}}{a} x'_2,\ x_4=x'_4,\]
where $a$ is a parameter such that $a(a-1)\ne 0$.

\noindent {\bf Limit of 2D potential}:
\[ V_{[1,1,1,1]}
\stackrel{\epsilon\ \to\ 0}{\Longrightarrow}\ V_{[31]},\]
where $V[31]$ is given by (\ref{V[31]}) 
and 
\[ a_1=\frac{c_1}{2\epsilon^2}+\frac{c_3}{4a^4\epsilon^4},a_2=\frac{c_2}{4\sqrt{2}(a^2-1)^2\epsilon^3}+\frac{c_3}{4 (a^2-1)^2\epsilon^4},\]
\[ a_3=\frac{c_2}{4\sqrt{2}(a^2-1)^2a^2\epsilon^3}+\frac{c_3}{4 (a^2-1)^2a^4\epsilon^4},\ a_4=c_4.\]

\noindent {\bf Basis of conformal symmetries for original system}: 
\[ H_0+V_{[1,1,1,1]},\ Q_{12},\ Q_{13},\]
where
\[Q_{jk}=(x_j\partial_{x_k}-x_k\partial_{x_j})^2+a_j\frac{x_k^2}{x_j^2}+a_k\frac{x_j^2}{x_k^2},\  1\le j<k\le 4.\]

\noindent {\bf Contracted  basis}: 
\[H_0+V_{[1,1,1,1]}\to H_0'+V_{[3,1]},\]
\[\epsilon^2\left( Q_{12}+\frac{c_3}{2a^2(a^2-1)\epsilon^4}
+\frac{\sqrt{2}c_2}{a^2(a^2-1)\epsilon^3}\right)\to -\frac{c_1}{2(a^2-1)}\]
\[-\frac{2c_3{x'_2}^2}{a^2(a^2-1)(x'_1+ix'_3)^2}-\frac{c_2}{2a^2(a^2-1)(x'_1+ix'_3)}-\frac{1}{2a^2(a^2-1)}(L'_{12}-iL'_{23})^2,\]
\[\epsilon\left(Q_{13}+a^2Q_{12}+\frac{(a^2-1)c_3}{2a^4\epsilon^4}+\frac{\sqrt{2}\ c_2}{8a^2\epsilon^3}+\frac{c_1(a^2-1)}{2\epsilon^2}\right)
\to \frac{\sqrt{2}\ c_1 x'_2}{x'_1+ix'_3}\]
\[ +\frac{\sqrt{2}\ c_2(4{x'_2}^2+{x'_4}^2)}{4(x'_1+ix'_3)^2}+\frac{2\sqrt{2}\ c_3 x'_2(2{x'_2}^2+{x'_4}^2)}{(x'_1+ix'_3)^3}
+\frac{i\sqrt{2}}{2}\{ L'_{13},L'_{12}-iL'_{23}\}.\]

\medskip

If we apply the $[1,1,1,1]\to [3,1]$  contraction to the $[2,1,1]$ system, the system contracts to one with potential $V[3,1]$, but with different parameters, or to $[0]$.

If we apply the same contraction to the $[2,2]$ system, the system again contracts to one with potential $V[0]$, but with different parameters.

If we apply the same contraction to the $[3,1]$ system, the system contracts to  itself, but with different parameters.

If we apply the same contraction to the $[4]$ system, the system contracts to the system with potential $V[0]$, (\ref{E3'}), but with altered parameters.

If we apply the same contraction to the $[0]$ system, the system contracts to the system with potential $V[0]$, (\ref{E3'}), but with altered parameters.

 If we apply this same contraction to the $(1)$ system, (\ref{V(1)norm}) it becomes a  potential conformally equivalent to (\ref{V(2)norm}).

If we apply this same contraction to the $(2)$ system, (\ref{V(2)norm}) it contracts to itself with altered parameters.

\subsubsection{[1,1,1,1] to [3,1] contraction and St\"ackel transforms}
For fixed $A_j$ we have the expansions
\[  V^A_{[1,1,1,1]}=\frac{A_1}{x_1^2}+\frac{A_2}{x_2^2}+\frac{A_3}{x_3^2}+
\frac{A_4}{x_4^2}= \]
\[ \frac{A_4}{x_4^2}+\frac{2\left(A_2+(A_1-A_2-A_3)a^2+A_3a^4\right)\epsilon^2}{(x_1+ix_3)^2}\]
\[+\frac{4\sqrt{2}a^2x_2(A_3-A_1-2A_3a^2+A_3a^4)\epsilon^3}{(x_1+ix_3)^3}\]
\[-\frac{4a^2\left(A_1a^2x_1^2+(-3A_1 +3A_3(1-a^2))x_2^2   +A_1a^2x_3^2\right)\epsilon^4}{(x_1+ix_3)^4}+O(\epsilon^5),\]
\[ V^A_{[2,1,1]}=\frac{A_1}{x_1^2}+\frac{A_2}{x_2^2}+\frac{A_3(x_3-ix_4)}{(x_3+ix_4)^3}+\frac{A_4}{(x_3+ix_4)^2}\]
\[ =\frac{2\left(A_1a^2+A_2(1-a^2)+(A_3+A_4)a^2(1-a^2)^2\right)}{(x'_1+ix'_3)^2}\epsilon^2+O(\epsilon^3),\]
\[ V^A_{[2,2]}=\frac{A_1}{(x_1+ix_2)^2}+\frac{A_2(x_1-ix_2)}{(x_1+ix_2)^3}+\frac{A_3}{(x_3+ix_4)^2}+\frac{A_4(x_3-ix_4)}{(x_3+ix_4)^3}\]
   \[ =\frac{k_1A_1+k_2A_2+k_3(A_3+A_4)}{(x'_1+ix'_3)^2}\epsilon^2+O(\epsilon^3),\ k_1,k_2,k_3\ {\rm generic},\]
\[ V^A_{[3,1]}=\frac{A_1}{(x_3+ix_4)^2}+\frac{A_2x_1}{(x_3+ix_4)^3}+\frac{A_3(4x_1^2+x_2^2)}{(x_3+ix_4)^4}+\frac{A_4}{x_2^2}\]
\[=\left[\frac{2A_1a^2-2A_2a^2\sqrt{a^2-1}+4A_3a^2(3a^2-4)-2A_4}{(x'_1+ix'_3)^2}\right](a^2-1)\epsilon^2+O(\epsilon^4),\]

\subsubsection{Conformal St\"ackel transforms of the (1) system}  We write potential $V(1)$ in the form
\be\label{V(1)norm} V(1)=a_1\frac{1}{(x_1+ix_2)^2}+a_2\frac{1}{(x_3+ix_4)^2}
+a_3\frac{(x_3+ix_4)}{(x_1+ix_2)^3}+a_4\frac{(x_3+ix_4)^2}{(x_1+ix_2)^4}\ee
and designate it  $(a_1,a_2,a_3,a_4)$, defining the conformally superintegrable system $[1]$. For every choice of $(a_1,a_2,a_3,a_4)$ the potential 
$V(1)$ generates  a conformal St\"ackel transform to 
a special case of $E15$, always flat.
 
\subsubsection{Conformal St\"ackel transforms of the (2) system}  We write potential $V(2)'$ in the normalized form
\be\label{V(2)norm} V(2)'=a_1\frac{1}{(x_3+ix_4)^2}+a_2\frac{(x_1+ix_2)}{(x_3+ix_4)^3}
+a_3\frac{(x_1+ix_2)^2}{(x_3+ix_4)^4}+a_4\frac{(x_1+ix_2)^3}{(x_3+ix_4)^5}\ee
and designate it  $(a_1,a_2,a_3,a_4)$, defining the conformally superintegrable system $[2]$. For every choice of $(a_1,a_2,a_3,a_4)$ 
the potential $V(2)'$ generates  a conformal St\"ackel transform to 
a special case of $E15$, always flat.

\section{Helmholtz contractions from B\^ocher contractions} We describe how B\^ocher contractions of conformal superintegrable systems induce contractions 
of Helmholtz  superintegrable systems. The basic idea here is that the procedure of taking a conformal St\"ackel transform of a conformal superintegrable system, followed
by a Helmholtz contraction yields the same result as taking a B\^ocher contraction followed by an ordinary St\"ackel transform: The 
diagrams commute. We illustrate with an example.

We consider the conformal St\"ackel transforms of the conformal system $[1,1,1,1]$ with potential $V_{[1,1,1,1]}$. 
The various possibilities are listed in subsection \ref{3.1.1}. Let $H$ be the initial Hamiltonian. In
terms of tetraspherical coordinates the conformal St\"ackel transformed potential will 
take the form 
\[ V=\frac{\frac{a_1}{x_1^2}+\frac{a_2}{x_2^2}+\frac{a_3}{x_3^2}+\frac{a_4}{x_4^2}}{\frac{A_1}{x_1^2}+\frac{A_2}{x_2^2}+\frac{A_3}{x_3^2}+\frac{A_4}{x_4^2}}
=\frac{V_{[1,1,1,1]}}{F({\bf x},{\bf A})},\]
where
\[ F({\bf x},{\bf A})=\frac{A_1}{x_1^2}+\frac{A_2}{x_2^2}+\frac{A_3}{x_3^2}+\frac{A_4}{x_4^2},\]
and the transformed Hamiltonian will be 
\[{\hat H}=\frac{1}{ F({\bf x},{\bf A})}H,\]
where the transform is determined by the fixed vector $(A_1,A_2,A_3,A_4)$. Now we apply the B\^ocher contraction $[1,1,1,1]\to [2,1,1]$ to this system.
In the limit as $\epsilon\to 0$ 
the potential $V_{[1,1,1,1]}\to V_{[2,1,1]}$, (\ref{V[211]}), and $H\to H'$ the $[2,1,1]$ system. Now consider
\[ F({\bf x}(\epsilon),{\bf A})= V'({\bf x}',A)\epsilon^\alpha+O(\epsilon^{\alpha+1}),\]
where the  the integer exponent $\alpha$ depends upon our choice of $\bf A$. We will provide the theory to show that the system defined by Hamiltonian
\[ {\hat H}'=\lim_{\epsilon\to 0}\epsilon^\alpha {\hat H}(\epsilon)=\frac{1}{V'({\bf x}',A)}H'\]
is a superintegrable system that arises from the  system $[2,1,1]$ by a conformal St\"ackel transform induced by the potential $V'({\bf x}',A)$.
Thus the Helmholtz superintegrable system with potential $V=V_{1,1,1,1}/F$ contracts to the Helmholtz superintegrable system with potential $V_{[2,1,1]}/V'$.
The contraction is induced by a generalized In\"on\"u-Wigner Lie algebra contraction of the conformal algebra $so(4,\C)$.
In this case the possibilities for $V'$ can be read off from the expression in subsection \ref{3.1.2}. Then the $V'$ can be identified with a 
$[2,1,1]$ potential from the list in subsection \ref{3.1.3}. The results  follow. For each $\bf A$
corresponding to a constant curvature or Darboux superintegrable system $O$ we list the contracted system $O'$ and $\alpha$. For Koenigs spaces we will not go into detail 
but merely give the contraction for a ``generic'' Koenigs system: One for which  there are no rational numbers $r_j$, not all $0$, such that 
$\sum_{j=1}^4r_jA_j=0$. This ensures that the contraction is also ``generic".

\begin{example}
 In Section \ref{3.1.2}, first equation, consider the St\"ackel transform  $(1,0,0,0)$, i.e.,  $1/x_1^2$ .
The transformed system is
\[H=\frac{1}{\frac{1}{x_1^2}}(\sum_{i=1}^4 \partial_{x_i}^2)+ \frac{1}{\frac{1}{x_1^2}}(\frac{a_1}{x_1^2}+\frac{a_2}{x_2^2}+\frac{a_3}{x_3^2}+\frac{a_4}{x_4^2})\]
which is $S9$. Now take the $[1,1,1,1]\to [2,1,1]$  Bocher contraction, equation (\ref{V[211]}). The sum of the derivatives in $H$ goes to $\sum_{i=1}^4 \partial_{x'_i}^2$ and the numerator of the potential goes to
equation (\ref{V[211]}). However, the denominator $1/x_1^2$ goes as
\[1/x_1^2=-2 \epsilon^2/((x_1'+ix_2')^2 +O(\epsilon^6)\]
from the first equation in Section \ref{3.1.2}, case $A_1=1$, $A_2=0$, $A_3=0$, $A_4=0$. 
 Thus, if we set
$H'=\epsilon^2  H$ and go to the limit as $\epsilon \to 0$, we get 
a contracted system with potential 
$b_1+b_2(x^2+y^2)+b_3/x^2+b_4/y^2$
in Cartesian coordinates, up to a scalar factor $-2$. This is $E1$.
\end{example}

\subsection{Contraction [1,1,1,1] to [2,1,1] applied to conformal St\"ackel transforms of system V[1,1,1,1].}
\begin{enumerate}
 \item \[ {\bf A}=(1,0,0,0),\ (0,1,0,0),\quad O=S_9\to O'=E_1,\quad \alpha =2,\]
 \[ {\bf A}=(0,0,1,0),\ (0,0,0,1)\quad  O=S_9\to O'=S_2,\quad \alpha =0,\]
 \item\[ {\bf A}=(1,1,1,1),\quad O=S_8\to O'=S_4,\quad \alpha =0,\]
  \[ {\bf A}=(0,1,0,1), \ (1,0,1,0)\quad  O=S_8\to O'=S_2,\quad \alpha =0,\]
 \item \[{\bf A}=(0,0,1,1),\quad O=S_7\to O'=S_4,\quad \alpha =0,\]
 \[ {\bf A}=(1,1,0,0,)\quad  O=S_7\to O'=E_{16},\quad \alpha =4,\]
 
  \item  \[ {\bf A}=(A_1,A_2,0,0),\ (A_1A_2\ne 0,A_1\ne A_2),\ O=D4B\to O'=E_1,\quad \alpha =2,\]
  \[ (0,0,A_1,A_2), \  O=D4B\to O'=D4A,\quad \alpha =0,\]
   \[{\bf A}= {\rm all\ other\ permutations}, \  O=D4B\to O'=S_2,\quad \alpha =0,\]
 \item \[{\bf A}=(1,1,A,A),\ (A,A,1,1),\ A\ne 0,\ O=D4C\to O'=S_4,\quad \alpha =0,\]
  \[{\bf A}= {\rm all\ other\ permutations}, \  O=D4C\to O'=D4A,\quad \alpha =0,\]
 \item \[{\bf A}=(A_1,A_2,A_3,A_4),\ O=K[1,1,1,1]\to O'=D4A,\quad \alpha=0.\]
 \end{enumerate}

  \begin{comment} Already in this example we are able to characterize contractions of Darboux systems in a manner 
  completely analogous to those of constant curvature systems. That wasn't possible before we extended our method to conformally superintegrable systems.
   
  \end{comment}

 \subsection{Contraction [1,1,1,1] to  [2,2] applied to conformal St\"ackel transforms of system V[1,1,1,1].}
 The target systems are conformal
 St\"ackel transforms of $V_{[2,2]}$. Partial results are: 
\begin{enumerate}
 \item \[ {\bf A}=(1,0,0,0)\ {\rm and\ all\ permutations},\quad O=S_9\to O'=E_7,\quad \alpha =2,\]
  \item\[ {\bf A}=(1,1,1,1),\ (0,1,1,0)\quad O=S_8\to O'=E_{19},\quad \alpha =4,\]
 \[ {\bf A}=(0,1,0,1), \ (1,0,1,0)\quad  O=S_8\to O'=E_7,\quad \alpha =2,\]
 \[ {\bf A}=(1,0,0,1)\quad  O=S_8\to O'=E_{17},\quad \alpha =2,\] 
 \item \[{\bf A}=(0,0,1,1),\quad O=S_7\to O'=E_{17},\quad \alpha =4,\]
 \[ {\bf A}=(1,1,0,0,)\quad  O=S_7\to O'=E_{19},\quad \alpha =4,\]
 \item \[{\bf A}=(0,0,A_3,A_4),\ A_3A_4\ne 0,A_3\ne A_4,\ {\rm and\ all\ permutations},\]
 \[O=D4B\to O'=E_7,\ \alpha =2,\]
 \item \[{\bf A}=(1,1,A,A),\ A\ne 0,\ {\rm and\ all\ permutations},\]
 \[O=D4C\to O'=E_{19},\quad \alpha =1,\]
 \item \[{\bf A}=(A_1,A_2,A_3,A_4),\ O=K[1,1,1,1]\to O'=E_7,\quad \alpha=2.\]
 \end{enumerate}
 Additional results can be obtained for this contraction and the following by permutiong the coordinate inidces  of the image 
 potential before appluying the contraction.

 \subsection{Contraction [1,1,1,1] to  [3,1] applied to conformal St\"ackel transforms of system V[1,1,1,1].}
 The target systems are conformal
 St\"ackel transforms of $V_{[3,1]}$. Partial results are (assuming generic $a$): 
\begin{enumerate}
 \item \[ {\bf A}=(1,0,0,0), (0,1,0,0),(0,0,1,0),\quad O=S_9\to O'=E_2,\quad \alpha =2,\]
 \[ {\bf A}=(0,0,0,1),\quad O=S_9\to O'=S_1,\quad \alpha =0,\]
  \item\[ {\bf A}=(1,1,1,1),(0,1,0,1),(1,0,0,1)\quad O=S_8\to O'=S_1,\quad \alpha =0,\]
  \[ {\bf A}=(1,0,1,0),(0,1,1,0)\quad  O=S_8\to O'=E_2,\quad \alpha =2,\] 
 \item \[{\bf A}=(0,0,1,1),\quad O=S_7\to O'=S_1,\quad \alpha =0,\]
 \[ {\bf A}=(1,1,0,0,)\quad  O=S_7\to O'=E_{2},\quad \alpha =2,\]
 \item {\small \[{\bf A}=(0,0,A_3,A_4),(A_3,0,0,A_4),(0,A_3,0,A_4),\ A_3A_4\ne 0,A_3\ne A_4,\]\[ O=D4B\to O'=S_1,\ \alpha =0,\]
  \[ {\bf A}=(A_1,A_2,0,0),(A_1,0,A_2,0),(0,A_1,A_2,0),\ A_1A_2\ne 0,A_1\ne A_2,\]
  \[O=D4B\to O'=E_2,\ \alpha =2,\] }
 \item \[{\bf A}=(1,1,A,A),\ {\rm and\ all\ permutations},\ A\ne 0,1,\]
 \[O=D4C\to O'=S_1,\quad \alpha =0,\]
 \item \[{\bf A}=(A_1,A_2,A_3,A_4),\ O=K[1,1,1,1]\to O'=S_1,\quad \alpha=0.\]
 \end{enumerate}

\subsection{Contraction [1,1,1,1] to [4] applied to conformal St\"ackel transforms of system V[1,1,1,1].} 
  The target systems are conformal
 St\"ackel transforms of $V_{[4]}$.  Partial results are (generic in the parameters $a,b$): 
\begin{enumerate}
 \item \[ {\bf A}=(1,0,0,0), \ {\rm and\ all\ permutations},\quad O=S_9\to O'=E_{10},\quad \alpha =6,\]
  \item\[ {\bf A}=(1,1,1,1),(0,1,0,1),(1,0,0,1),\quad O=S_8\to O'=E_{10},\quad \alpha =6,\]
  \[ {\bf A}=(1,0,1,0),(0,1,1,0)\quad  O=S_8\to O'=E_{10},\quad \alpha =6,\] 
  \item \[{\bf A}=(0,0,1,1),(1,1,0,0,),\quad O=S_7\to O'=E_{10},\quad \alpha =6,\]
 \item {\small \[{\bf A}=(0,0,A_3,A_4),  \ {\rm and\ all\ permutations},\ A_3A_4\ne 0,A_3\ne A_4,\]
  \[ O=D4B\to O'=E_{10},\ \alpha =6,\]}
 \item \[{\bf A}=(1,1,A,A), \ {\rm and\ all\ permutations},\ A\ne 0,1,\]
  \[O=D4C\to O'=E_{10},\quad \alpha =6,\]
 \item \[{\bf A}=(A_1,A_2,A_3,A_4),\ O=K[1,1,1,1]\to O'=E_{10},\quad \alpha=6.\]
 \end{enumerate}

\subsection{Contraction [2,2] to [4] applied to conformal St\"ackel transforms of system V[1,1,1,1].} 
  The target systems are conformal
 St\"ackel transforms of $V_{[4]}$.  Partial results are: 
\begin{enumerate}
 \item \[ {\bf A}=(1,0,0,0),\ {\rm and\ all\ permutations},\quad O=S_9\to O'=E_{10},\quad \alpha =4,\]
  \item\[ {\bf A}=(1,1,1,1),\quad O=S_8\to O'=E_{9},\quad \alpha =6,\]
\[   {\bf A}=(0,1,0,1),(1,0,1,0),\quad O=S_8\to O'=E_{10},\quad \alpha =4,\]
\[   {\bf A}=(0,1,1,0),(1,0,0,1),\quad O=S_8\to O'=E_{9},\quad \alpha =5,\]
  \item \[{\bf A}=(0,0,1,1),(1,1,0,0,),\quad O=S_7\to O'=E_{10},\quad \alpha =5,\]
 \item {\small\[{\bf A}=(0,0,A_3,A_4),\ {\rm and\ all\ permutations},\ A_3A_4\ne 0,A_3\ne A_4,\]
 \[ O=D4B\to O'=E_{10},\ \alpha =4,\]
 \item \[{\bf A}=(1,1,A,A),{\rm and\ all\ permutations}\ A\ne 0,1,\]
 \[O=D4C\to O'=E_{10},\quad \alpha =5,\]}
 \item \[{\bf A}=(A_1,A_2,A_3,A_4),\ O=K[1,1,1,1]\to O'=E_{10},\quad \alpha=4.\]
 \end{enumerate}
 Note that, although the values of $\alpha$ differ, the target systems agree with those for $[1,1,1,1]\to [4]$ contractions of $V_{[1,1,1,1]}$, except in the single case
 $S_8\to E_9$.

\subsection{Contraction [2,1,1] to [3,1] applied to conformal St\"ackel transforms of system V[1,1,1,1].} 
  The target systems are conformal
 St\"ackel transforms of $V_{[3,1]}$.  Partial results are: 
\begin{enumerate}
 \item \[ {\bf A}=(1,0,0,0), (0,1,0,0),\quad O=S_9\to O'=E_{2},\quad \alpha =6,\]
 \[ {\bf A}= (0,0,1,0),\quad O=S_9\to O'=E_{2},\quad \alpha =4,\]
 \[ {\bf A}= (0,0,0,1),\quad O=S_9\to O'=S_1,\quad \alpha =0,\]
  \item\[ {\bf A}=(1,1,1,1),(0,1,0,1),(1,0,0,1)\quad O=S_8\to O'=S_1,\quad \alpha =0,\]
\[   {\bf A}=(1,0,1,0),(0,1,1,0)\quad O=S_8\to O'=E_{2},\quad \alpha =4,\]
  \item \[{\bf A}=(0,0,1,1),\quad O=S_7\to O'=S_1,\quad \alpha =0,\]
  \[{\bf A}=(1,1,0,0,),\quad O=S_7\to O'=E_2,\quad \alpha =8,\]
 \item {\small \[{\bf A}=(0,0,A_3,A_4),(A_3,0,0,A_4),(0,A_3,0,A_4),\ A_3A_4\ne 0,A_3\ne A_4,\]
 \[O=D4B\to O'=S_1,\ \alpha =0,\]
\[{\bf A}= (A_1,A_2,0,0),(0,A_1,A_2,0)\ A_1A_2\ne 0,A_1\ne A_2,\]
\[O=D4B\to O'=E_{2},\quad \alpha =6,\]
\[ {\bf A}=(A_1,0,A_3,0),\ A_1A_3\ne 0, A_1\ne A_3,\]
\[O=D4B\to O'=E_{2},\quad \alpha =4,\]}
 \item \[{\bf A}=(1,1,A,A),\ {\rm and\ all\ permutations},\ A\ne 0,1,\]
 \[O=D4C\to O'=S_1,\quad \alpha =0,\]
 \item \[{\bf A}=(A_1,A_2,A_3,A_4),\ O=K[1,1,1,1]\to O'=S_1,\quad \alpha=0.\]
 \end{enumerate}
 Note that, although the values of $\alpha$ differ, the target systems agree with those for $[1,1,1,1]\to [3,1]$ contractions of $V_{[1,1,1,1]}$.

\subsection{Contraction [1,1,1,1] to [2,1,1] applied to conformal St\"ackel transforms of system V[2,1,1].} 
  The target systems are conformal
 St\"ackel transforms of $V_{[2,1,1]}$.  Partial results are: 
\begin{enumerate}
 \item \[ {\bf A}=(1,1,0,0),,\quad O=S_4\to O'=S_4,\quad \alpha =0,\]
 \item \[ {\bf A}= (1,0,0,0),(0,1,0,0),\quad O=S_2\to O'=S_{2},\quad \alpha =0,\]
  \item\[ {\bf A}=(0,0,0,1),\quad O=E_1\to O'=E_1,\quad \alpha =2,\]
\item \[   {\bf A}=(0,0,1,0),\quad O=E_{16}\to O'=E_{16},\quad \alpha =4,\]
  \item \[{\bf A}=(A_3,A_4,0,0), (A_3A_4\ne 0,A_3\ne A_4),\quad \]
  \[O=D4A\to O'=D4A,\quad \alpha =0,\]
 \item \[{\bf A}=(0,0,A_1,A_2,),(A_1A_2\ne 0),\quad O=D3B\to O'=E_1,\quad \alpha =2,\]
 \item \[{\bf A}=(A,0,0,1),(0,A,0,1)\ A\ne 0,\quad O=D2B\to O'=S_2,\quad \alpha =0,\]
 \item \[{\bf A}=(1,1,A,0),\ A\ne 0,\ O=D2C\to O'=S_4,\quad \alpha =0,\]
 \item \[{\bf A}=(A_3,A_4,A_2,A_1),\ O=K[2,1,1]\to O'=S_4,\quad \alpha=0.\]
 \end{enumerate}

\subsection{Contraction [1,1,1,1] to [2,2] applied to conformal St\"ackel transforms of system V[2,1,1].} 
  The target systems are conformal
 St\"ackel transforms of $V_{[2,2]}$.  Partial results are: 
\begin{enumerate}
 \item \[ {\bf A}=(1,1,0,0),\quad O=S_4\to O'=E_{19},\quad \alpha =4,\]
 \item \[ {\bf A}= (1,0,0,0),(0,1,0,0),\quad O=S_2\to O'=E_7,\quad \alpha =2,\]
  \item\[ {\bf A}=(0,0,0,1),\quad O=E_1\to O'=E_8,\quad \alpha =2,\]
\item \[   {\bf A}=(0,0,1,0),\quad O=E_{16}\to O'=E_{17},\quad \alpha =4,\]
  \item \[{\bf A}=(A_1,A_2,0,0), (A_1A_2\ne 0,A_1\ne A_2),\quad\]
  \[O=D4A\to O'=E_7,\quad \alpha =2,\]
 \item \[{\bf A}=(0,0,A_3,A_4,),(A_3A_4\ne 0),\quad O=D3B\to O'=E_8,\quad \alpha =2,\]
 \item \[{\bf A}=(A,0,0,1),(0,A,0,1)\ A\ne 0,\quad O=D2B\to O'=E_7,\quad \alpha =2,\]
 \item \[{\bf A}=(1,1,A,0),\ A\ne 0,\ O=D2C\to O'=E_{19},\quad \alpha =4,\]
 \item \[{\bf A}=(A_1,A_2,A_3,A_4),\ O=K[2,1,1]\to O'=E_7,\quad \alpha=2,\]
 \end{enumerate}

\subsection{Contraction [1,1,1,1] to [3,1] applied to conformal St\"ackel transforms of system V[2,1,1].} 
  The target systems are conformal
 St\"ackel transforms of $V_{[3,1]}$. Generically in $a$,  partial results are: 
\begin{enumerate}
 \item \[ {\bf A}=(1,1,0,0),\quad O=S_4\to O'=E_{2},\quad \alpha =2,\]
 \item \[ {\bf A}= (1,0,0,0),(0,1,0,0),\quad O=S_2\to O'=E_2,\quad \alpha =2,\]
  \item\[ {\bf A}=(0,0,0,1),\quad O=E_1\to O'=E_2,\quad \alpha =2,\]
\item \[   {\bf A}=(0,0,1,0),\quad O=E_{16}\to O'=E_{2},\quad \alpha =2,\]
  \item \[{\bf A}=(A_1,A_2,0,0), (A_1A_2\ne 0,A_1\ne A_2),\quad\]
  \[O=D4A\to O'=E_2,\quad \alpha =2,\]
 \item \[{\bf A}=(0,0,A_3,A_4,),(A_3A_4\ne 0),\quad O=D3B\to O'=E_2,\quad \alpha =2,\]
 \item \[{\bf A}=(A,0,0,1),(0,A,0,1)\ A\ne 0,\quad O=D2B\to O'=E_2,\quad \alpha =2,\]
 \item \[{\bf A}=(1,1,A,0),\ A\ne 0,\ O=D2C\to O'=E_{2},\quad \alpha =2,\]
 \item \[{\bf A}=(A_1,A_2,A_3,A_4),\ O=K[2,1,1]\to O'=E_2,\quad \alpha=2,\]
 \end{enumerate}

\subsection{Contraction [1,1,1,1] to [4] applied to conformal St\"ackel transforms of system V[2,1,1].} 
  The target systems are conformal
 St\"ackel transforms of $V_{[4]}$.  Partial results are: 
 
 St\"ackel transforms of $V_{[3,1]}$. Generically in $a$, the results are: 
\begin{enumerate}
 \item \[ {\bf A}=(1,1,0,0),\quad O=S_4\to O'=E_{10},\quad \alpha =6,\]
 \item \[ {\bf A}= (1,0,0,0),(0,1,0,0),\quad O=S_2\to O'=E_{10},\quad \alpha =6,\]
  \item\[ {\bf A}=(0,0,0,1),\quad O=E_1\to O'=E_{10},\quad \alpha =6,\]
\item \[   {\bf A}=(0,0,1,0),\quad O=E_{16}\to O'=E_{10},\quad \alpha =6,\]
  \item \[{\bf A}=(A_1,A_2,0,0), (A_1A_2\ne 0,A_1\ne A_2),\quad\]
  \[O=D4A\to O'=E_{10},\quad \alpha =6,\]
 \item \[{\bf A}=(0,0,A_3,A_4,),(A_3A_4\ne 0),\quad O=D3B\to O'=E_{10},\quad \alpha =6,\]
 \item \[{\bf A}=(A,0,0,1),(0,A,0,1)\ A\ne 0,\quad O=D2B\to O'=E_{10},\quad \alpha =6,\]
 \item \[{\bf A}=(1,1,A,0),\ A\ne 0,\ O=D2C\to O'=E_{10},\quad \alpha =6,\]
 \item \[{\bf A}=(A_1,A_2,A_3,A_4),\ O=K[2,1,1]\to O'=E_{10},\quad \alpha=6,\]
 \end{enumerate}

\subsection{Contraction [2,2] to [4] applied to conformal St\"ackel transforms of system V[2,1,1].} 
  The target systems are conformal
 St\"ackel transforms of $V_{[4]}$.  Partial results are: 
 
\begin{enumerate}
 \item \[ {\bf A}=(1,1,0,0),\quad O=S_4\to O'=E_{10},\quad \alpha =5,\]
 \item \[ {\bf A}= (1,0,0,0),(0,1,0,0),\quad O=S_2\to O'=E_{10},\quad \alpha =4,\]
  \item\[ {\bf A}=(0,0,0,1),\quad O=E_1\to O'=E_{10},\quad \alpha =4,\]
\item \[   {\bf A}=(0,0,1,0),\quad O=E_{16}\to O'=E_{10},\quad \alpha =5,\]
  \item \[{\bf A}=(A_1,A_2,0,0), (A_1A_2\ne 0,A_1\ne A_2),\quad\]
  \[O=D4A\to O'=E_{10},\quad \alpha =4,\]
 \item \[{\bf A}=(0,0,A_3,A_4,),(A_3A_4\ne 0),\quad O=D3B\to O'=E_{10},\quad \alpha =4,\]
 \item \[{\bf A}=(A,0,0,1),(0,A,0,1)\ A\ne 0,\quad\]
 \[O=D2B\to O'=E_{10}, ({\rm generically})\quad \alpha =4,\]
 \[\qquad  O=D2B\to O'=E_{9}, ({\rm special\ case})\quad \alpha =5,\]
 \item \[{\bf A}=(1,1,A,0),\ A\ne 0,\ O=D2C\to O'=E_{10}, ({\rm generically})\quad \alpha =5,\]
 \item \[{\bf A}=(A_1,A_2,A_3,A_4),\ O=K[2,1,1]\to O'=E_{10},\quad \alpha=4,\]
 \end{enumerate}

\subsection{Contraction [2,1,1] to [3,1] applied to conformal St\"ackel transforms of system V[2,1,1].} 
  The target systems are conformal
 St\"ackel transforms of $V_{[3,1]}$.  Partial results are:

\begin{enumerate}
 \item \[ {\bf A}=(1,1,0,0),\quad O=S_4\to O'=E_{2},\quad \alpha =8,\]
 \item \[ {\bf A}= (1,0,0,0),(0,1,0,0),\quad O=S_2\to O'=E_{2},\quad \alpha =6,\]
  \item\[ {\bf A}=(0,0,0,1),\quad O=E_1\to O'=E_{2},\quad \alpha =4,\]
\item \[   {\bf A}=(0,0,1,0),\quad O=E_{16}\to O'=E_{2},\quad \alpha =4,\]
  \item \[{\bf A}=(A_1,A_2,0,0), (A_1A_2\ne 0,A_1\ne A_2),\quad\]
  \[O=D4A\to O'=E_{2},\quad \alpha =6,\]
 \item \[{\bf A}=(0,0,A_3,A_4,),(A_3A_4\ne 0),\quad\]
 \[O=D3B\to O'=E_{2}, ({\rm generic})\quad \alpha =4,\]
 \item \[{\bf A}=(A,0,0,1),(0,A,0,1)\ A\ne 0,\quad\]
 \[O=D2B\to O'=E_{2}, ({\rm generically})\quad \alpha =6,\]
 \item \[{\bf A}=(1,1,A,0),\ A\ne 0,\ O=D2C\to O'=E_{2}\quad \alpha =4,\]
 \item \[{\bf A}=(A_1,A_2,A_3,A_4),\ O=K[2,1,1]\to O'=E_{2},\quad \alpha=4,\]
 \end{enumerate}

 \subsection{Contraction [1,1,1,1] to [2,1,1] applied to conformal St\"ackel transforms of system V[2,2].}
  The target systems are conformal
 St\"ackel transforms of $V_{[2,2]}$.  Partial results are: 
 \begin{enumerate}
 \item \[ {\bf A}=(0,0,1,0),\quad O=E_8\to O'=E_{8},\quad \alpha =0,)\]
  \item\[ {\bf A}=(0,0,0,1),\quad O=E_{17}\to O'=E_{17},\quad \alpha =0,\]
 \item \[ {\bf A}= (1,0,A_3,0),\quad O=E_7\to O'=E_{7},\quad \alpha =0,  ({\rm generically}\]
 \item \[   {\bf A}=(0,1,0,A_4),\quad O=E_{19}\to O'=E_{19},\quad \alpha =0, ({\rm generically})\]
  \item \[{\bf A}=(0,0,A_3,A_4,),(A_3A_4\ne 0),\quad O=D3C\to O'=D3C,\quad \alpha =0,\]
  \item \[{\bf A}=(A_1,A_2,0,0), (A_1A_2\ne 0),\quad O=D3D\to O'=E_{7},\quad \alpha =2,\]
  \item \[{\bf A}=(A_1,A_2,A_3,A_4),\ O=K[2,2]\to O'=D3C,\quad \alpha=0,\]
 \end{enumerate}

 \subsection{Contraction [1,1,1,1] to [2,2] applied to conformal St\"ackel transforms of system V[2,2].}
 The target systems are conformal
 St\"ackel transforms of $V_{[2,2]}$.  Partial results are: 
 \begin{enumerate}
 \item \[ {\bf A}=(0,0,1,0),\quad O=E_8\to O'=E_{8},\quad \alpha =2,)\]
  \item\[ {\bf A}=(0,0,0,1),\quad O=E_{17}\to O'=E_{17},\quad \alpha =2,\]
 \item \[ {\bf A}= (1,0,A_3,0),\quad O=E_7\to O'=E_{7},\quad \alpha =2,  ({\rm generically}\]
 \item \[   {\bf A}=(0,1,0,A_4),\quad O=E_{19}\to O'=E_{19},\quad \alpha =4, ({\rm generically})\]
  \item \[{\bf A}=(0,0,A_3,A_4,),(A_3A_4\ne 0),\quad O=D3C\to O'=E_8,\quad \alpha =2,\]
  \item \[{\bf A}=(A_1,A_2,0,0), (A_1A_2\ne 0),\quad O=D3D\to O'=E_{7},\quad \alpha =2,\]
  \item \[{\bf A}=(A_1,A_2,A_3,A_4),\ O=K[2,2]\to O'=E_7,\quad \alpha=2,\]
 \end{enumerate}

 \subsection{Contraction [1,1,1,1] to [3,1] applied to conformal St\"ackel transforms of system V[2,2].}
 The target systems are conformal
 St\"ackel transforms of $V_{[3,1]}$.  Partial results are: 
 \begin{enumerate}
 \item \[ {\bf A}=(0,0,1,0),\quad O=E_8\to O'=E_{3}',\quad \alpha =2,\]
  \item\[ {\bf A}=(0,0,0,1),\quad O=E_{17}\to O'=E_{3}',\quad \alpha =2, \]
 \item \[ {\bf A}= (1,0,A_3,0),\quad O=E_7\to O'=E_{3}',\quad \alpha =2,  ({\rm generically})\]
 \item \[   {\bf A}=(0,1,0,A_4),\quad O=E_{19}\to O'=E_{3}',\quad \alpha =2, ({\rm generically})\]
  \item \[{\bf A}=(0,0,A_3,A_4,),(A_3A_4\ne 0),\quad\]
  \[O=D3C\to O'=E_3',\quad \alpha =2, ({\rm generically})\]
  \[ \qquad O=D3C\to O'=D1C,\quad \alpha =3, ({\rm special \ case})\]
  \item \[{\bf A}=(A_1,A_2,0,0), (A_1A_2\ne 0),\quad\]
  \[O=D3D\to O'=E_{3}',\quad \alpha =2,  ({\rm generically})\]
  \item \[{\bf A}=(A_1,A_2,A_3,A_4),\ O=K[2,2]\to O'=E_3',\quad \alpha=2,\]
 \end{enumerate}

 \subsection{Contraction [1,1,1,1] to [4] applied to conformal St\"ackel transforms of system V[2,2].}
 Partial results:
 \begin{enumerate}
 \item \[ {\bf A}=(0,0,1,0),\quad O=E_8\to O'=E'_{3},\quad \alpha =6,\]
  \item\[ {\bf A}=(0,0,0,1),\quad O=E_{17}\to O'=E'_{3},\quad \alpha =6, \]
 \item \[ {\bf A}= (1,0,A_3,0),\quad O=E_7\to O'=E'_{3},\quad \alpha =6, \] 
 \item \[   {\bf A}=(0,1,0,A_4),\quad O=E_{19}\to O'=E'_{3},\quad \alpha =6, \]
   \item \[{\bf A}=(0,0,A_3,A_4,),(A_3A_4\ne 0),\quad O=D3C\to O'=E'_{3},\quad \alpha =6,\]
    \item \[{\bf A}=(A_1,A_2,0,0), (A_1A_2\ne 0),\quad O=D3D\to O'=E'_{3},\quad \alpha =6, \]
  \item \[{\bf A}=(A_1,A_2,A_3,A_4),\ O=K[2,2]\to O'=E'_{3},\quad \alpha=6,\]
 \end{enumerate}

 \subsection{Contraction [2,2] to [4] applied to conformal St\"ackel transforms of system V[2,2].}
 The target systems are conformal
 St\"ackel transforms of $V_{[4]}$. Partial results:
 \begin{enumerate}
 \item \[ {\bf A}=(0,0,1,0),\quad O=E_8\to O'=E_{10},\quad \alpha =4,\]
  \item\[ {\bf A}=(0,0,0,1),\quad O=E_{17}\to O'=E_{10},\quad \alpha =5, \]
 \item \[ {\bf A}= (1,0,A_3,0),\quad O=E_7\to O'=E_{10},\quad \alpha =4,  ({\rm generically})\]
 \[\qquad O=E_7\to O'=E_{9},\quad \alpha =5,  ({\rm special\ case})\]
 \item \[   {\bf A}=(0,1,0,A_4),\quad O=E_{19}\to O'=E_{10},\quad \alpha =5, ({\rm generically})\]
 \[\qquad O=E_{19}\to O'=E_{9},\quad \alpha =6,  ({\rm special\ case})\]
  \item \[{\bf A}=(0,0,A_3,A_4,),(A_3A_4\ne 0),\quad O=D3C\to O'=E_{10},\quad \alpha =4,\]
    \item \[{\bf A}=(A_1,A_2,0,0), (A_1A_2\ne 0),\quad O=D3D\to O'=E_{10},\quad \alpha =4, \]
  \item \[{\bf A}=(A_1,A_2,A_3,A_4),\ O=K[2,2]\to O'=E_{10},\quad \alpha=4,\]
 \end{enumerate}
 
 \subsection{Contraction [2,1,1] to  [3,1] applied to conformal St\"ackel transforms of system V[2,2].}
 The target systems are conformal
 St\"ackel transforms of $V_{[0]}$.Partial results:                         
 \begin{enumerate}
 \item \[ {\bf A}=(0,0,1,0),\quad O=E_8\to O'=E_{3}',\quad \alpha =4,\]
  \item\[ {\bf A}=(0,0,0,1),\quad O=E_{17}\to O'=E_{3}',\quad \alpha =4, \]
 \item \[ {\bf A}= (1,0,A_3,0),\quad O=E_7\to O'=E_{3}',\quad \alpha =2,  \]
 \item \[   {\bf A}=(0,1,0,A_4),\quad O=E_{19}\to O'=E_{3}',\quad \alpha =0, \]
  \item \[{\bf A}=(0,0,A_3,A_4,),(A_3A_4\ne 0),\quad\]
  \[O=D3C\to O'=E_{3}',\quad \alpha =4, ({\rm generically})\]
  \[\qquad O=D3C\to O'=D1C,\quad \alpha =6, ({\rm special\ case})\]
    \item \[{\bf A}=(A_1,A_2,0,0), (A_1A_2\ne 0),\quad O=D3D\to O'=E_{3}',\quad \alpha =0, \]
  \item \[{\bf A}=(A_1,A_2,A_3,A_4),\ O=K[2,2]\to O'=E_{3}',\quad \alpha=0,\]
 \end{enumerate}
 
 \subsection{Contraction [1,1,1,1] to  [2,1,1] applied to conformal St\"ackel transforms of system V[3,1].}
  The target systems are conformal
 St\"ackel transforms of  the singular system $V(1)$. All systems are flat space and St\"ackel equivalent to special cases of $E15$.

\subsection{Contraction [1,1,1,1] to  [2,2] applied to conformal St\"ackel transforms of system V[3,1].}
 The target systems are conformal
 St\"ackel transforms of  the singular system $V(1)$. All systems are flat space and St\"ackel equivalent to special cases of $E15$.

 \subsection{Contraction [1,1,1,1] to [3,1] applied to conformal St\"ackel transforms of system V[3,1].}
 The target systems are conformal
 St\"ackel transforms of $V_{[3,1]}$. Partial results are: 
\begin{enumerate}
 \item \[ {\bf A}=(0,0,0,1),\quad O=S_1\to O'=E_2,\quad \alpha =2,\]
  \item\[ {\bf A}=(1,0,0,0),\quad O=E_{2}\to O'=E_{2},\quad \alpha =2, \]
 \item \[ {\bf A}= (a,1,0,0),\quad O=D1B\to O'=E_{2},\quad \alpha =2,  \]
 \item \[   {\bf A}=(0,0,1,0),\quad O=D2A\to O'=E_2, (\rm generically)\quad \alpha =2, \]
   \item \[{\bf A}=(A_1,A_2,A_3,A_4),\ O=K[3,1]\to O'=E_2,\quad \alpha=2.\]
 \end{enumerate}

\subsection{Contraction [1,1,1,1] to [4] applied to conformal St\"ackel transforms of system V[3,1].}
 The target systems are conformal
 St\"ackel transforms of $V_{[4]}$. Partial results are: 
\begin{enumerate}
 \item \[ {\bf A}=(0,0,0,1),\quad O=S_1\to O'=E_{10},\quad \alpha =6,\]
  \item\[ {\bf A}=(1,0,0,0),\quad O=E_{2}\to O'=E_{10},\quad \alpha =6, \]
 \item \[ {\bf A}= (a,1,0,0),\quad O=D1B\to O'=E_{10},\quad \alpha =6,  \]
 \item \[   {\bf A}=(0,0,1,0),\quad O=D2A\to O'=E_{10}, \quad \alpha =6, \]
   \item \[{\bf A}=(A_1,A_2,A_3,A_4),\ O=K[3,1]\to O'=E_{10},\quad \alpha=6.\]
 \end{enumerate}

 \subsection{Contraction [2,2] to [4] applied to conformal St\"ackel transforms of system V[3,1].}
 The target systems are conformal
 St\"ackel transforms of the singular system $V_{[2]}$. All systems are flat space and St\"ackel equivalent to special cases of $E15$.

 \subsection{Contraction [2,1,1] to [3,1] applied to conformal St\"ackel transforms of system V[3,1].}
 The target systems are conformal
 St\"ackel transforms of $V_{[3,1]}$. Partial results:
\begin{enumerate}
 \item \[ {\bf B}=(0,0,0,1),\quad O=S_1\to O'=S_1,\quad \alpha =0,\]
  \item\[ {\bf B}=(1,0,0,0),\quad O=E_{2}\to O'=E_{2},\quad \alpha =6, \]
 \item \[ {\bf B}= (a,1,0,0),\quad O=D1B\to O'=E_{2},\quad \alpha =6,  \]
 \item \[   {\bf B}=(0,0,1,0),\quad O=D2A\to O'=E_2,\quad \alpha =6, \]
   \item \[{\bf B}=(B_1,B_2,B_3,B_4),\ O=K[3,1]\to O'=S_1,\quad \alpha=0.\]
 \end{enumerate}

 \subsection{Contraction [1,1,1,1] to [2,1,1] applied to conformal St\"ackel transforms of system V[4].}
  The target systems are conformal
 St\"ackel transforms of $V_{[0]}$. Partial results are: 
 \begin{enumerate}
 \item \[ {\bf D}=(1,D_2,0,0),\quad O=E_{10}\to O'=E_{3}',\quad \alpha =2,)\]
  \item\[ {\bf D}=(0,1,0,0),\quad O=E_{9}\to O'=E_{11},\quad \alpha =3,\]
 \item \[ {\bf D}= (0,0,0,1),\quad O=D1A\to O'=E_{20},\quad \alpha =4, \]
  \item \[{\bf D}=(D_1,D_2,D_3,D_4),\ O=K[4]\to O'=E_3',\quad \alpha=2,\]
 \end{enumerate}

\subsection{Contraction [1,1,1,1] to [2,2] applied to conformal St\"ackel transforms of system V[4].}
 The target systems are conformal
 St\"ackel transforms of  the singular system $V(2)$. All systems are flat space and St\"ackel equivalent to special cases of $E15$.

 \subsection{Contraction [1,1,1,1] to [3,1] applied to conformal St\"ackel transforms of system V[4].}
 The target systems are conformal
 St\"ackel transforms of $V_{[0]}$. Partial results are: 
\begin{enumerate}
 \item \[ {\bf C}=(1,C_2,0,0),\quad O=E_{10}\to O'=E_{3}',\quad \alpha =2,\]
  \item\[ {\bf C}=(0,1,0,0),\quad O=E_{9}\to O'=E_{3}',\quad \alpha =2,\]
 \item \[ {\bf C}= (0,0,0,1),\quad O=D1A\to O'=E_{3}',\quad \alpha =2, \]
  \item \[{\bf C}=(C_1,C_2,C_3,C_4),\ O=K[4]\to O'=E_3',\quad \alpha=2,\]
 \end{enumerate}

\subsection{Contraction [1,1,1,1] to [4] applied to conformal St\"ackel transforms of system V[4].}
 The target systems are conformal
 St\"ackel transforms of $V_{[0]}$. Partial results are: 
\begin{enumerate}
 \item \[ {\bf C}=(1,C_2,0,0),\quad O=E_{10}\to O'=E_{3}',\quad \alpha =6,\]
  \item\[ {\bf C}=(0,1,0,0),\quad O=E_{9}\to O'=E_{3}',\quad \alpha =6,\]
 \item \[ {\bf C}= (0,0,0,1),\quad O=D1A\to O'=E_{3}',\quad \alpha =6, \]
  \item \[{\bf C}=(C_1,C_2,C_3,C_4),\ O=K[4]\to O'=E_{3}',\quad \alpha=6,\]
 \end{enumerate}

 \subsection{Contraction [2,2] to [4] applied to conformal St\"ackel transforms of system V[4].}
 The target systems are conformal
 St\"ackel transforms of the singular system $V_{(2)}$. All systems are flat space and St\"ackel equivalent to special cases of $E15$.

 \subsection{Contraction [2,1,1] to [3,1] applied to conformal St\"ackel transforms of system V[4].}
 The target systems are conformal
 St\"ackel transforms of $V_{[0]}$. Partial results:
 
\begin{enumerate}
 \item \[ {\bf C}=(1,C_2,0,0),\quad O=E_{10}\to O'=E_{3}',\quad \alpha =1,(a\ne0), 0,(a=0)\]
  \item\[ {\bf C}=(0,1,0,0),\quad O=E_{9}\to O'=E_{3}',\quad \alpha =1,\]
 \item \[ {\bf C}= (0,0,0,1),\quad O=D1A\to O'=E_{3}',\quad \alpha =-1, \]
  \item \[{\bf C}=(C_1,C_2,C_3,C_4),\ O=K[4]\to O'=E3',\quad \alpha=-1,\]
 \end{enumerate}

 \subsection{Contraction [1,1,1,1] to [2,1,1] applied to conformal St\"ackel transforms of system V[0].}
  The target systems are conformal
 St\"ackel transforms of $V_{[0]}$. Partial results are: 
 \begin{enumerate}
 \item \[ {\bf C}=((C_2^2+C_3^2)/4,C_2,C_3,1),\quad O=E_{20}\to O'=E_{3}',\]
 \[(C_2^2+C_3^2\ne 0)\quad \alpha =2,\]
 \[ \qquad O=E_{20}\to O'=E_{11}, (C_2^2+C_3^2= 0,C_2C_3\ne 0)\quad \alpha =3,\]
 \[ \qquad O=E_{20}\to O'=D3A, (C_2= C_3 =0)\quad \alpha =4,\]
  \item\[ {\bf C}=(C_1,1,\pm i,0),\quad O=E_{11}\to O'=E_{3}', (C_1\ne 0)\quad \alpha =2,\]
  \[\qquad O=E_{11}\to O'=E_{11}, (C_1= 0)\quad \alpha =3,\]
 \item \[ {\bf C}= (1,0,0,0), \quad O=E_{3}'\to O'=E_{3}',\quad \alpha =2\]
 \item \[   {\bf C}=(C_1,C_2,C_3,0),\ (C_2^2+C_3^2\ne 0),\quad\]
 \[O=D1C\to O'=E_{3}', (C_1\ne 0)\quad \alpha =2, \]
 \[ \qquad O=D1C\to O'=D1C, (C_1= 0)\quad \alpha =3, \]
  \item \[{\bf C}=(C_1,C_2,C_3,1),\,(4C_1\ne C_2^2+C_3^2),\quad\]
  \[O=D3A\to O'=E_{3}',\ (C_1\ne 0)\quad \alpha =2,\]
  \[ \qquad O=D3A\to O'=D1C,\ (C_1= 0)\quad \alpha =3,\]
  \item \[{\bf C}=(C_1,C_2,C_3,C_4), \ O=K[0]\to O'=E_{3}',\quad \alpha=2,\]
 \end{enumerate}

\subsection{Contraction [1,1,1,1] to [2,2] applied to conformal St\"ackel transforms of system V[0].}
  The target systems are conformal
 St\"ackel transforms of $V_{[0]}$. Partial results are: 
 \begin{enumerate}
 \item \[ {\bf C}=((C_2^2+C_3^2)/4,C_2,C_3,1),\quad O=E_{20}\to O'=E_{3}',\]
 \[(C_2^2+C_3^2\ne 0)\quad \alpha =2,\]
 \[ \qquad O=E_{20}\to O'=E_{20}, (C_3=-iC_2\ne 0)\quad \alpha =4,\]
 \[ \qquad O=E_{20}\to O'=E_{3}', ( C_3 =iC_2\ne 0)\quad \alpha =2,\]
  \item\[ {\bf C}=(C_1,1,\pm i,0),\quad O=E_{11}\to O'=E_{3}', (C_1\ne 0,C_3=-i)\quad \alpha =2,\]
  \[\qquad O=E_{11}\to O'=E_{11}, (C_3= i)\quad \alpha =2,\]
   \[\qquad O=E_{11}\to O'=E_{11}, (C_1= 0,C3=-i)\quad \alpha =4,\] 
 \item \[ {\bf C}= (1,0,0,0), \quad O=E_{3}'\to O'=E_{3}',\quad \alpha =2\]
 \item \[   {\bf C}=(C_1,C_2,C_3,0),\ (C_2^2+C_3^2\ne 0),\quad O=D1C\to O'=D1C, \quad \alpha =2, \]
  \item \[{\bf C}=(C_1,C_2,C_3,1),\,(4C_1\ne C_2^2+C_3^2),\ O=D3A\to O'=E_{11},\]
  \[(C_1\ne 0,C_2^2+C_3^2=0)\ \alpha =2,\]
  \[ \qquad O=D3A\to O'=D1C,\ (C_2^2+C_3^2\ne 0)\quad \alpha =2,\]
  \item \[{\bf C}=(C_1,C_2,C_3,C_4), \ O=K[0]\to O'=D1C,\quad \alpha=2,\]
 \end{enumerate}
 
 \subsection{Contraction [1,1,1,1] to [3,1] applied to conformal St\"ackel transforms of system V[0].}
 The target systems are conformal
 St\"ackel transforms of $V_{[0]}$. Partial  results are: 
 \begin{enumerate}
 \item \[ {\bf C}=((C_2^2+C_3^2)/4,C_2,C_3,1),\quad O=E_{20}\to O'=E_{3}',\quad \alpha =2,\]
  \item\[ {\bf C}=(C_1,1,\pm i,0),\quad O=E_{11}\to O'=E_{3}', ({\rm generic})\quad \alpha =2,\]
  \[\qquad O=E_{11}\to O'=D1C, ({\rm spacial\ case})\quad \alpha =3,\]
   \item \[ {\bf C}= (1,0,0,0), \quad O=E_{3}'\to O'=E_{3}',\quad \alpha =2\]
 \item \[   {\bf C}=(C_1,C_2,C_3,0),\ (C_2^2+C_3^2\ne 0),\quad\]
 \[O=D1C\to O'=E_3',({\rm generic}) \quad \alpha =2, \]
 \[ \qquad O=D1C\to O'=D1C,({\rm special\ case}) \quad \alpha =3, \]
  \item \[{\bf C}=(C_1,C_2,C_3,1),\,(4C_1\ne C_2^2+C_3^2),\]
  \[O=D3A\to O'=E_{3}',\ ({\rm generic})\ \alpha =2,\]
   \item \[{\bf C}=(C_1,C_2,C_3,C_4), \ O=K[0]\to O'=E_3',\quad \alpha=2,\]
 \end{enumerate}

 \subsection{Contraction [1,1,1,1] to [4] applied to conformal St\"ackel transforms of system V[0].}
 The target systems are conformal
 St\"ackel transforms of $V_{[0]}$. Partial results are: 
 \begin{enumerate}
 \item \[ {\bf C}=((C_2^2+C_3^2)/4,C_2,C_3,1),\quad O=E_{20}\to O'=E_{3}',\quad \alpha =6,\]
  \item\[ {\bf C}=(C_1,1,\pm i,0),\quad O=E_{11}\to O'=E_{3}', \quad \alpha =6,\]
    \item \[ {\bf C}= (1,0,0,0), \quad O=E_{3}'\to O'=E_{3}',\quad \alpha =6\]
 \item \[   {\bf C}=(C_1,C_2,C_3,0),\ (C_2^2+C_3^2\ne 0),\quad O=D1C\to O'=E_3', \quad \alpha =6, \]
  \item \[{\bf C}=(C_1,C_2,C_3,1),\,(4C_1\ne C_2^2+C_3^2),\ O=D3A\to O'=E_{3}',\alpha =6,\]
   \item \[{\bf C}=(C_1,C_2,C_3,C_4), \ O=K[0]\to O'=E_3',\quad \alpha=6,\]
 \end{enumerate}

 \subsection{Contraction [2,2] to [4] applied to conformal St\"ackel transforms of system V[0].}
 The target systems are conformal
 St\"ackel transforms of  $V_{[0]}$.  
 Partial results are: 

 \begin{enumerate}
 \item \[ {\bf C}=((C_2^2+C_3^2)/4,C_2,C_3,1),\quad O=E_{20}\to O'=E_{3}',\quad \alpha =4,\]
  \item\[ {\bf C}=(C_1,1,\pm i,0),\quad O=E_{11}\to O'=E_{3}', ({\rm generic})\quad \alpha =4,\]
  \[ \qquad O=E_{11}\to O'=E_{3}', ({\rm special case})\quad \alpha =5,\]
    \item \[ {\bf C}= (1,0,0,0), \quad O=E_{3}'\to O'=E_{3}',\quad \alpha =4\]
 \item \[   {\bf C}=(C_1,C_2,C_3,0),\ (C_2^2+C_3^2\ne 0),\quad\]
                                                               \[
 O=D1C\to O'=E_3',({\rm generic}) \quad \alpha =4, \]
  \item \[{\bf C}=(C_1,C_2,C_3,1),\,(4C_1\ne C_2^2+C_3^2),\]
  \[O=D3A\to O'=E_{3}',\ ({\rm generic})\ \alpha =4,\]
   \item \[{\bf C}=(C_1,C_2,C_3,C_4), \ O=K[0]\to O'=E_3',\quad \alpha=4,\]
 \end{enumerate}

 \subsection{Contraction [2,1,1] to [3,1] applied to conformal St\"ackel transforms of system V[0].}
 The target systems are conformal
 St\"ackel transforms of $V_{[0]}$. Partial results:
  \begin{enumerate}
 \item \[ {\bf C}=((C_2^2+C_3^2)/4,C_2,C_3,1),\quad O=E_{20}\to O'=E_{3}'\quad \alpha =6,\]
  \item\[ {\bf C}=(C_1,1,\pm i,0),\quad O=E_{11}\to O'=E_{3}', ({\rm generic})\quad \alpha =6,\]
  \item \[ {\bf C}= (1,0,0,0), \quad O=E_{3}'\to O'=E_{3}',\quad \alpha =6\]
 \item \[   {\bf C}=(C_1,C_2,C_3,0),\ (C_2^2+C_3^2\ne 0),\quad\]
 \[O=D1C\to O'=E_3', ({\rm generic})\quad \alpha =6, \]
  \item \[{\bf C}=(C_1,C_2,C_3,1),\,(4C_1\ne C_2^2+C_3^2),\]
  \[O=D3A\to O'=E_{3}',\ ({\rm generic})\ \alpha =6,\]
   \item \[{\bf C}=(C_1,C_2,C_3,C_4), \ O=K[0]\to O'=E_3',\quad \alpha=6,\]
 \end{enumerate}
 
\section{Summary of the 8 Laplace superintegrable systems with nondegenerate potentials}
All systems are  of the form $\left(\sum_{j=1}^4\partial_{x_j}^2+V({\bf x})\right)\Psi=0$, or $\left(\partial_x^2+\partial_y^2+{\tilde V}\right)\Psi=0$
as a flat space system in Cartesian coordinates. The potentials are:

\be\label{V[1111norm']} V_{[1,1,1,1]}=\frac{a_1}{x_1^2}+\frac{a_2}{x_2^2}+\frac{a_3}{x_3^2}+\frac{a_4}{x_4^2},\ee

 \[{\tilde V}_{[1,1,1,1]}=\frac{a_1}{x^2}+\frac{a_2}{y^2}+\frac{4a_3}{(x^2+y^2-1)^2}-\frac{4a_4}{(x^2+y^2+1)^2},\]

\be\label{V211norm'} V_{[2,1,1]}=\frac{a_1}{x_1^2}+\frac{a_2}{x_2^2}+\frac{a_3(x_3-ix_4)}{(x_3+ix_4)^3}+\frac{a_4}{(x_3+ix_4)^2},\ee
\[{\tilde  V}_{[2,1,1]}=\frac{a_1}{x^2}+\frac{a_2}{y^2}-a_3(x^2+y^2)+a_4,\]

\be\label{V[22norm']} V_{[2,2]}=\frac{a_1}{(x_1+ix_2)^2}+\frac{a_2(x_1-ix_2)}{(x_1+ix_2)^3}
+\frac{a_3}{(x_3+ix_4)^2}+\frac{a_4(x_3-ix_4)}{(x_3+ix_4)^3},\ee
\[{\tilde  V}_{[2,2]}=\frac{a_1}{(x+iy)^2}+\frac{a_2(x-iy)}{(x+iy)^3}
+a_3-a_4(x^2+y^2),\]

\be\label{V[31]norm'} V_{[3,1]}=\frac{a_1}{(x_3+ix_4)^2}+\frac{a_2x_1}{(x_3+ix_4)^3}
+\frac{a_3(4{x_1}^2+{x_2}^2)}{(x_3+ix_4)^4}+\frac{a_4}{{x_2}^2},\ee
\[ {\tilde V}_{[3,1]}=a_1-a_2x
+a_3(4x^2+{y}^2)+\frac{a_4}{{y}^2},\]

\be\label{V[4]norm'} V_{[4]}=\frac{a_1}{(x_3+ix_4)^2}+a_2\frac{x_1+ix_2}{(x_3+ix_4)^3}
+a_3\frac{3(x_1+ix_2)^2-2(x_3+ix_4)(x_1-ix_2)}{(x_3+ix_4)^4}\ee
\[+a_4\ \frac{4(x_3+ix_4)(x_3^2+x_4^2)+2(x_1+ix_2)^3}{(x_3+ix_4)^5},\]
\[ {\tilde V}_{[4]}=a_1-a_2(x+iy)
+a_3\left(3(x+iy)^2+2(x-iy)\right)
-a_4\left(4(x^2+y^2)+2(x+iy)^3\right),\]

\be\label{V[0]norm'} V_{[0]}=\frac{a_1}{(x_3+ix_4)^2}+\frac{a_2x_1+a_3x_2}{(x_3+ix_4)^3}+a_4\frac{x_1^2+x_2^2}{(x_3+ix_4)^4},\ee
\[ {\tilde V}_{[0]}=a_1-(a_2x+a_3y)+a_4(x^2+y^2),\]

\be\label{Varb'} V_{arb}=\frac{1}{(x_3+ix_4)^2}f(\frac{-x_1-ix_2}{x_3+ix_4}),\ee
\[ {\tilde V}_{arb}=f({x+iy}),\ f\ {\rm arbitrary}\]

\be\label{V[1]norm'}V(1)=a_1\frac{1}{(x_1+ix_2)^2}+a_2\frac{1}{(x_3+ix_4)^2}
+a_3\frac{(x_3+ix_4)}{(x_1+ix_2)^3}+a_4\frac{(x_3+ix_4)^2}{(x_1+ix_2)^4},\ee
\[{\tilde V}(1)=\frac{a_1}{(x+iy)^2}+a_2
-\frac{a_3}{(x+iy)^3}+\frac{a_4}{(x+iy)^4},\]
This is a special case of (\ref{Varb'}).
\be\label{V[2]norm'}
V(2)'=a_1\frac{1}{(x_3+ix_4)^2}+a_2\frac{(x_1+ix_2)}{(x_3+ix_4)^3}
+a_3\frac{(x_1+ix_2)^2}{(x_3+ix_4)^4}+a_4\frac{(x_1+ix_2)^3}{(x_3+ix_4)^5},\ee
\[ {\tilde V}(2)'=a_1+a_2(x+iy)
+a_3(x+iy)^2+a_4(x+iy)^3.\]
This is a special case of (\ref{Varb'}).

\section{Summary of St\"ackel equivalence classes of Helmholtz superintegrable systems}
\begin{enumerate}
\item{$[1,1,1,1]$}:  \[ S9,S8,S7,D4B,D4C, K[1,1,1,1]\]
\item{$[2,1,1]$}:   \[ S4,S2,E1,E16,D4A,D3B,D2B,D2C,K[2,1,1]\]
\item{$[2,2]$}:\[E8,E17,E7,E19,D3C,D3D,K[2,2]\]
\item{$[3,1]$}: \[ S1,E2,D1B,D2A,K[3,1]\]
\item{$[4]$}: \[E10,E9,D1A,K[4]\]
\item{$[0]$}: \[ E20,E11,E3',D1C,D3A,K[0]\]
\item{$(1)$}:\[{\rm special\  cases\ of}\ E15\]
\item{$(2)$}: \[{\rm special\  cases\ of}\ E15\]
\end{enumerate}

\subsection{Summary of B\^ocher contractions of Laplace  systems}\label{4} This is a summary of the results of applying each 
of the B\^ocher contractions to each of the Laplace conformally superintegrable systems.
{\small
 \begin{enumerate}
\item{$[1,1,1,1]\to [2,1,1]$ contraction}: \[ V_{[1,1,1,1]}\downarrow V_{[2,1,1]};\ V_{[2,1,1]}\downarrow V_{[2,1,1]},V_{[2,2]},V_{[3,1]};\
V_{ [2,2]}\downarrow V_{ [2,2]},V_{[0]};\
V_{[3,1]}\downarrow V_{(1)},V_{[3,1]};\]
\[ V_{[4]}\downarrow V_{[0]},V_{(2)};\  V_{[0]}\downarrow V_{[0]};\  V_{(1)}\downarrow V_{(1)},V_{(2)};\   V_{(2)}\downarrow V_{(2)}. \]
\item{$[1,1,1,1]\to [2,2]$ contraction}: \[ V_{[1,1,1,1]}\downarrow V_{[2,2]};\ V_{[2,1,1]}\downarrow V_{ [2,2]},\,{\rm\ special\ case\ of\ } E15;\
V_{ [2,2]}\downarrow V_{[2,2]},V_{[0]};\
V_{[3,1]}\downarrow V_{(1)},\, {\rm special\ case\ of\ } E_{15};\]
\[V_{[4]}\downarrow V_{ (2)};\  V_{[0]}\downarrow V_{[0]};\   V_{(1)}\downarrow V_{(1)},{\rm\ special\ case\ of\ } E15;\   V_{(2)}\downarrow V_{(2)}. \]
\item{$[2,1,1]\to [3,1]$ contraction}: \[ V_{[1,1,1,1]}\downarrow  V_{[3,1]};\, V_{[2,1,1]}\downarrow  V_{[3,1]},V_{[0]};\,
 V_{[2,2]}\downarrow  V_{ [0]},\quad
 V_{[3,1]}\downarrow  V_{[3,1]},V_{[0]};\,
 V_{[4]}\downarrow  V_{[0]};\]
\[  V_{ [0]}\downarrow  V_{[0}];\,    V_{(1)}\downarrow  V_{(2)};\,    V_{(2)}\downarrow  V_{(2)}. \]
\item{$[1,1,1,1]\to [4]$ contraction}: \[  V_{[1,1,1,1]}\downarrow  V_{[4]};\,  V_{[2,1,1]}\downarrow  V_{[4]};\,
 V_{ [2,2]}\downarrow  V_{[0]};\,
 V_{[3,1]}\downarrow  V_{[4]};\,
 V_{[4]}\downarrow  V_{[0]},V_{[4]};\,   V_{[0]}\downarrow  V_{[0]};\]
 \[V_{(1)}\downarrow  V_{ (2)};\,    V_{(2)}\downarrow  V_{(2)}; \]
\item{$[2,2]\to [4]$ contraction}: \[  V_{ [1,1,1,1]}\downarrow  V_{[4]};\,  V_{[2,1,1]}\downarrow  V_{[4]},V_{(2)};\,
  V_{[2,2]}\downarrow  V_{[4]},V_{[0]};\,
 V_{[3,1]}\downarrow  V_{(2)};\,
 V_{[4]}\downarrow  V_{(2)};\] \[  V_{[0]}\downarrow  V_{[0]},V_{(2)};\,  V_{ (1)}\downarrow  V_{(2)};\,    V_{(2)}\downarrow  V_{(2)}; \]
\item{$[1,1,1,1]\to [3,1]$ contraction}: \[  V_{[1,1,1,1]}\downarrow  V_{[3,1]},\  V_{[2,1,1]}\downarrow  V_{[3,1]},V_{[0]};\,
  V_{[2,2]}\downarrow  V_{ [0]};\,
 V_{[3,1]}\downarrow  V_{[3,1]},V_{[0]};\,
 V_{[4]}\downarrow  V_{[0]},\   V_{[0]}\downarrow  V_{[0]},\]
 \[V_{ (1)}\downarrow  V_{(2)},\    V_{(2)}\downarrow  V_{(2)}. \]
\end{enumerate} }

\section{Summary of Helmholtz contractions} The superscript for each targeted Helmholtz system is the value of $\alpha$. In each table, corresponding to a single Laplace equation equivalence class, the top line is
a list of the Helmholtz systems in the class, and the lower lines are the target systems under the B\^ocher contraction.
{\small 
\bigskip
Contractions of  systems:
\be\label{Table1} \ba{clllllll}& $[1,1,1,1]$&{\rm equivalence}&{\rm class\ }\ & {\rm contractions}& &\\
\hline\\
{\rm contraction} &{S_9}&S_7&S_8&D_4B&D_4C&K[1111]\\
\hline\\
{[}1111]\downarrow[211]&E_1^2&S_4^0&S_4^0&E_1^2&S_4^0&D_4A^0\\
&S_2^0&S_2^0&E_{16}^0&D_4A^0&D_4A^0\\
&&&&S_2^0&\\
\hline\\
{[}1111]\downarrow[22]&E_7^2 &E_{19}^4 &E_{17}^4&E_7^2&E_{19}^1&E_7^2\\
&&E_7^2&E_{19}^4\\
& &E_{17}^2 &\\
\hline\\
{[}1111]\downarrow[31]&E_2^2&S_1^0&S_1^0&S_1^0&S_1^0&S_1^0\\
&S_1^0&E_2^2&E_2^2&E_2^2\\
\hline\\
{[}1111]\downarrow[4]&E_{10}^6&E_{10}^6&E_{10}^6&E_{10}^6&E_{10}^6&E_{10}^6\\
\hline\\
{[}22]\downarrow[4]&E_{10}^4&E_9^6&E_{10}^5&E_{10}^4&E_{10}^5&E_{10}^4\\
&&E_{10}^4&-\\
&&E_9^5&\\
\hline\\
{[}211]\downarrow[31]&E_2^6&S_1^0&S_1^0&S_1^0&S_1^0&S_1^0\\
&E_2^4&E_2^4&E_2^8&E_2^6\\
&S_1^0&&&E_2^4\\
\hline
\ea
\ee

\be\label{Table2} \ba{clllllllll}& $[2,1,1]$&{\rm equivalence}&{\rm class\ }\ & {\rm contractions}& &\\
\hline\\
{\rm contraction} &{S_4}&S_2&E_1&E_{16}&D_4A&D_3B&D_2B&D_2C&K[211]\\
\hline\\
{[}1111]\downarrow[211]&S_4^0&S_2^0&E_1^2&E_{16}^4&D_4A^0&E_1^2&S_2^0&S_4^0&S_4^0\\
&E_{17}^4&E_8^2&E_8^0&E_{17}^0&E_8^2&D_3C^0&E_8^0&E_{17}^0&D_3C^0\\
&S_1^0&S_1^0&E_2^2&E_2^2&S_1^0&E_2^2&S_1^0&S_1^0&S_1^0\\
&&E_2^2&&&&D_1B^3&E_2^2&&\\
\hline\\
{[}1111]\downarrow[22]&E_{17}^4 &E_{8}^2 &E_{8}^2&E_{17}^4&E_{7}^2&E_8^2&E_7^2&E_{19}^4&E_7^2\\
&&&&&E_8^2&E_{17}^2&E_8^2&E_{17}^4&\\
\hline\\
{[}1111]\downarrow[31]&S_1^0&S_1^0&E_2^2&E_2^2&S_1^0&E_2^2&E_1^2&S_1^0&S_1^0\\
&&&&&&D_1B^3&&&\\
&{E_3'}^2&{E_3'}^2&{E_3'}^2&{E_3'}^2&{E_3'}^2&{E_3'}^2&{E_3'}^2&{E_3'}^2&{E_3'}^2\\
&&&&&D_1C^3&D_1C^3&D_1C^3&&\\
\hline\\
{[}1111]\downarrow[4]&E_{10}^6&E_{10}^6&E_{10}^6&E_{10}^6&E_{10}^6&E_{10}^6&E_{10}^6&E_{10}^6&E_{10}^6\\
&&&&&E_9^8&E_9^8&E_9^8&E_9^8&\\
\hline\\
{[}22]\downarrow[4]&E_{10}^5&E_{10}^4&E_{10}^4&E_{10}^5&E_{10}^4&E_{10}^4&E_{10}^4&E_{10}^4&E_{10}^4\\
&&&&&&E_{10}^5&E_{10}^5&&\\
&\qquad {\rm \mbox{St\"{a}ckel}}& {\rm  transforms}& {\rm of}& V(2)&\\
\hline\\
{[}211]\downarrow[31]&S_1^0&S_1^0&E_2^6&E_2^8&S_1^0&E_2^6&S_1^0&S_1^0&S_1^0\\
&&E_2^5&&&&&E_2^5&&\\
&{E_3'}^8&{E_3'}^6&{E_3'}^4&{E_3'}^4&{E_3'}^6&{E_3'}^6&{E_3'}^4&{E_3'}^4&{E_3'}^4\\
\hline
\ea
\ee

\be\label{Table3} \ba{clllllll}& $[2,2]$&{\rm equivalence}&{\rm class\ }\ & {\rm contractions}& &\\
\hline\\
{\rm contraction} &E_8&E_{17}&E_7&E_{19}&D_3C&D_3D&K[22]\\
\hline\\
{[}1111]\downarrow[211]&E_8^0&E_{17}^0&E_7^0&E_{19}^0&D_3C^0&E_7^2&D_3C^0\\
&{E_3'}^2&{E_3'}^2&{E_3'}^2&{E_3'}^2&{E_3'}^2&{E_3'}^2&{E_3'}^2\\
\hline\\
{[}1111]\downarrow[22]&E_{8}^2 &E_{17}^4 &E_{7}^2&E_{19}^4&E_{8}^2&E_8^2&E_7^2\\
&{E_3'}^2&E_{11}^2&{E_3'}^2&E_{11}^2&E_{11}^2&E_{11}^2&E_{11}^2\\
\hline\\
{[}1111]\downarrow[31]&{E_3'}^2&{E_3'}^2&{E_3'}^2&{E_3'}^2&{E_3'}^2&{E_3'}^2&{E_3'}^2\\
&&&&E_{11}^4,E_{20}^4&D_1C^3&D_1C^3&\\
\hline\\
{[}1111]\downarrow[4]&{E'}_{3}^6&{E'}_{3}^6&{E'}_{3}^6&{E'}_{3}^6&{E'}_{3}^6&{E'}_{3}^6&{E'}_{3}^6\\
&&&E_{11}^8&E_{11}^8&E_{11}^8&E_{11}^8&\\
\hline\\
{[}22]\downarrow[4]&E_{10}^4&E_{10}^5&E_{10}^4&E_{10}^5&E_{10}^4&E_{10}^4&E_{10}^4\\
&&&E_9^5&E_9^6&&&\\
&{E_3'}^2&E_{11}^1&{E_3'}^2&E_{11}^1&E_{11}^1&E_{11}^1&E_{11}^1\\
&&&E_{11}^3&E_{20}^4&&&\\
\hline\\
{[}211]\downarrow[31]&{E'_3}^4&{E'}_3^4&{E'_3}^2&{E_3'}^2&{E'_3}^4&D_1C^2&D_1C^2\\
&{E_3'}^6&{E_3'}^6&{E_3'}^6&{E_{20}}^4&{E_3'}^6&{E_3'}^6&{E_3'}^6\\
&&&&&D_1C^9&&\\
\hline
\ea
\ee

\be\label{Table4} \ba{clllllll}& $[3,1]$&{\rm equivalence}&{\rm class\ }\ & {\rm contractions}& &\\
\hline\\
{\rm contraction} &S_1&E_{2}&D_1B&D_2A&K[31]\\
\hline\\
{[}1111]\downarrow[211]&\qquad {\rm \mbox{St\"{a}ckel}}& {\rm  transforms}& {\rm of}& V(1)&\\
&S_1^0&E_2^2&E_2^2&E_2^2&S_1^0\\
&&&D_1B^3&D_2A^4&\\
\hline\\
{[}1111]\downarrow[22]&\qquad {\rm \mbox{St\"{a}ckel}}& {\rm  transforms}& {\rm of}& V(1)\\
\hline\\
{[}1111]\downarrow[31]&S_1^0&E_2^2&E_2^2&E_2^2&S_1^0\\
&&&D_1B^3&&&\\
&{E_3'}^2&{E_3'}^2&{E_3'}^2&{E_3'}^2&{E_3'}^2\\
&&&D_1C^3&&\\
\hline\\
{[}1111]\downarrow[4]&{E}_{10}^6&{E}_{10}^6&{E}_{10}^6&{E}_{10}^6&{E}_{10}^6\\
&&&E_9^8&&\\
\hline\\
{[}22]\downarrow[4]&\qquad {\rm \mbox{St\"{a}ckel}}& {\rm  transforms}& {\rm of}& V(2)\\
\hline\\
{[}211]\downarrow[31]&{S_1}^0&{E}_2^6&{E_2}^6&{E_2}^6&{S_1}^0\\
&&E_2^2&S_1^1&S_1^0&\\
&{E_3'}^4&{E_3'}^6&{E_3'}^6&{E_3'}^6&{E_3'}^4\\
\hline
\ea
\ee

\be\label{Table5} \ba{clllllll}& $[4]$&{\rm equivalence}&{\rm class\ }\ & {\rm contractions}& &\\
\hline\\
{\rm contraction} &E_{10}&E_{9}&D_1A&K[4]\\
\hline\\
{[}1111]\downarrow[211]&{E_3'}^2&E_{11}^2&E_{20}^4&{E_3'}^2\\
&&{E_3'}^2&{E_3'}^2&&\\
&&\qquad {\rm \mbox{St\"{a}ckel}}& {\rm  transforms}& {\rm of}& V(2)\\
\hline\\
{[}1111]\downarrow[22]&\qquad {\rm \mbox{St\"{a}ckel}}& {\rm  transforms}& {\rm of}& V(2)\\
&{E_3'}^2&{E_3'}^2&D_1C^2&D_3A^2\\
\hline\\
{[}1111]\downarrow[31]&{E_3'}^2&{E_3'}^2&{E_3'}^2&{E_3'}^2\\
&E_{11}^2&&&\\
\hline\\
{[}1111]\downarrow[4]&{E_3'}^6&{E_3'}^6&{E_3'}^6&{E_3'}^6\\
&E_{11}^8&&&\\
&E_{10}^6&E_{10}^6&E_{10}^6&E_{10}^6\\
&E_9^8&&&\\
\hline\\
{[}22]\downarrow[4]&\qquad {\rm \mbox{St\"{a}ckel}}& {\rm  transforms}& {\rm of}& V(2)\\
\hline\\
{[}211]\downarrow[31]&{E_3'}^1&{E_3'}^1&{E_3'}^{-1}&{E_3'}^{-1}\\
&{E_3'}^4&{E_3'}^5&{E_3'}^4&{E_3'}^3\\
&{E_3'}^6&{E_3'}^6&{E_3'}^6&{E_3'}^6\\
\hline
\ea
\ee

\be\label{Table6} \ba{clllllll}& $[0]$&{\rm equivalence}&{\rm class\ }\ & {\rm contractions}& &\\
\hline\\
{\rm contraction} &E_{20}&E_{11}&E_3'&D_1C&D_3A&K[0]\\
\hline\\
{[}1111]\downarrow[211]&{E_3'}^2&{E_{3}'}^2&{E_3'}^2&{E_3'}^2&{E_3'}^2&{E_3'}^2\\
&E_{11}^3&E_{11}^3&&D_1C^3&D_1C^3&&\\
\hline\\
{[}1111]\downarrow[22]&E_{11}^2&{E_{11}}^2&{E_3'}^2&E_{11}^2&E_{11}^2&E_{11}^2\\
&&&&&{E_3'}^2&{E_3'}^2\\
\hline\\
{[}1111]\downarrow[31]&{E_3'}^2&{E_3'}^2&{E_3'}^2&{E_3'}^2&{E_3'}^2&{E_3'}^2\\
&&&&D_1C^3&D_1C^3&\\
\hline\\
{[}1111]\downarrow[4]&{E_3'}^6&{E_3'}^6&{E_3'}^6&{E_3'}^6&{E_3'}^6&{E_3'}^6\\
&E_{11}^8&E_{11}^8&&E_{11}^8&E_{11}^8&\\
\hline\\
{[}22]\downarrow[4]&{E_3'}^4&{E_3'}^4&{E_3'}^4&{E_3'}^4&{E_3'}^4&{E_3'}^4\\
&{E_{11}}^5&E_{11}^5&&E_{11}^5&\\
\hline\\
{[}211]\downarrow[31]&{E_3'}^6&{E_3'}^6&{E_3'}^6&{E_3'}^6&{E_3'}^6&{E_3'}^6\\
&&&&D_1C^9&&\\
\hline
\ea
\ee }

\section{Acknowledgement}
This work was partially supported by a grant from the Simons Foundation (\# 208754 to Willard Miller, Jr).


\begin{thebibliography}{99}
%-----------------------------------------------------------------------

\bibitem{CK2014} Capel J.J. and  Kress J.M., Invariant Classification of Second-order Conformally Flat Superintegrable Systems,
J. Phys.A: Math. Theor. {\bf 47} (2014), 495202.

\bibitem{Kress2007}
J. M. Kress, Equivalence of Superintegrable Systems in Two Dimensions, {\it Physics of Atomic Nuclei},  {\bf 70}, No. 3, pp. 560–566, (2007).

\bibitem{Bromwich} J.T.A. Bromwich, Quadratic forms and their classification by means of invariant factors, Cambridge Tracts \# 3, 
Cambridge University Press, 1904.

\bibitem{Koenigs}
Koenigs, G., Sur les g\'eod\'esiques a int\'egrales quadratiques. A note
appearing in ``Lecons sur la th\'eorie g\'en\'erale des
surfaces''. G. Darboux. Vol 4, 368-404, {\it Chelsea Publishing} 1972.
%-----------------------------------------------------------------------

\bibitem{KM2014}   Kalnins E. G. and Miller, W. Jr., Quadratic algebra contractions and 2nd order superintegrable systems,  {\it Anal. Appl.}
{\bf 12}, 583-612,  (2014). DOI: 10.1142/S0219530514500377.

\bibitem{Post2011a} Post S., Coupling Constant Metamorphosis, the St\"ackel Transform and Superintegrability, in Symmetries in Nature: Symposium in Memoriam Marcos Moshinsky (Cuernavaca, Mexico, August 9-14, 2010),
Vol. 1323, Editors L. Benet,P. Hess, J. Torres, K. Wolf, AIP Conference Proceedings, 2011, 265?274.

\bibitem{Post2011b}
 Post S., Models of quadratic algebras generated by superintegrable systems in 2D, SIGMA {\bf 7}
(2011), 03

\bibitem{CKP2015} J. Capel, J. Kress, S. Post, Invariant Classification and Limits of Maximally Superintegrable Systems in 3D, 
arXiv:1501.06601 [math-ph], 2015

\bibitem{KKMW}
Kalnins E.\ G., Kress J.\ M., Miller, W.\ Jr. and Winternitz P.,
{\it Superintegrable systems in Darboux spaces.}
{\it J.~Math.~Phys.},\  V.44, 5811--5848,  (2003).
% 
%-----------------------------------------------------------------------

\bibitem{KKM20051}
E.~G.~Kalnins, J.~M.~Kress, and  W.~Miller,~Jr. 
Second  order superintegrable systems in conformally
flat spaces.  III: 3D classical structure theory. {\it
  J. Math. Phys.} {\bf 46}, 103507
(2005).
 
\bibitem{Kalnins} E.G. Kalnins, 
Separation of Variables for Riemannian Spaces of Constant
Curvature,
 Pitman, Monographs and Surveys in Pure and Applied Mathematics
{\bf 28},
 Longman, Essex, England,
1986
 
 


 \bibitem{MPW2013}  Miller, W. Jr.,   Post, S. and  Winternitz, P..  Classical and Quantum Superintegrability with Applications , 
 {\it  J. Phys. A: Math. Theor.} {\bf 46}, (2013) 423001. 

%-----------------------------------------------------------------------
 

\bibitem{Bocher} B\^ocher, M., Ueber die  Reihenentwickelungender Potentialtheorie, B. G. Teubner, Leipzig 1894.

\bibitem{KMR1984}  E.G. Kalnins, W, Miller, Jr,  and G.J. Reid, Separation of variables for complex Riemannian spaces of constant curvature. I. 
Orthogonal separable coordinates for Snc and Enc, {\it Proc. R. Soc. Lond. A }, {\bf 394}, (1984), pp. 183-206.

\bibitem{KKMP2011}  E.  G. Kalnins, J. M. Kress, W. Miller, Jr.  and S. Post. 
Laplace-type  equations as conformal superintegrable systems, {\it  Adv. Appl.  Math.} 2011.


\bibitem{Miller1977} W. Miller, Jr.,
 Symmetry and Separation of Variables,
 Addison-Wesley,
 Reading, Massachusetts,
1977


\bibitem{KKM20042}
Kalnins E.G., Kress J.M, and  Miller W.Jr., 
Second  order superintegrable systems in conformally
flat spaces.  II: The classical 2D St\"ackel transform. {\it J. Math. Phys.},
2005, {\bf V.46}, 053510.
%-----------------------------------------------------------------------



\bibitem{KKM20061}
E.~G.~Kalnins, J.~M.~Kress and  W.~Miller Jr. 
Second  order superintegrable systems in conformally
flat spaces.  V: 2D and 3D quantum systems. {\it
  J. Math. Phys.}, 2006, V.47, 093501.
%-----------------------------------------------------------------------

\bibitem{KMP2010} E. G. Kalnins, W. Miller, Jr  and S. Post,  
Coupling constant metamorphosis and Nth order symmetries in classical and quantum mechanics, 
{\it  J. Phys. A: Math. Theor.}, {\bf 43} (2010) 035202.


\bibitem{4} E.\ G\ Kalnins, J.\ M.\ Kress and W.\ Miller Jr. Second order superintegrable systems in conformally flat spaces I. 2D 
classical structure theory. {\it J.  Math.  Phys.}, {\bf 46}, 053509, (2005).

\bibitem{5} E.\ G.\ Kalnins, J.\ M.\ Kress and W.\ Miller Jr. Second order superintegrable systems in conformally flat spaces 
II. The classical 2D St\"ackel  transform. {\it J.
 Math. Phys,}, {\bf 46}, 053510, (2005).

\bibitem{7} E.\ G.\ Kalnins, J.\ M.\ Kress,W.\ Miller Jr.\  and G.\ S.\ Pogosyan. Completeness of superintegrability in two dimensional constant curvature 
spaces. {\it J. Phys. A Math Gen.} {\bf  34}, 4705, (2001).


\end{thebibliography}
\end{document}